\documentclass{aa}  
\usepackage[colorlinks=true]{hyperref}
\hypersetup{linkcolor=blue, citecolor=blue}
\usepackage[draft]{todonotes}
\usepackage{graphicx}
\usepackage{txfonts}
\usepackage[draft]{todonotes}
\usepackage{amsmath}

\begin{document} 
\newcommand{\vdag}{(v)^\dagger}
\newcommand\aastex{AAS\TeX}
\newcommand\latex{La\TeX}
%
%
\newcommand{\ColDens}{cm$^{-2}$}
\newcommand{\kms}{km~s$^{-1}$}
\newcommand{\mum}{$\mu$m}
%
%
\newcommand{\NCO}{$N_{CO}$}
\newcommand{\NHI}{$N_{{H{\sc I}}}$}
\newcommand{\NHmolecular}{$N_{{H}_{2}}$}
\newcommand{\NHtotal}{$N_{H}$}
%
%
\newcommand{\Av}{$A_{V}$}
\newcommand{\AvTotal}{$A_{V} ({H})$}
\newcommand{\AvAtomic}{$A_{V} ({H}{\sc I})$}
\newcommand{\AvMolecular}{$A_{V} ({H}_{2})$}
%
%
\newcommand{\IRASlong}{I$_{100}$}
\newcommand{\IRASshort}{I$_{60}$}
\newcommand{\ImidInfrared}{$I (\nu_{12})$}
%
%
\newcommand{\EBV}{$E(B - V)$}
\newcommand{\EBVstar}{$E(B - V)_{\star}$}
\newcommand{\Td}{$T_{d}$}
\newcommand{\TauDust}{$\tau_{353}$}
%
%
\newcommand{\TK}{$T_{K}$}
\newcommand{\TauHI}{$\tau_{{H}\,{\sc I}}$}
\newcommand{\WHI}{$W_{{H}\,_{\sc I}}$}
\newcommand{\WCO}{$W_{CO}$}
\newcommand{\MolecAbund}{$f_{H_{2}}$}
\newcommand{\XCO}{$X_{CO}$}
\newcommand{\fhmol}{$f_{{H}_{2}}$}
%
%
\newcommand{\form}{H$_2$CO}
\newcommand{\water}{H$_2$O}
\newcommand{\ammonia}{\mbox{{\rm NH}$_3$}}
\newcommand{\co}{$^{12}$CO}
\newcommand{\coo}{$^{13}$CO}
\newcommand{\cooo}{C$^{18}$O}
\newcommand{\coooo}{C$^{17}$O}
\newcommand{\hcop}{HCO$^+$}
\newcommand{\hcopi}{H$^{13}$CO$^+$}
\newcommand{\dcop}{DCO$^+$}
\newcommand{\nthp}{N$_2$H$^+$}
\newcommand{\ntdp}{N$_2$D$^+$}
\newcommand{\otwo}{O$_2$}
\def \18OI{[$^{18}$O\,{\sc i}]}
\def \17OI{[$^{18}$O\,{\sc i}]}
\def \HI{H\,{\sc i}}
\def \HII{H\,{\sc ii}}
\newcommand{\Hmol}{H$_{2}$}
%
%
\newcommand{\cplus}{C$^+$}
%
%
\def \Cplus{C$^{+}$}
\def \Catomic{C\,{\sc i}}
\def \CI{[C\,{\sc i}]}
\def \CII{[C\,{\sc ii}]}
\def \13CII{[$^{13}$C\,{\sc ii}]}
\def \NII{[N\,{\sc ii}]}
\def\NIII{[N\,{\sc iii}]}
\def \OI{[O\,{\sc i}]}
    \title{Constraining the \Hmol\ column densities in the diffuse interstellar medium using dust extinction and \HI\ data}
  

    \titlerunning{\Hmol\ column densities obtained from dust extinction and \HI\ data}

   \author{R. Skalidis \inst{1}\fnmsep\inst{2}, P. F. Goldsmith\inst{2}, P. F. Hopkins\inst{3} \and S. B. Ponnada \inst{3}
          }

   \institute{
   Owens Valley Radio Observatory, California Institute of Technology, MC 249-17, Pasadena, CA 91125, USA \\
    \email{skalidis@caltech.edu} 
    \and
    Jet Propulsion Laboratory, California Institute of Technology, 4800 Oak Grove Drive, Pasadena, CA 91109-8099, USA \\
    \email{paul.f.goldsmith@jpl.nasa.gov}
    \and
    California Institute of Technology, TAPIR, Mailcode 350-17, Pasadena, CA 91125, USA
    }

   \date{Received; accepted}


  \abstract
   {Carbon monoxide (CO) is a poor tracer of \Hmol\ in the diffuse interstellar medium (ISM), where most of the carbon is not incorporated into CO molecules unlike the situation at higher extinctions.}
   {We present a novel, indirect method to constrain \Hmol\ column densities (\NHmolecular) without employing CO observations. We show that previously--recognized nonlinearities in the relation between the extinction, \AvMolecular, derived from dust emission and the \HI\ column density (\NHI) are due to the presence of molecular gas.}
  {We employ archival \NHmolecular\ data, obtained from the UV spectra of stars, and calculate \AvMolecular\ towards these sight lines using 3D extinction maps. The following relation fits the data: log \NHmolecular\  = $1.38742~\left( \log{A_{V} (H_{2}}) \right)^{3} - 0.05359~\left( \log{A_{V} (H_{2}}) \right)^{2}  + 0.25722~\log{A_{V} (H_{2})} - 20.67191$. This relation is useful for constraining \NHmolecular\ in the diffuse ISM as it requires only \NHI\ and dust extinction data, which are both easily accessible. In $95\%$ of the cases, the estimates produced by the fitted equation have deviations under a factor of $3.5$.  We construct a \NHmolecular\ map of our Galaxy and compare it to the CO integrated intensity (\WCO) distribution. }
   {We find that the average ratio (\XCO) between \NHmolecular\ and \WCO\ is approximately equal to $2 \times 10^{20}$~\ColDens~(K~km s$^{-1}$)$^{-1}$, consistent with previous estimates. However, we find that the \XCO\ factor varies by orders of magnitude on arcminute scales between the outer and the central portions of molecular clouds. For regions with \NHmolecular~$\gtrsim 10^{20}$~\ColDens, we estimate that the average \Hmol\ fractional abundance,  \fhmol=~2~\NHmolecular~/~(2\NHmolecular\ + \NHI), is $0.25$. Multiple (distinct) largely atomic clouds are likely found along high--extinction sightlines (\Av~$\geq 1$~mag), hence limiting \fhmol\ in these directions.}
   {More than $50 \%$ of the lines of sight with \NHmolecular~$\geq 10^{20}$~\ColDens\ are untraceable by CO with a $J$ = 1--0 sensitivity limit \WCO~=~$1$~K~\kms.}
   

   \maketitle
%
\section{Introduction}

Molecular hydrogen is the most abundant molecule in the Universe, rendering it one of the most, if not the most, important chemical constituents. Molecular hydrogen's lack of a permanent electric dipole moment results in its rotational levels being connected only by quadrupole transitions. Their weakness, together with the large spacing of its rotational levels, results in rotational emission from \Hmol\ being extremely weak at the temperatures typically encountered in molecular regions and is only rarely observed in infrared emission.

An alternative method to observe molecular hydrogen is through absorption lines; molecular hydrogen has several Lyman--Werner absorption lines at UV wavelengths (11.2 - 13.6 eV). The column density of \Hmol\ can be determined by measuring the depths of the Lyman--Werner absorption lines against UV--bright background sources \citep{savage_1977,savage_sembach_1996.review.on.ISM.abundances.using.HST, cartledge_2004.abundance.using.HST, shull_2021, mangum_2015.how.to.observe.H2}.

The direct measurement of \NHmolecular, either through emission or absorption, requires space observatories. The MIRI spectrometer onboard JWST \citep{rieke_2015.miri.intro, bouchet_2015.miri.instrumentation.methods, kendrew_2015.miri.low.res.spectrometer, boccaletti_2015.miri.predcted.performance, wells_2015.medium.res.spectrometer, glasse_2015.sensitivity, gordon_2015.data.reduction,wright_2023.miri.inflight.performance} is sensitive to \Hmol\ rotational transitions \citep[e.g.,][]{armus_2023}, while the absorption lines can be detected in spectra obtained with FUSE or HST \citep{savage_1977, 1978ApJ_BSD, shull_2000.ISM.NH2.using.UV.spectra, sofia_2005.galactic.sightlines, sheffer_2008,rachford_2002, 2009ApJS_Rachford, Gillmon2006, shull_2021}. The necessity of space observatories makes \NHmolecular\ measurements expensive and thus relatively sparse. For this reason, various indirect techniques have been developed to estimate \NHmolecular. We summarize the most widely applied indirect methods in the following sections. 

\subsection{Dust intensity at far infrared wavelengths}

Emission from dust at far--infrared wavelengths ($\lambda \gtrsim 100$ \mum) traces cold molecular gas; the dust in dark and giant molecular clouds emits thermal radiation as a gray--body. Using multi--wavelength observations of dust emission, the dust spectral energy distribution (SED) can be obtained and fitted analytically with the following free parameters \citep[e.g.,][]{hildebrand_1983.review.determination.ISM.mass.using.far.infrared.data, draine_2007, paradis_2010.spectral.index.variations, paradis_2023, kirk_2010, martinPG_2012, planck_collaboration_2014.all_sky.dust.model.miville,schnee_2006.NH2.using.dust.far.infrared, juvela_2018.SED.fitting}: 1) dust temperature, and 2) dust grain emissivity, which, assuming a dust-to-gas mass ratio, can be converted to \NHmolecular. Dust SED fitting  was the most common strategy applied to analyze Herschel images to obtain \NHmolecular\ towards several nearby interstellar clouds \citep[e.g.,][]{, andre_2010, miville_2010_polaris_PS, palmeirim_2013_striations, konyves_2010.presterllar.cores.Aquilla, konyves_2015.dense.cores.Aquila, cox_2016}.

\subsection{CO - \Hmol\ conversion factor (\XCO)}

Carbon monoxide is an abundant molecule in the ISM. It is abundant in regions where \NHmolecular~$\ge 10^{20}$~\ColDens, as self-shielding there is sufficient to prevent rapid photo-dissociation of CO by the background ISM radiation field. Thus, when there is CO, there is also \Hmol, although the opposite is not always true \citep[e.g.,][]{Grenier_2005, planck_2011.dark.gas.estiamtes.bernard,barriault_2010.UrsaMajorCirrus.OH.traces.H2.but.not.CO, langer_2010, langer_2014.co.dark.GOTC+, langer_2015.CO.dark.CMZ, velusamy_2010, pineda_2013, goldsmith_2013, skalidis_2022, madden_2020.method.for.tracing.CO.dark.in.galaxies.using.cloudy.simulations, kalberla_2020, murray_2018, lebouteiller_2019.co.dark.magellanic.clouds, glover_2016.temperature.co.dark, seifried_2020.SILC.CO_dark.properties, DiLi_2015.dark.gas}.  

The CO integrated intensity (\WCO) can be used as a proxy for  \NHmolecular\ by employing the 
\XCO\ factor, defined as
\begin{equation}
    \label{eq:xco_factor}
    N_{H_{2}} = X_{CO}W_{CO}.
\end{equation}	

In our Galaxy, the average value of \XCO\ is $\langle X_{CO} \rangle \approx 2\times 10^{20}$~\ColDens~(K~\kms)$^{-1}$ \citep{bolato_2013}. When \XCO\ is measured, factor of two deviations are expected from the average value due to the assumptions employed to estimate \NHmolecular\ by various methods \citep{Dame2001,bolato_2013}.  

The conversion of \WCO\ to \NHmolecular\ using the Galactic average \XCO\ is meaningful only in a statistical sense, and when it is applied on Galactic scales. Within individual clouds in the interstellar medium (ISM), \XCO\ can vary by orders of magnitude \citep[e.g.,][]{pineda_2008.perseus.XCO.variations, lee_2014.XCO.perseus.molecular.cloud, pineda_2013, ripple_heyer_2013.XCO.variations} due to the sensitivity of \XCO\ to local ISM conditions, including gas density, turbulent line width, and the interstellar radiation field \citep{visser_2009,glover_2016,gong_2017}. 

\subsection{\Hmol\ not traced by CO (CO-dark \Hmol)}

Diffuse regions of the ISM (defined as having \Av~$\lesssim 1$ mag, and density $n \lesssim 100$~cm$^{-3}$) may contain
a significant fraction of molecular hydrogen but relatively little CO compared to the abundance of carbon \citep{Grenier_2005}. The reason is that the formation of \Hmol\ formation takes place at smaller extinctions than does that of CO due to the rapid onset of self--shielding at lower values of the extinction. 

In the transition phase where the \Hmol\ density has started rising but that of CO is still small, oxygen and carbon are largely in atomic form. In this region, then, \Hmol\ is primarily associated with ionized (\Cplus) or neutral (\Catomic) carbon, and not with CO. Molecular hydrogen in regions that are undetectable in CO are referred to as CO-dark \Hmol. The mass of CO-dark \Hmol\ is estimated to be $\sim 30 \%$ of the total (primarily \Hmol) mass \citep{pineda_2010,kalberla_2020}. 

\subsection{Dust extinction residuals}

Dust in the ISM is well mixed with gas \citep{boulanger_1996,2009ApJS_Rachford}. Thus, the extinction and reddening produced by dust is proportional to the total (atomic plus molecular hydrogen) gas column along the LOS \citep{1978ApJ_BSD}. All established methodologies (Sect.\ref{sec:dust-to-gas_ratio}) concur that in regions where the gas is atomic, the dust reddening, \EBV, is linearly correlated with \NHI\ \citep{liszt_2023.H2.properties.diffuse.ISM.CO, lenz_2017, Nguyen_2018.dgr.variation, shull_panopoulou_2023}; changes in the dust or hydrogen content cause \NHI~/~\EBV\ to vary \citep{liszt_2014.ratio.NHI.EBV.high.latitudes, shull_panopoulou_2023}. At high Galactic latitudes, \cite{lenz_2017} found that the linear correlation between \EBV\ and \NHI\ holds for \NHI\ $\lesssim 4 \times 10^{20}$ \ColDens. At higher \HI\ column densities, \EBV\ increases non-linearly with \NHI, behavior plausibly attributed to the presence of molecular gas. This implies that the residuals between the total dust extinction (or reddening) and the extinction induced by dust mixed with \HI\ gas should probe \NHmolecular\ \citep{planck_2011.dark.gas.estiamtes.bernard, paradis_2012, lenz_2015, kalberla_2020}.

This strategy for estimating \NHmolecular\ can only be applied if two conditions are satisfied: 1) the \NHI~/~\EBV\ ratio must be constrained, and 2) a mapping function that converts the extinction residuals to \NHmolecular\ exists. \NHI~/~\EBV\ has been constrained in the past by various authors \citep[e.g.,][]{liszt_2014.ratio.NHI.EBV.high.latitudes,lenz_2017, Nguyen_2018.dgr.variation, shull_2021}. The major challenge is to find an accurate way to convert the extinction residuals to \NHmolecular. This is the focus of this work.

\subsection{The present work}

In this paper we present a novel method to convert the dust extinction residuals to \NHmolecular\ and then construct a full-sky \NHmolecular\ map of our Galaxy without using CO observations. We use our \NHmolecular\ estimates to constraint some of the molecular gas properties of our Galaxy. 

We calculate the residuals between the total extinction and the extinction expected from dust mixed with \HI\ gas towards lines of sight (LOSs) with \NHmolecular\ measurements; these \NHmolecular\ measurements have been obtained from absorption lines (Sect.~\ref{sec:AvH2_NH2}). We find that the extinction residuals are well correlated with \NHmolecular, and fit them with a polynomial function. Then, we calculate the extinction residuals using full-sky \NHI\ and \EBV\ maps (Sect.~\ref{sec:full_sky_NH2_map}). Finally, we convert the extinction residuals to \NHmolecular\ using our fitted function in order to construct  a full--sky \NHmolecular\ map at $16\arcmin$ resolution. 

We compare our constructed \NHmolecular\ map with the CO intensity map of \cite{Dame2001} (Sect.~\ref{sec:comparison_pas_obs}), reproducing the previously--reported global value of \XCO. Our \NHmolecular\ estimates are consistent with a previous work, which combines dust extinction residuals and the \XCO\ factor (Sect.~\ref{sec:comparison_kalberla_map}). We construct a full-sky map of the molecular fractional abundance, and show that molecular hydrogen is less abundant than atomic hydrogen (Sect.~\ref{sec:molecular_gas_properties}). In Sect.~\ref{sec:co_dark_H2}, we determine the proportion of CO--dark molecular hydrogen in relation to the total amount of molecular hydrogen. We explore the \NHtotal~/~\Av\ ratio and compare against previous results (Sect.~\ref{sec:dust-to-gas_ratio}). In Sects.~\ref{sec:discussion}, and~\ref{sec:conclusions} we discuss our results in the context of the existing literature and present our conclusions. 

\begin{figure*}
    \centering
    \includegraphics[width=\hsize]{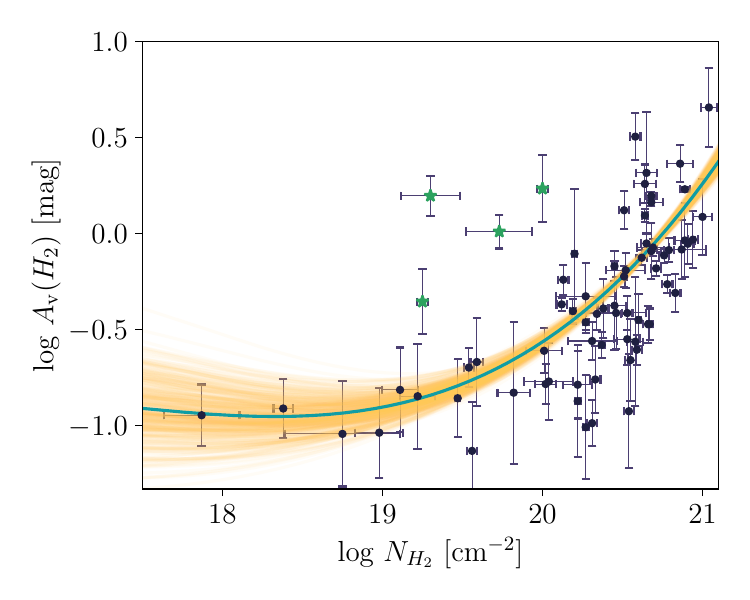}
    \caption{Extinction of dust mixed with molecular gas as a function of \NHmolecular. The \Hmol\ column densities were determined from UV observations towards stars, and the \AvMolecular\ were determined from the non-linearities of the \EBV\ versus \NHI\ relation (Eq.~\ref{eq:Av_molecular}). Green stars identify outlying points which were not considered in the fitting. At high column densities, \AvMolecular\ is strongly correlated with \NHmolecular, indicating that the nonlinear increase of dust reddening with respect to \NHI\ is induced by the presence of molecular gas. The orange curves were randomly sampled from the posterior distribution of the MCMC fitting, representing the fitting uncertainties. The green curve corresponds to the mean posterior profile that we employ to convert \AvMolecular\ to \NHmolecular\ (Eq.~\ref{eq:NH2_Av_fit}).}
    \label{fig:AvH2_NH2}
\end{figure*}

\begin{figure}
    \centering
    \includegraphics[width=\hsize]{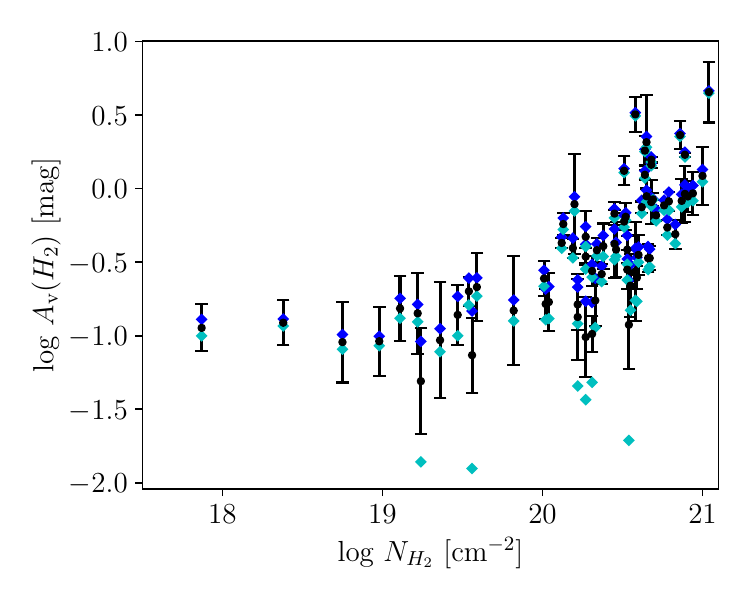}
    \caption{Same as Fig.~\ref{fig:AvH2_NH2} but for different \NHI~/\EBV\ ratios. The black points with errorbars correspond to our formal measurements using the \NHI~/\EBV\ constraint of \cite{lenz_2017}. Blue and cyan points correspond to measurements obtained when we assume that \NHI~/~\EBV~$\times 10^{-21}$ is equal to 10 and 8 \ColDens~mag$^{-1}$ respectively. Approximately all data points show statistically consistent results.}
    \label{fig:gas_dust_variations}
\end{figure}

\begin{figure*}
   \centering
   \includegraphics[width=0.97\hsize]{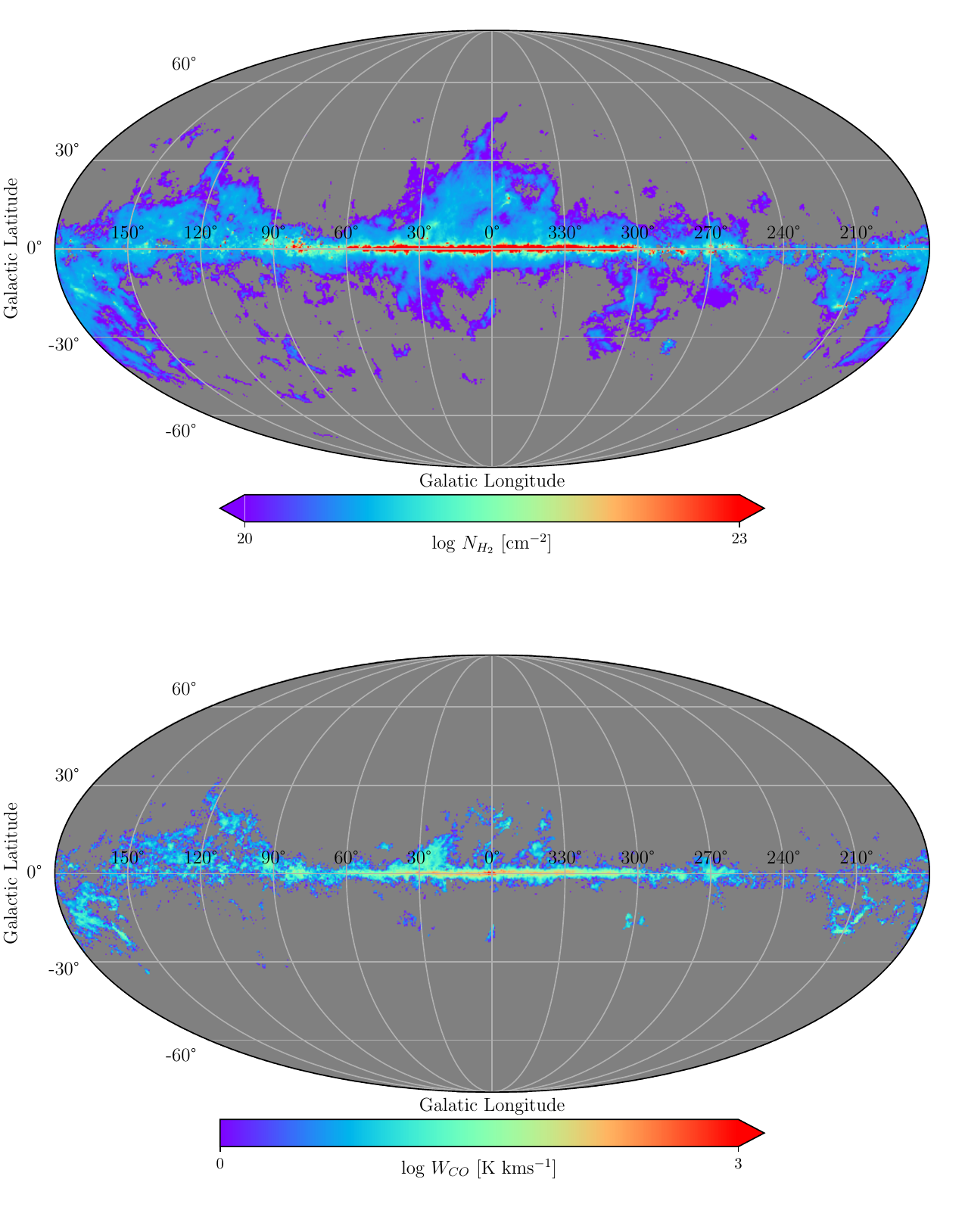}
    \caption{\textbf{Top panel}: Mollweide projection of our \NHmolecular\ map. The dark blue color indicates regions with $19 \lesssim \log$~\NHmolecular~(~\ColDens)~$\lesssim 20$. \textbf{Bottom panel}: \WCO\ map from \cite{Dame2001}. Our \NHmolecular\ map was constructed by calculating \AvMolecular, using \EBV\ and \NHI\ data, and then converting to \NHmolecular\ using the relation given in  Eq.~\ref{eq:Av_H2_fit}. The \NHmolecular\ map shows several relatively large angular scale structures.  As these are CO-dark \Hmol\ regions, they are undetectable in the \WCO\ map.  (Sect.~\ref{sec:co_dark_H2}). In both maps, longitude increases to the left from the central (vertical) line at $\ell = 0\degr$, the grid spacing, shown by the white solid lines, is $30\degr$, the angular resolution is $16\arcmin$, and NSIDE=1024.}
    \label{fig:NH2_XCO_full_sky_maps}
\end{figure*}

\section{Probing molecular gas using dust extinction}
\label{sec:AvH2_NH2}

For atomic gas, \cite{lenz_2017} found that dust reddening correlates with the column density of the atomic hydrogen as \NHI~/~\EBV~$= 8.8 \times 10^{21}$~\ColDens~mag$^{-1}$. This is our reference value and we show that our results remain statistically the same when we consider variations about this ratio (Sect.~\ref{sec:intrinsic_variations_dust_to_gas_ratio}). When the gas becomes molecular, \EBV\ increases non-linearly with respect to \NHI. 

We assume a constant total-to-selective extinction $R_{V}=$~\Av~/~\EBV~=~3.1 \citep{cardelli_1989.selective.extinction, schlafly_finkbeiner_2011} and calculate the visual extinction of dust mixed with atomic gas,
\begin{equation}
    \label{eq:Av_atomic}
    A_{V} ({H}\, {\sc I}) = 0.35 \times 10^{-21} ~ N_{H_{\sc I}}.
\end{equation}
$R_{V}$ fluctuates throughout our Galaxy \citep[e.g.,][]{peek_stanimirovic_2013, zhang_2023.Rv.2d.full.sky.map}. However, the uniformity of $R_{V}$ is implicit in our method, because we employ dust reddening maps that have been constructed using far-infrared multi-wavelength dust intensity data (Sect.~\ref{sec:full_sky_NH2_map}, Appendix~\ref{sec:appendix_nh2_different_av_maps}). These maps directly constrain the dust emission properties, such as dust temperature and opacity, when some dust model is fitted to the data, and indirectly \EBV, by assuming some extinction law and $R_{V} = 3.1$ \citep[e.g.,][]{schlegel_1998}. In addition, most of the existing constraints of \NHI~/~\EBV\ also rely on these dust reddening maps. Therefore, the uniformity of  $R_{V}$ inevitably propagates in our methods and accounts for some of our uncertainties, which are calculated in Sect.~\ref{sec:confidence_intervals_our_method}.

We define the residuals between the total dust extinction and \AvAtomic\ as
\begin{equation}
    \label{eq:Av_molecular}
    A_{V} \left (H_{2} \right) = A_{V} ({H}) - A_{V} (H\, {\sc I}),
\end{equation}
where \AvTotal\ denotes the total extinction from dust associated with both atomic and molecular hydrogen. On the left hand side, we use \Hmol\ in the parenthesis because we hypothesize, and verify below (Fig.~\ref{fig:AvH2_NH2}), that these residuals trace the visual extinction of dust mixed with molecular hydrogen. 

Eq.~\ref{eq:Av_molecular} is the cornerstone of our methodology for estimating \NHmolecular. \AvAtomic\ can be estimated using \NHI\ measurements (Eq.~\ref{eq:Av_atomic}) from surveys such as HI4PI \citep{hi4pi}, while \AvTotal\ can be extracted from publicly available full-sky dust extinction maps \citep[e.g.,][]{schlegel_1998,planck_collaboration_2016}. The next step in our strategy is to find an empirical relation for \NHmolecular\ as a function of \AvMolecular. Before proceeding we discuss some pathologies in the definition of \AvMolecular.

\subsection{Pathologies in the definition of \AvMolecular}

Dust extinction is a positive definite quantity, but \AvMolecular\ can become negative if \AvAtomic~$>$~\AvTotal. This situation can be encountered when either the signal-to-noise ratio (S/N) of the measurements is low or if the value of \NHI~/~\EBV\ deviates from its global value; variations of this ratio propagate to \AvMolecular, due to Eq.~\ref{eq:Av_molecular} (Sect.~\ref{sec:intrinsic_variations_dust_to_gas_ratio}). We present a thought experiment to demonstrate how the use of a global \NHI~/~\EBV\ (Eq.~\ref{eq:Av_atomic}) can affect the estimation of \AvMolecular.

Consider an ISM cloud with a local \NHI~/~\EBV\ ratio smaller than the Galactic value. We consider that the imaginary cloud has a given \AvTotal, and gas is atomic. Thus, the total extinction of the cloud will be produced by dust mixed with \HI\ gas. If we used the local \NHI~/~\EBV\ of the cloud, we would derive that \AvTotal~$=$~\AvAtomic, hence \AvMolecular~$= 0$. However, in Eq.~\ref{eq:Av_molecular} we use the global value from \cite{lenz_2017}, which would yield for the cloud of our thought experiment that \AvTotal~$<$~\AvAtomic, and hence \AvMolecular~$<0$. 

Thus, \AvMolecular\ can become negative due to the use of a global \NHI~/~\EBV, which limits the detectability of our method. However, this is something that we cannot overcome because local measurements are limited. For this reason, for every measurement with \AvMolecular~$<0$, we consider gas to be fully atomic, and it is not considered in the analysis below.

\subsection{Mapping \AvMolecular\ to \NHmolecular} 

We calculated \AvMolecular\ using Eq.~\ref{eq:Av_molecular} towards several lines of sight with reliable direct \NHmolecular\ determinations. The direct \NHmolecular\ measurements were obtained by observing the Lyman--Werner absorption lines towards background sources in our Galaxy. We extracted the \NHmolecular\ measurements, with their corresponding observational uncertainties, from the catalogues of \cite{sheffer_2008}, \cite{Gillmon2006}, \cite{Gudennavar_2012}, and \cite{shull_2021}; these catalogues also provide estimates for \NHI, which allow us to obtain \AvAtomic\ (Eq.~\ref{eq:Av_atomic}). For measurements with asymmetric observational uncertainties, we calculated their average value. 

The majority of the sources used for estimating \NHmolecular\ are nearby -- at distances less than 1 kpc -- and close to the Galactic plane. Therefore, the column densities obtained from the absorption lines of these stars trace the integrated gas abundances up to the distance of each object, and not the total gas column. Therefore, we also need to calculate the dust extinction up to the distance of each star ($d_{\star}$). Below, we explain how we obtained fractional extinctions up to the distance of each star using the publicly available 3D extinction map of \cite{green_bayestar_2019.latest.update}.

For each object with \NHmolecular\ measurement, we extracted $d_{\star}$ from the catalogue of \cite{bailer_jones_2021}, which uses parallaxes from the Gaia Data Release 3 \citep{gaia_dr3}. We then calculated the dust extinction integral of each star as:
\begin{equation}
    \label{eq:ebv_star}
    E(B - V)_{\star} = \int_{0}^{d_{\star}} dz~E(B-V)(z) ,
\end{equation}
where $z$ is the line-of-sight distance between the observer and the star at distance $d_{\star}$, and $E(B-V)(z)$ denotes the dust reddening at a given distance $z$. We extracted the $E(B-V)(z)$ values from the 3D map (Bayestar19) of \cite{green_bayestar_2019.latest.update} \footnote{\url{http://argonaut.skymaps.info/}}. This map provides a sample of $E(B-V)(z)$ profiles drawn from a posterior distribution; for the calculation of $E(B - V)_{\star}$ we used the most probable extinction profile.  Dust reddening values from the map of \cite{green_bayestar_2019.latest.update} are calibrated for the SDSS bands. We converted to dust reddening by multiplying the output of their map by 0.884, which is the standard color correction recommended by those authors. Then, we converted \EBVstar\ to \AvTotal\ by multiplying by $R_{V}$. 

Both \AvTotal\ and \AvAtomic\ are constrained for the LOSs where \NHmolecular\ measurements exist from spectroscopic observations. This allows us to calculate \AvMolecular\ for each LOS and compare with \NHmolecular. Several objects with spectroscopically constrained \NHmolecular\ are located in the anti-center of our Galaxy, which are not covered by the extinction map of \cite{green_bayestar_2019.latest.update}. These measurements were not considered in the analysis below. 

Fig.~\ref{fig:AvH2_NH2} displays \AvMolecular\ versus \NHmolecular. We do not show measurements with log \AvMolecular~(mag)~$\lesssim - 1.2$ because they are non-detections. We also note that some measurements, mostly with low S/N, yielded negative \AvMolecular.  These were also discarded.

For $\log$~\NHmolecular~(\ColDens)~$\lesssim 19$, \AvMolecular\ remains constant with respect to \NHmolecular. Most of the measurements there have S/N~$<$~3 \footnote{The uncertainties are shown in logarithmic scale, while S/N refers to a linear scale.}, and hence we cannot confidently determine whether the flatness of the profile is physical or induced by noise in the data. We observe a strong correlation between \AvMolecular\ and \NHmolecular\ for $\log$~\NHmolecular~$\geq$~20. This correlation strongly supports the initial hypothesis that the extinction residuals, found in the total dust extinction when the extinction induced by dust mixed with \HI\ is subtracted, yield the extinction of dust mixed with \Hmol\ (as given by Eq.~\ref{eq:Av_molecular}). 

The various quantities involved in our analysis (\EBV, \NHmolecular, \NHI) are measured with beams of differing size: a) \NHmolecular\ is measured from UV absorption lines with a ``pencil beam'' defined by the star, b) several \NHI\ measurements have been constrained by fitting the Ly$\alpha$ profiles \citep[e.g., as in][]{shull_2021}, which are also characterized by pencil beams, while some of the \NHI\ measurements were obtained from \HI\ emission line data, where the resolution is $16\arcmin$ \citep{hi4pi}, and c) the resolution of the 3D extinction map of \cite{green_bayestar_2019.latest.update} varies from $3.4\arcmin$ to $13.7\arcmin$. Some of the scatter of the points in Fig.~\ref{fig:AvH2_NH2} is a result of this effect, but we note that the beam mismatch seems to be less important for diffuse and extended ISM clouds \citep[e.g.,][]{pineda_2017, murray_2018}. 
Although we cannot make a correction for this issue, we include it when we calculate the confidence levels of our method (Sect.~\ref{sec:confidence_intervals_our_method}).

\subsubsection{Observational uncertainties in \AvMolecular}
\label{sec:Av_H2_uncertainties}

There are two different sources of uncertainty that affect the estimates of \AvMolecular: 1) Uncertainties coming from the 3D extinction map of \cite{green_bayestar_2019.latest.update}, and 2) distance uncertainties of stars.

To evaluate the dust extinction uncertainties, we sampled 300 random values from the posterior distribution of the \cite{green_bayestar_2019.latest.update} map at the most probable distance of each star ($d_{\star}$). We calculated the extinction spread ($\sigma_{ext}$) at $d_{\star}$. 

Parallax uncertainties affect the distance estimates of a star. We consider the distance upper and lower limits denoted  as $d_{\star, +}$ and $d_{\star, -}$ respectively. From the map of \cite{green_bayestar_2019.latest.update}, we calculated the most probable extinction at $d_{\star, +}$, and at $d_{\star, -}$, denoted as \EBV\ ($d_{\star, +}$) and \EBV\ ($d_{\star, -}$),respectively. We then calculated the differences $\sigma_{d+}$ = \EBV\ ($d_{\star, +}$) -  \EBV\ ($d_{\star}$) and $\sigma_{d-}$ = \EBV\ $d_{\star}$ -  \EBV\ ($d_{\star, -}$). The average ($\sigma_{d}$) between $\sigma_{d+}$ and $\sigma_{d-}$ measures the extinction uncertainties due to parallax uncertainties. We take the total dust extinction uncertainty to be the quadratic sum of $\sigma_{ext}$ and $\sigma_{d}$. These uncertainties are shown as the error bars in Fig.~\ref{fig:AvH2_NH2} and discussed in the following.

\subsubsection{Uncertainties from \NHI~/~\EBV\ variations}
\label{sec:intrinsic_variations_dust_to_gas_ratio}

At high Galactic latitudes $\left ( |b| \geq 60\degr \right)$, \cite{lenz_2017} found that the correlation between \EBV\ and \NHI\ is very tight, with intrinsic variations only $0.015$~mag. However, as shown recently by \cite{shull_panopoulou_2023}, the uncertainties in the \EBV\ versus \NHI\ relation are larger than what is inferred by \cite{lenz_2017}, with variations being dominated by systematic uncertainties in the employed datasets.

\cite{lenz_2017} used the \EBV\ map of \cite{schlegel_1998}, and \NHI\ estimated using only emission line data from \cite{hi4pi}, assuming that the line is optically thin. However, the \EBV\ values of \cite{planck_collaboration_2014.all_sky.dust.model.miville} are systematically larger than the values of \cite{schlegel_1998} at high Galactic latitudes \citep{shull_panopoulou_2023}. Thus, if the Planck map were employed, the ratio between \NHI\ and \EBV\ would be biased  toward smaller values than what would have been inferred from the \cite{schlegel_1998} map. In addition, beam effects in the \HI\ emission data can induce significant scatter in the obtained \NHI\ constraints. Several of these uncertainties are summarized in \cite{shull_panopoulou_2023}, who found that \NHI~/~\EBV~$\times 10^{-21}$ ranges from 8 to 10 \ColDens~mag$^{-1}$; this range is orders of magnitude larger than what is claimed by \cite{lenz_2017}, and represents the formal uncertainty in \NHI~/~\EBV. We explored how \NHI~/~\EBV\ variations, which propagate to our calculations through Eq.~\ref{eq:Av_atomic}, affect our results.

Fig.~\ref{fig:gas_dust_variations} shows \AvMolecular\ versus \NHmolecular\ for every measurement, except for the outliers (green stars), shown in Fig.~\ref{fig:AvH2_NH2}. Black points correspond to \AvMolecular\ when we use the \cite{lenz_2017} relation, \NHI~/~\EBV~$\times 10^{-21}$~= $8.8$~\ColDens~mag$^{-1}$; the errorbars are calculated as explained in Sect.~\ref{sec:Av_H2_uncertainties}. The blue and cyan points correspond to the obtained \AvMolecular\ when we assume that \NHI~/~\EBV~$\times 10^{-21}$ is equal to 10 and 8 \ColDens~mag$^{-1}$ respectively.

For $\log$~\NHmolecular~$> 20.56$, the colored (black and cyan) points are very close to the black points. In this regime, gas is mostly molecular, and thus the contribution of the dust mixed with atomic gas to the total dust extinction is minor. Thus, the uncertainty of \NHI~/~\EBV\ has a weak effect in high-extinction regions. When \NHmolecular\ is low, deviations between black and colored points become more prominent, because the relative abundance of atomic gas is higher there. However, even there, the offset between the black and the colored points is statistically insignificant for the vast majority of the points (Fig.~\ref{fig:gas_dust_variations}). There are only six exceptions, but this is only a minor fraction of the sample.

We conclude that our analysis remains robust in relation to variations of \NHI~/~\EBV, which implies that uncertainties coming from the 3D extinction maps  \citep{green_bayestar_2019.latest.update} or Gaia distances (Sect.~\ref{sec:Av_H2_uncertainties}) are more important. However, variations in \NHI~/~\EBV\ likely contribute to the scatter of the observed \AvMolecular\ -- \NHmolecular\ relation (Fig.~\ref{fig:AvH2_NH2}). Our estimated confidence intervals include all the aforementioned uncertainties (Sect.~\ref{sec:confidence_intervals_our_method}).

\subsection{Fitting the data}
\label{spec:Av_H2_fit}

We fitted a polynomial to the \AvMolecular\ -- \NHmolecular\ relation using Bayesian analysis. We assumed uniform priors, and calculated the following likelihood:
\begin{equation}
    \label{eq:log_likelihood}
    \ln{\mathcal{L}} = - \sum_{i=1}^{N}  \left [  \ln{\sqrt{2 \pi}} + \ln{\sigma} + \frac{ \left ( y_{i} - \tilde{y} \right)^{2}}{2\sigma^{2}}   \right ],
\end{equation}
where $\sigma$ is the quadratic sum of the observational uncertainties of $\log$ \AvMolecular, and $\log$ \NHmolecular, $y_{i}$ corresponds to the measured \AvMolecular, $\tilde{y}$ is the intrinsic model, which we assume to be a polynomial, and $i$ is the measurement index.

We sampled the parameter space with the "emcee" Markov Chain Monte Carlo (MCMC) sampler \citep{emcee}. Measurements shown as green stars in Fig.~\ref{fig:AvH2_NH2} were not included in the fit because they are variable sources (Sect.~\ref{sec:Av_H2_uncertainties}). However, the fit changes only slightly even if they are included.

The best--fit polynomial equation is 
\begin{equation}
    \label{eq:NH2_Av_fit}
    \begin{aligned}
    \log A_{V}(H_{2}) = ~ & 0.03077~\left(\log N_{H_{2}} \right)^{3} - 1.60469~\left(\log N_{H_{2}}\right)^{2} \\
    & + 27.80868~\log N_{H_{2}} - 161.06073,           
    \end{aligned}
\end{equation}
with uncertainties for the fitted coefficients (from highest to smallest power of the logarithm) being $0.00769$, $0.42823$, $7.88171$, and $49.91439$. The spread of the fit, shown by the orange lines in Fig.~\ref{fig:AvH2_NH2}, grows for $\log$~\NHmolecular~(\ColDens)~$< 19$, making our fit unreliable in that range. This determines the sensitivity threshold of our method.

In practice, \NHmolecular\ is usually the unknown, while \AvMolecular\ can be measured from dust extinction and \NHI\ observational data. \NHmolecular\ can be estimated from \AvMolecular\ using the following analytical relation, which also fits our data well, 
\begin{equation}
    \label{eq:Av_H2_fit}
    \begin{aligned}
    \log~N_{H_{2}} = ~ & 1.38742~\left( \log A_{V}({H_{2}}) \right)^{3} - 0.05359~\left( \log A_{V}({H_{2}}) \right)^{2} + \\ 
    & + 0.25722~\log  A_{V}({H_{2}}) - 20.67191,             
    \end{aligned}
\end{equation}

The fitting uncertainties of the coefficients, from highest to smallest power of the logarithm, are $0.17822$, $0.17277$, $0.05248$, and $0.01555$. The above equation maps \AvMolecular\ to \NHmolecular\ and we use it to construct a full-sky \NHmolecular\ map of our Galaxy (Sect.~\ref{sec:full_sky_NH2_map}).

\subsubsection{Fitting different polynomials}

We experimented with the degree of the polynomial used to fit Eq.~\ref{eq:Av_H2_fit}.   For even polynomials, we found that \NHmolecular\ is reduced at high \AvMolecular, because the coefficient of the highest power of the fitted polynomial is always negative, hence making the curve concave downward. When we forced the coefficient of the highest-power term to be positive (by employing this constraint in the prior distribution), the reduction of \NHmolecular\ disappeared. 
However, we wish to keep a flat distribution of priors over the entire domain range (uninformative priors).  We thus decided to work with polynomials whose highest-power term is odd because they did not show any reduction of \NHmolecular\ at high \AvMolecular. 

A linear polynomial would not reproduce the non-linear behaviour of the data at $\log$~\NHmolecular(\ColDens)~$\lesssim 20$, and $\log$~\AvMolecular~(mag)~$\lesssim - 1$ (Fig~\ref{fig:AvH2_NH2}). As a result, linear models would underestimate \NHmolecular\ in low-extinction regions. We conclude that a cubic polynomial (as shown in Eq.~\ref{eq:Av_H2_fit}) is the simplest function that accurately represents the data, while higher-degree polynomials overfit the data. 

\subsubsection{Outliers in the \AvMolecular\ - \NHmolecular\ relation}
\label{spec:Av_H2_outliers}

In the fitting process, we excluded the four measurements shown as green stars (Fig.~\ref{fig:AvH2_NH2}). According to the SIMBAD database, one of the outliers (identified as HD200120) is a Be star -- this object also appears in the Be star catalogue of the IAU \citep{IAU_Be.stars.caltagoue}, hence further supporting the validity of the spectral classification of this object -- while there is a T-Tauri (HD40111) and a $\beta$ Cephei variable (HD172140) star. Both Be and T-Tauri stars can show some variability in both photometric and spectroscopic observations: Be stars eject material, due to their rapid rotation, while the brightness of T-Tauri objects can vary significantly, due to their high accretion rate, even within months. $\beta$ Cephei are B-type stars whose surface pulsates owing to their rapid rotation. The last object that was considered an outlier (HD 164816) is a Be star according to the SIMBAD database. However, accurate spectroscopic observations suggest that it is a binary system \citep{sota_2014.goss.survey}. In either case, some degree of variability is expected, which explains why this measurement is an outlier. We consider the \NHmolecular\ measurements towards these stars to be untrustworthy and did not include them in the fit. But even if we include these measurements, the fit changes insignificantly.

\subsection{Confidence intervals in the estimated \NHmolecular}
\label{sec:confidence_intervals_our_method}

Our fitted model (Eq.~\ref{eq:Av_H2_fit}) allows us to estimate \NHmolecular\ using \HI\ and \EBV\ data. But, the estimates do not come without uncertainties. 

The uncertainties of the fit (shown by the orange lines in Fig.~\ref{fig:AvH2_NH2}) are much smaller than the variance of the points about the fitted curve. These measured variations could be induced by several factors, such as variations in \NHI~/~\EBV\ (Sect.~\ref{sec:intrinsic_variations_dust_to_gas_ratio}), $R_{V}$ variations, metallicity gradients, \HI\ emission line optical depth effects (Sect.\ref{sec:discussion}), or due to the beam mismatch of the employed data, unaccounted for in our model. Therefore, the fitting uncertainties alone overestimate our confidence on the \NHmolecular\ estimates derived by Eq.~\ref{eq:Av_H2_fit}.

To place reasonable confidence levels, we assumed that the depicted \AvMolecular\ spread (Fig.~\ref{fig:AvH2_NH2}) represents the intrinsic spread in our Galaxy. We calculated the linear scale ratio ($r$) between the observed \AvMolecular\ with the value obtained from Eq.~\ref{eq:Av_H2_fit} for every data point with S/N~$\geq 3$. From the cumulative distribution function of $r$ -- the distribution of $r$ is asymmetric with skewness equal to 1.38 -- we obtained the following probabilities: for $68\%$ of the measurements $r~\epsilon~[0.75, 2.48]$, for $95\%$ $r~\epsilon~[0.46, 3.54]$, and for $99\%$ $r~\epsilon~[0.30, 4.78]$. This implies that in $95\%$ of the cases, the "true" \NHmolecular\ should not deviate by a factor greater than $3.5$ from the values obtained from the fit (Eq.~\ref{eq:Av_H2_fit}). 

\begin{figure*}
   \centering
   \includegraphics[width=0.97\hsize]{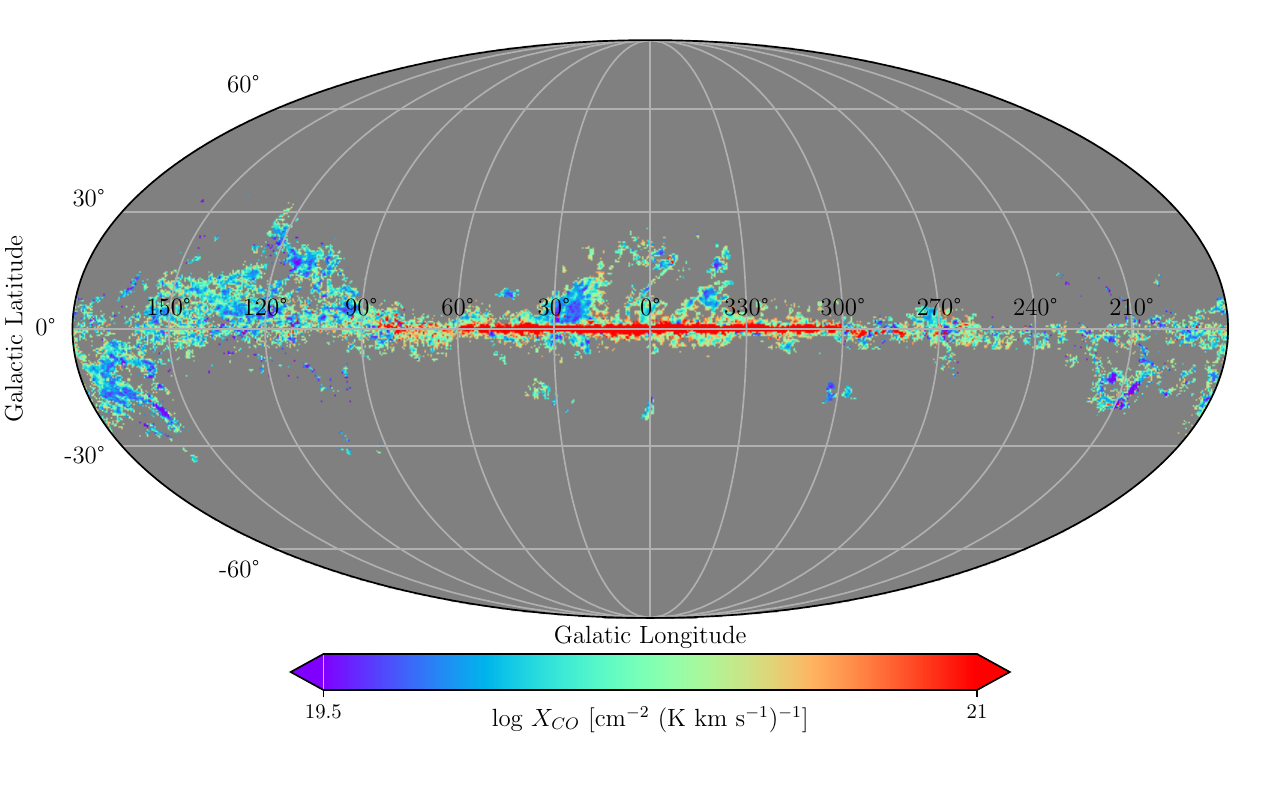}
    \caption{Mollweide projection of our constructed \XCO\ map. Our global value for \XCO\ is $2 \times 10^{20}$ K~\kms~\ColDens, which is consistent with previous estimates. \XCO\ varies by orders of magnitude within a few tens of arcminutes. In the surroundings of molecular regions, \XCO~$\sim 10^{21}$ \ColDens~(K~\kms)$^{-1}$, while in the inner parts of clouds it decreases to \XCO~$\sim 5 \times 10^{19}$~\ColDens~(K~\kms)$^{-1}$. In the Galactic center, \XCO\ increases significantly due to the high optical depths in that region. The angular resolution is $16\arcmin$, and NSIDE=1024.
    \label{fig:XCO_map}}
\end{figure*}

\begin{figure*}
   \centering
   \includegraphics[width=\hsize]{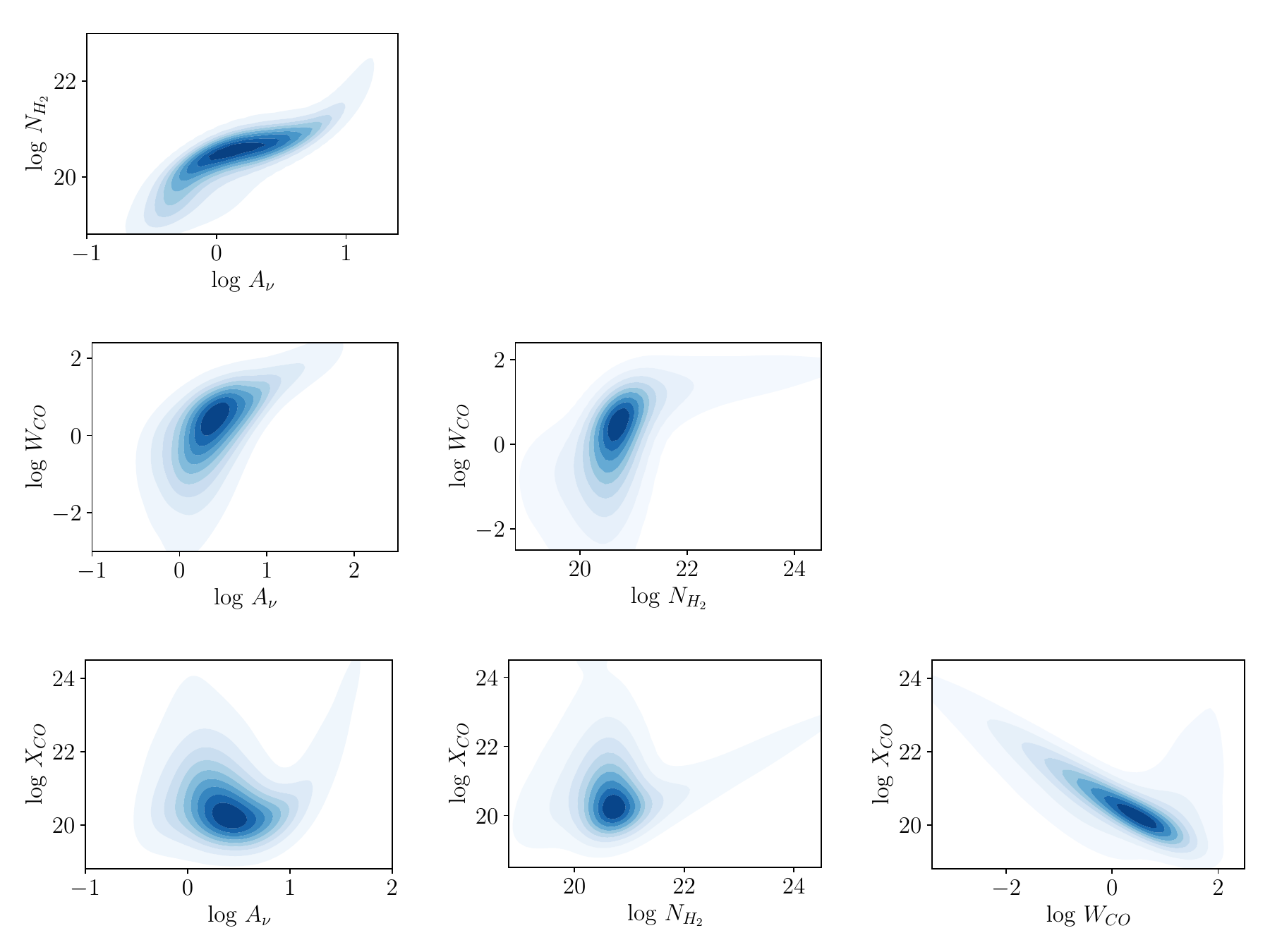}
    \caption{2D probability density functions relating the different parameters studied in this work (Sect.~\ref{sec:XCO_comparison}).}
    \label{fig:gas_correlations_CO}
\end{figure*}

\begin{figure*}
   \centering
   \includegraphics[width=0.97\hsize]{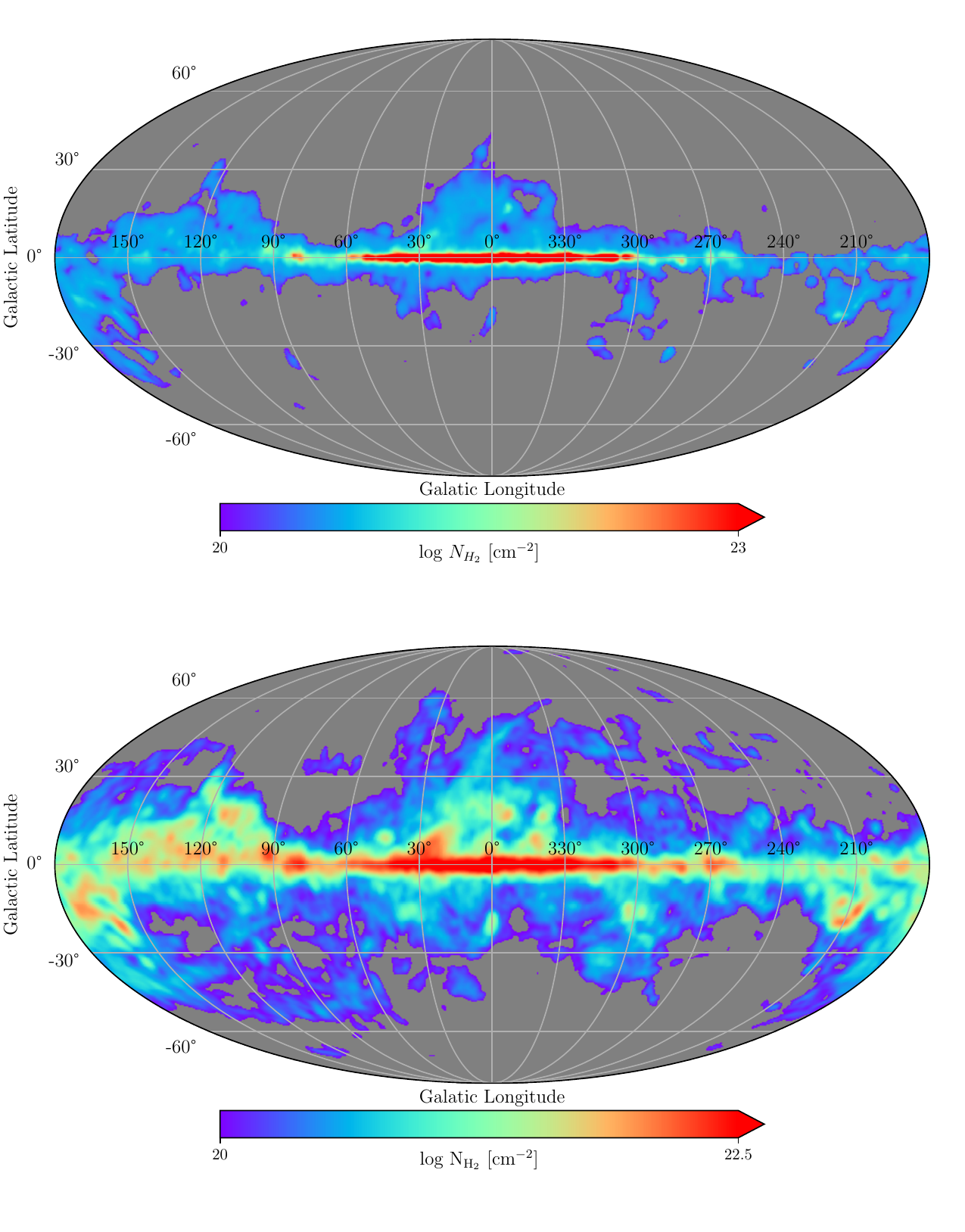}
    \caption{\NHmolecular\ map constructed using the method presented here (\textbf{upper panel}) and using that in \cite{kalberla_2020} (\textbf{lower panel}). The resolution of both maps is $2\degr$ and NSIDE=1024.}
    \label{fig:NH2_full_sky_maps_kalberla}
\end{figure*}

\section{Full-sky \NHmolecular\ map of our Galaxy}
\label{sec:full_sky_NH2_map}

We use the \NHI\ data from the HI4PI survey \citep{hi4pi} and the dust extinction map of \cite{schlegel_1998} -- the values of this map are multiplied by 0.884 to account for systematic uncertainties \citep{schlafly_2010.renormalizing.schlegel.map, schlafly_finkbeiner_2011} -- to construct a full-sky \NHmolecular\ map of our Galaxy. For comparison, we use the Planck extinction map. Overall, the Planck map shows enhanced extinction at high Galactic latitudes; this is also noted by \cite{shull_panopoulou_2023}. The various estimated quantities for the molecular hydrogen do not change by more than a factor of two when the Planck map is used (Appendix~\ref{sec:appendix_nh2_different_av_maps}).

For the construction of the full-sky \NHmolecular\ map, we calculated \AvMolecular\ using the \NHI\ and the \EBV\ maps (Eq.~\ref{eq:Av_molecular}), and then converted to \NHmolecular\ (Eq.~\ref{eq:Av_H2_fit}). The resolution of the extinction map ($6.1\arcmin$) is higher than that of the \NHI\ data ($16.2\arcmin$). We thus smoothed the extinction map to the resolution of the \HI\ data. Fig.~\ref{fig:NH2_XCO_full_sky_maps} (top panel) shows a mollweide projection of our constructed \NHmolecular\ map for the Milky Way. We only visualize regions with log~\NHmolecular~$\geq 19$ because below this limit our calibration relation is dominated by noise (Fig.~\ref{fig:AvH2_NH2}).

We identify several molecular clouds of the Gould Belt, including the Polaris Flare $\left ( \ell, b \sim 125\degr, +30 \degr  \right)$, Taurus $\left ( \ell, b  \sim 170\degr, -15\degr  \right)$, and Orion $\left ( \ell, b \sim 209\degr, -19\degr  \right)$. The North Polar Spur appears as a large-scale molecular feature centered at Galactic coordinates $\ell, b \sim 15\degr, 15\degr$ and extending up to $b \sim 45\degr$. We observe many small-angular-scale diffuse molecular clouds in the Southern Galactic hemisphere, while in the Northern hemisphere molecular structures seem to be coherent over scales of many degrees. 

We hardly see any molecular gas structures with $\log$~\NHmolecular~(~\ColDens)~$\geq 19$ above the Galactic plane, for $|b| \geq 60 \degr$. This, however, seems to be the case only when we use the \cite{schlegel_1998} extinction map. When we construct the \NHmolecular\ map using extinctions from \cite{planck_coll_2020_ebv}, we see some molecular structures extending up to the Northern and Southern poles, and weaker emission from both extended and small-scale clouds at higher Galactic latitudes (Fig.~\ref{fig:NH2_map_planck}). In both \NHmolecular\ maps (Figs.~\ref{fig:NH2_XCO_full_sky_maps}, and \ref{fig:NH2_map_planck}), the majority of the molecular gas seems to lie within $|b| \lesssim 45\degr$. %

In the Galactic plane, extinctions are larger than the extinction range used for the fitting of the \NHmolecular\ -- \AvMolecular\ relation (Eq.~\ref{eq:Av_H2_fit}). Therefore, the estimated \Hmol\ column densities in the Galactic plane were derived by extrapolating the fitted relation to larger values of \Av.

\section{Comparison with previous results}
\label{sec:comparison_pas_obs}

\subsection{Comparing \NHmolecular\ and \WCO}
\label{sec:XCO_comparison}

Fig.~\ref{fig:NH2_XCO_full_sky_maps} shows our \NHmolecular\ map (top panel) and the CO (J=1-0) integrated intensity, \WCO, map (bottom panel) from \cite{Dame2001}. The two maps are well correlated, but the observed structures in the \NHmolecular\ map are more extended than in \WCO. In this section our analysis is focused on CO-bright \Hmol\ regions, while in Sect.~\ref{sec:co_dark_H2} we study the properties of CO-dark \Hmol, which extends to higher Galactic latitudes than CO-bright \Hmol. 

Fig.~\ref{fig:XCO_map} shows the full-sky \XCO\ (Eq.~\ref{eq:xco_factor}) map obtained with our \NHmolecular\ measurements and \WCO\ from \cite{Dame2001}. We calculated \XCO\ toward LOSs with \WCO~$\geq 1$~K~\kms, which is the noise level in the CO data. We find that the average value of \XCO\ is approximately equal to 2 $\times$ 10$^{20}$ \ColDens~(K~km s$^{-1})^{-1}$, which matches exactly with the previously reported value \citep{bolato_2013, liszt_2016.XCO}. 

We observe that in many regions \XCO\ varies by orders of magnitude within a few arcminutes. The small-scale variations \XCO\ are more prominent in regions above the Galactic plane and in the Galactic plane but only for $270\degr > \ell > 90 \degr$. \XCO\ tends to be high ($\log$~\XCO~$\gtrsim 20.5$) in the surroundings of molecular regions, while it decreases significantly ($\log$~\XCO~$\approx 19.5$) in the inner parts of molecular clouds. 

\XCO\ is enhanced in the surroundings of clouds because CO molecules are more effectively photodissociated there. The low CO abundance implies that \WCO\ is small, hence \XCO\ becomes high. On the other hand, in the inner parts of the molecular clouds, the total column density is greater, reducing the photodestruction rate of molecules due to both shielding by dust and by self-shielding, and the CO abundance increases. The enhanced CO abundance implies that \WCO\ increases and thus \XCO\ decreases. In the Galactic plane, \XCO\ becomes large because \NHmolecular\ is maximum there and \WCO\ saturates, due to the high optical depth of the $J = 1-0$ transition usually employed to determine \WCO. 

The observed behaviour is consistent with past observations, and numerical simulations \citep{bolato_2013}. Our results demonstrate that CO is a good tracer of \NHmolecular\ only for LOSs where CO is the dominant form of carbon, and the CO rotational line is not saturated.   

We explore quantitatively the correlation of the CO observables (\XCO, \WCO) with \Av\ and \NHmolecular. Fig.~\ref{fig:gas_correlations_CO} shows the 2D probability density functions of all possible combinations of \Av, \XCO, \WCO, and \NHmolecular. Measurements with \NHmolecular~$> 10^{21}$~\ColDens, and  \Av~$>5.5$~mag have been derived by extrapolating our fitted equation, hence no strong conclusions are made for this regime. The following paragraphs summarize the results of the comparison.

\Av\ -- \XCO:  The majority of points are clustered around \XCO~$\approx 10^{20}$ \ColDens~(K~km s$^{-1})^{-1}$. The relation looks relatively flat in this regime, although there is a slight anticorrelation; this is more evident in the low-probability contours, which are concave upward. The anticorrelation is due to the presence of CO-dark \Hmol\ gas in low-extinction regions \citep{seifried_2020.SILC.CO_dark.properties,borchert_walch_2022.synthetic.CO.emission.XCO}. For $\log$~\Av~$\gtrsim 1$, \XCO\ increases with \Av\ because the CO line saturates. The critical extinction that marks the CO line saturation is log \Av~$\approx 0.6 \Rightarrow A_{V} \approx 4$~mag, which is consistent with previous estimates toward the Perseus molecular cloud \citep{pineda_2008.perseus.XCO.variations,lee_2014.XCO.perseus.molecular.cloud}. The local ISM conditions, such as density and intensity of the background radiation field, can change the value of \Av\ where CO saturates. For this reason, there is an extinction range where a non-linear increase of \XCO\ is observed \citep[e.g.,][]{lombardi_2006.pipe.nebula.XCO}. Similar behaviour between \XCO\ and \Av\ is also observed in numerical simulations \citep{borchert_walch_2022.synthetic.CO.emission.XCO}.                  

\Av\ -- \WCO: Most of the points lie in the range $0.3 \lesssim$~$\log$~\WCO~(K~km s$^{-1}$) $\lesssim 1$. \WCO\ varies by orders of magnitude for $\log$~\Av~$\lesssim 1$, while the dispersion in \WCO\ decreases significantly for $\log$~\Av~$\geq 1$. The scaling between \Av\ and \WCO\ transitions at log \Av~$\approx 0.6$, and \WCO\ converges to a maximum value \WCO~$\approx 100$~K~\kms. This convergence is due to the increase in the optical depth of the CO line, which becomes optically thick, and is consistent with previous observations \citep{pineda_2008.perseus.XCO.variations,lee_2014.XCO.perseus.molecular.cloud}.

\Av\ -- \NHmolecular: \NHmolecular\ increases non-linearly with respect to \Av\ for $\log$~\Av~$\lesssim 0$, because \HI\ is transformed to \Hmol; the transition takes place for log~\NHtotal\ (\ColDens)~in the range~$\approx$~20.1 - 20.8 \citep{savage_1977, Gillmon2006, bellomi_2020}. If we assume that \Av~/~\NHtotal~$= 5.34 \times 10^{-22}$~cm$^{2}$~mag \citep{1978ApJ_BSD}, then we obtain that $-1.2 \lesssim$~$\log$~\Av\ (mag)~$\lesssim -0.5$, which is consistent with the extinction range where we observe the nonlinear increase in \NHmolecular. For $\log$~\Av~$\gtrsim 0$, the correlation between \NHmolecular\ and \Av\ becomes quasi-linear. The positive correlation between \Av\ and \NHmolecular\ reflects the fact that we see more gas when there is more dust and that the hydrogen is almost entirely molecular. We explore the dust-to-gas ratio in more detail in Sect.~\ref{sec:dust-to-gas_ratio}. In agreement with previous studies \citep{liszt_2023.H2.properties.diffuse.ISM.CO}, we find that only a few sightlines have appreciable molecular gas content, $\log$~\NHmolecular~(~\ColDens)~$\gtrsim 19$ when $\log$~\Av~(mag)~$\lesssim -0.5$.
     
\NHmolecular\ -- \WCO: The observed behavior in this figure is similar to that of \Av\ -- \WCO. For $\log$~\NHmolecular~(\ColDens)~$> 22$, \WCO\ converges to a maximum value because the line saturates. For log \NHmolecular~(\ColDens)~$< 21.5$, the two quantities are weakly correlated: \WCO\ varies by four orders of magnitude for $ 20 \lesssim $ log \NHmolecular\ $\lesssim 21$. The rapid increase in \WCO\ as function of \NHmolecular\ reflects the conversion of atomic carbon to CO and the resulting rise in fractional CO abundance. The most commonly-found sightlines have \WCO~$\approx 10$~K~\kms\ and \NHmolecular~$\approx 10^{21}$ \ColDens, which explains why the global \XCO\ is close to $10^{20}$ \ColDens~(K~\kms)$^{-1}$.   

\WCO\ -- \XCO:  For $\log$~\WCO~$\lesssim 0.6$, \XCO\ is anticorrelated with \WCO\ due to Eq.~\ref{eq:xco_factor}. For $\log$~\WCO~$\approx 0.6$, the CO line saturates and \XCO\ becomes independent of \WCO, which converges to 100~K~\kms.

\begin{figure*}
   \centering
   \includegraphics[width=0.97\hsize]{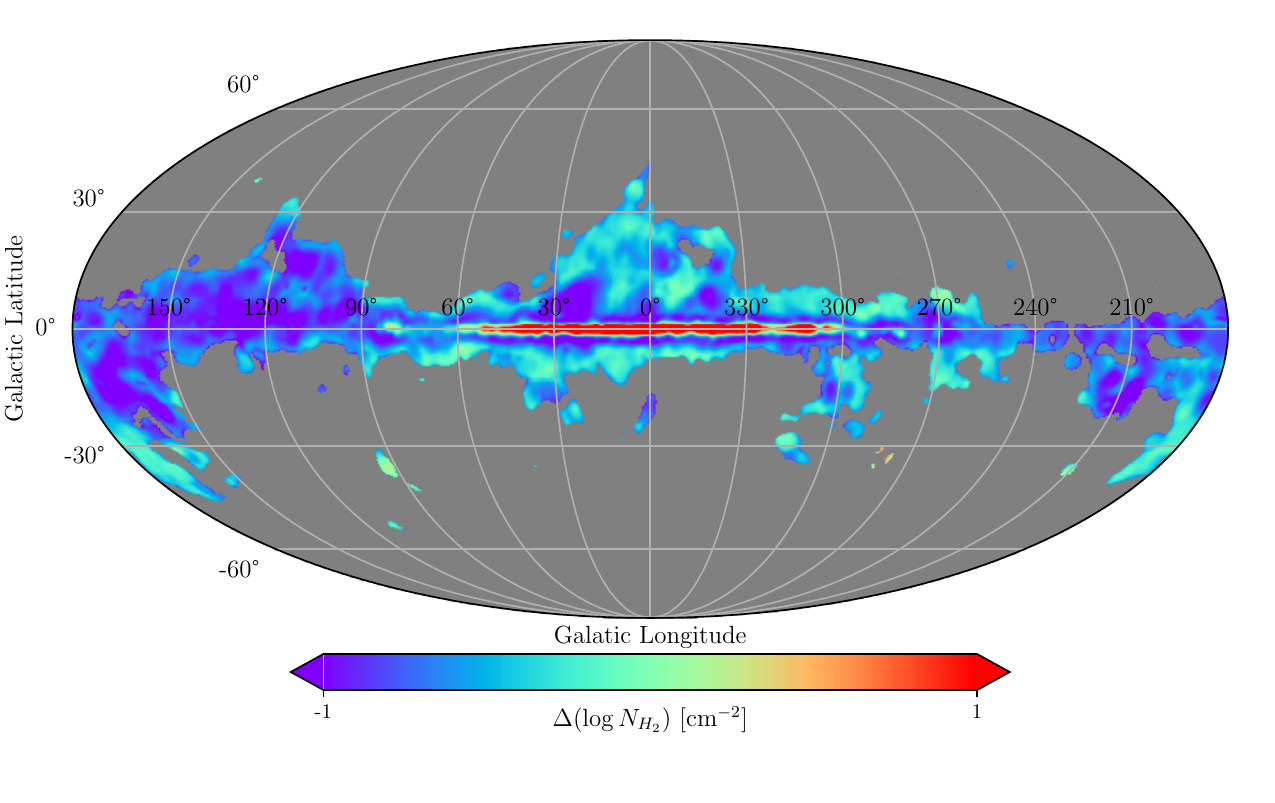}
   \caption{\label{fig:NH2_diff_full_sky_map} 
    Difference between our \NHmolecular\ estimates and those of \cite{kalberla_2020}.}
\end{figure*}

\subsection{Comparison to the \NHmolecular\ map of \cite{kalberla_2020}}
\label{sec:comparison_kalberla_map}

\cite{kalberla_2020} used the non-linear deviations in the \EBV\ -- \NHI\ relation to estimate \NHmolecular. Our method is similar to that of \cite{kalberla_2020}, although with some key differences that we discuss below. 

\subsubsection{Summary of \cite{kalberla_2020} method}

\HI\ emits at various Doppler-shifted frequencies (velocities). Each velocity component is considered to be a distinct cloud along the LOS. \cite{kalberla_2020} decomposed the \HI\ emission spectra into distinct Gaussian components \footnote{The decomposition of the emission spectrum into Gaussians has been widely applied in the past \citep[e.g.,][]{miville_2002, miville_deschenes2017.H2.properties.of.Galaxy.disk, heiles_troland_2003.ISM.phases.Gaussian.decomposition, murray_2014.Texc.lyman.alpha.scattering, kalberla_2015.treatment.of.GASS.systematics, kalberla_2018.phase.properties.HI.decomposition}. There are some uncertainties that can affect the decomposition of the emission spectra: 1) Contributions from non-Gaussian intensity peaks cannot be adequately captured. 2) Decomposing a spectrum into Gaussians is an ill-defined problem because different numbers of Gaussians can fit a spectrum equally well. 3) Distinct clouds that emit at  frequencies with small Doppler shifts can appear as a single component in emission. Some of the aforementioned problems have been treated with the development of sophisticated decomposition techniques that require minimum user input \citep{Lindner_2015, rohsa_2019, gausspy_2019}.  
}. 
Each Gaussian is characterized by an amplitude and a dispersion: the amplitude is proportional to the total \HI\ emitting gas, while the spread ($\sigma_{u}$) represents the internal velocity dispersion of the \HI\ gas, consisting of a thermal and turbulent component.

\cite{kalberla_2020} assigned an effective temperature ($T_{\rm D}$) - in their paper this is referred to as the Doppler temperature - to each decomposed Gaussian component, where $T_{\rm D}$ represents the total width of the Gaussian component. They calculated the dust extinction expected from each \HI\ Gaussian component, by assuming a constant dust-to-gas ratio \citep{lenz_2017}. Then, they added the \EBV\ contribution from all Gaussian components to estimate the total dust extinction induced by dust mixed with \HI\ gas. They compared their estimated \HI\ - based extinction to the total extinction, using the map of \cite{schlegel_1998}, and attributed the residuals to the presence of molecular gas. 

\subsubsection{Similarities and differences with our method}

The similarities between our method and that of \cite{kalberla_2020} are the following: 1) both methods rely on the assumption of a universal dust-to-gas ratio and on the hypothesis that nonlinear deviations in the \EBV\ / \NHI\ relation trace the molecular gas content. 2) Both methods employ a mapping function that converts the \EBV\ / \NHI\ non-linearities to \NHmolecular. 

The major difference between our method and that of \cite{kalberla_2020} lies on how the mapping function was obtained. \cite{kalberla_2020} derived their mapping function by minimizing the extinction nonlinearities from the linear relation through bootstrapping. We derived our mapping function from independent data by using \NHmolecular\ measurements from UV spectra of background objects (Sect.~\ref{sec:AvH2_NH2}). In addition, \cite{kalberla_2020} fitted their calibration function in regions where CO is undetectable. In CO-emitting regions, they estimated \NHmolecular\ by assuming a constant \XCO. On the other hand, our method does not require any assumption concerning \XCO. Finally, we note that \cite{kalberla_2020} decomposed the \HI\ components based on their spread ($T_{D}$), while we treated all the \HI\ components uniformly.

\subsubsection{Comparison of the maps}
\label{sec:comparison_of_kaleberla_and_our_map}

Fig.~\ref{fig:NH2_full_sky_maps_kalberla} shows the \NHmolecular\ map of \cite{kalberla_2020} together with our  map at $2\degr$ resolution. We only include regions with log \NHmolecular~(\ColDens)~$\geq {20}$, because below that threshold both methods might be susceptible to noise artefacts. 

The \NHmolecular\ structures are more extended in the map of \cite{kalberla_2020} than ours. This difference could be due to degeneracies in the definition of $T_{\rm D}$, which consists of a non--thermal (turbulent) and thermal component. The method of \cite{kalberla_2020} employs $T_{\rm D}$ as a proxy for the gas kinetic temperature, which is correlated with the molecular abundance. This approximation is accurate because the distribution of sonic Mach numbers in the diffuse ISM has a well-defined peak \footnote{The sonic Mach number ($\mathcal{M}_{s}$) is defined as $\mathcal{M}_{s} = \delta u_{turb} / c_{s}$, where $\delta u_{turb}$, and $c_{s}$ correspond to the turbulent and thermal sound speed respectively. The Doppler temperature is $T_{D} = 21.86~\delta u^{2}$ \citep{payne_1980.doppler.temperature}, where $\delta u$ represents the broadening of the \HI\ emission line, which consists of a thermal and turbulent component, hence $T_{D} \propto \delta u^{2}_{turb} + \delta u^{2}_{thermal}$. In the diffuse ISM, the average sonic Mach number is $\langle \mathcal{M}_{s} \rangle = 3.1$ \citep{heiles_troland_2003.diffuse.ISM.sonic.Mach.number}. The above equations suggest that $\langle T_{D} \rangle \propto 3.1~\langle c^{2}_{s} \rangle  + \langle \delta u^{2}_{thermal} \rangle$. Both $c_{s}$, and $u_{thermal}$ vary with the square root of the gas temperature, and thus $\langle T_{D} \rangle \propto \langle T \rangle$. This implies that the average $T_{D}$ probes the average gas temperature in the diffuse ISM. In the absence of a characteristic sonic Mach number, the turbulent and thermal components of $T_{D}$ become inseparable.} \citep{heiles_troland_2003.diffuse.ISM.sonic.Mach.number}. Although $T_{\rm D}$ represents the average gas temperature in the diffuse ISM, deviations toward individual regions are inevitably present.

We define the difference in the column densities of our map and the map of \cite{kalberla_2020} as: 
\begin{equation}
	\label{eq:Delta_NH2}
	\Delta \left ( \log N_{H_{2}} \right ) = \log N_{H_{2}}^{\rm here} - \log N_{H_{2}}^{\rm Kalberla+ 2020}.
\end{equation}

Fig.~\ref{fig:NH2_diff_full_sky_map} shows a full-sky map of $\Delta \left ( \rm log N_{H_{2}} \right )$, calculated only in regions where both maps yield \NHmolecular~$\geq 10^{20}$~\ColDens. The majority of the pixels have $\Delta \left ( \rm log N_{H_{2}} \right ) < 0$, while in the Galactic plane we find that $\Delta \left ( \rm log N_{H_{2}} \right ) > 0$.

\begin{figure}
   \centering
   \includegraphics[width=\hsize]{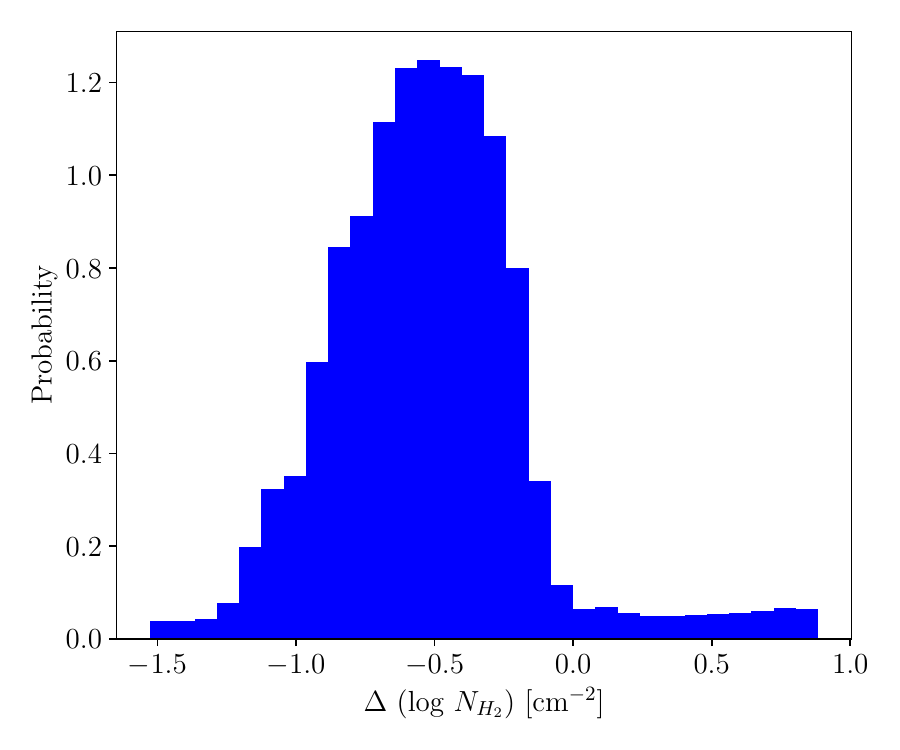}
    \caption{Distribution of the difference of our \NHmolecular\ estimates and those of \cite{kalberla_2020} (Eq.~\ref{eq:Delta_NH2}) for regions with $|b| \geq 10 \degr$. Our \NHmolecular\ estimates are, on average, three times smaller than those inferred by \cite{kalberla_2020}. This difference likely results in large part from the \XCO\ value employed in the method of \cite{kalberla_2020}.}
    \label{fig:NH2_diff_full_sky_distribution}
\end{figure}

We focus on Galactic latitudes $|b| \geq 10\degr$, because typical extinctions in the Galactic plane exceed the \Av\ range used for the calibration of the two methods. Fig.~\ref{fig:NH2_diff_full_sky_distribution} shows the distribution of $\Delta \left ( \rm log N_{H_{2}} \right )$ of all pixels  at $|b| \geq 10 \degr$. The distribution peaks at  -0.5, which implies that the \NHmolecular\ estimates in the map of \cite{kalberla_2020} are, on average, $\sim 3$ times larger than ours. Their estimates are consistent within the uncertainties of our method (Sect.~\ref{sec:confidence_intervals_our_method}). 

The method of \cite{kalberla_2020} was calibrated in CO-dark \Hmol\ regions, while in CO-bright regions they estimated \NHmolecular\ by assuming that \XCO~=~$4 \times 10^{20}$~\ColDens~(K~\kms)$^{-1}$. Although this value is consistent with the range of local \XCO\ variations, $1.7 - 4.1 \times 10^{20}$~\ColDens~(K~\kms)$^{-1}$, it is a factor of two larger than the Galactic average \citep{bolato_2013}. If \cite{kalberla_2020} had adopted the global value of \XCO, then our estimates would differ, on average, by less than a factor of two. A careful visual comparison of the $\Delta \left ( \rm log N_{H_{2}} \right )$ (Fig.~\ref{fig:NH2_diff_full_sky_map}) and the \WCO\ map (Fig.~\ref{fig:NH2_XCO_full_sky_maps}, bottom panel) shows that the \NHmolecular\ estimates of the two methods agree well, $\Delta \left ( \rm log N_{H_{2}} \right ) \approx 0$, in regions where CO is undetected. 

\begin{figure*}
   \centering
   \includegraphics[width=0.97\hsize]{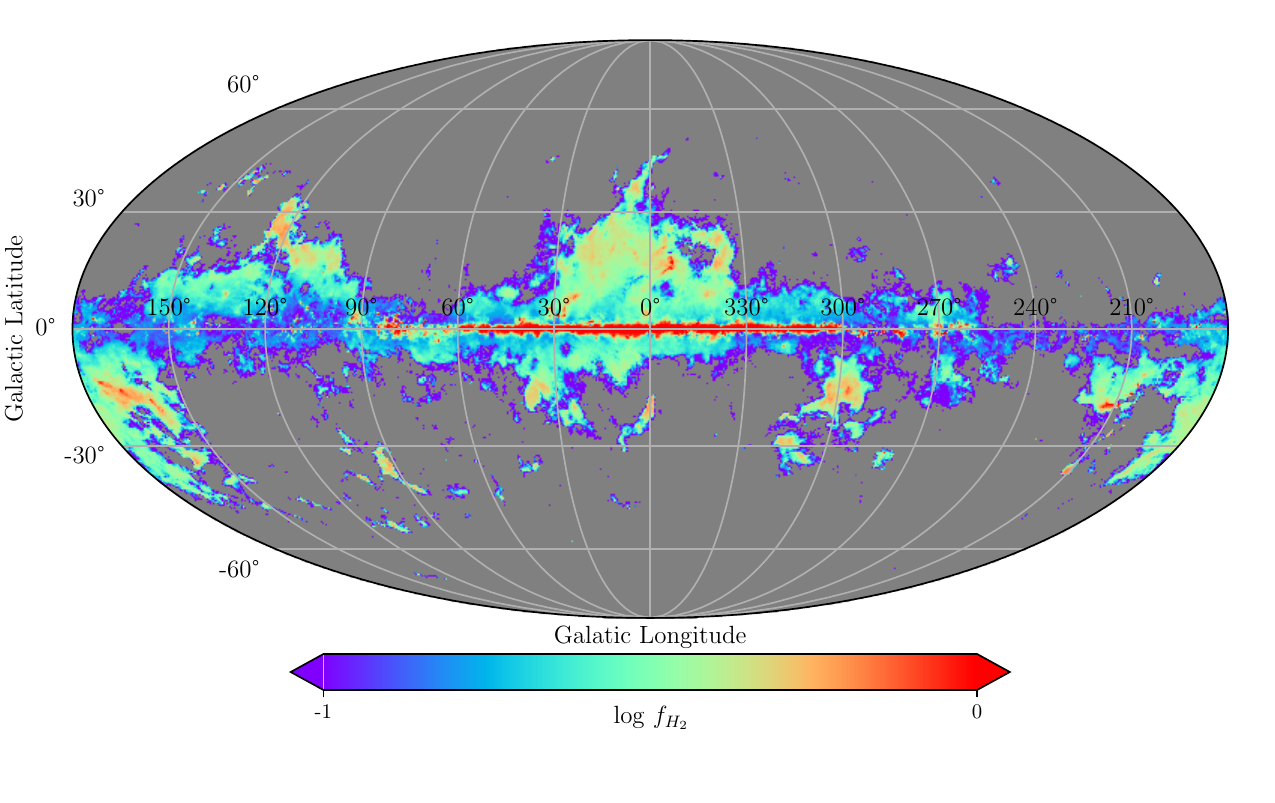}
    \caption{All-sky image of logarithm of fractional \Hmol\ abundance, \fhmol.  The mean value of log \fhmol is equal to -0.6, with standard deviation equal to 0.23}
    \label{fig:fh2_map}
\end{figure*}

\begin{figure}
   \centering
   \includegraphics[width=\hsize]{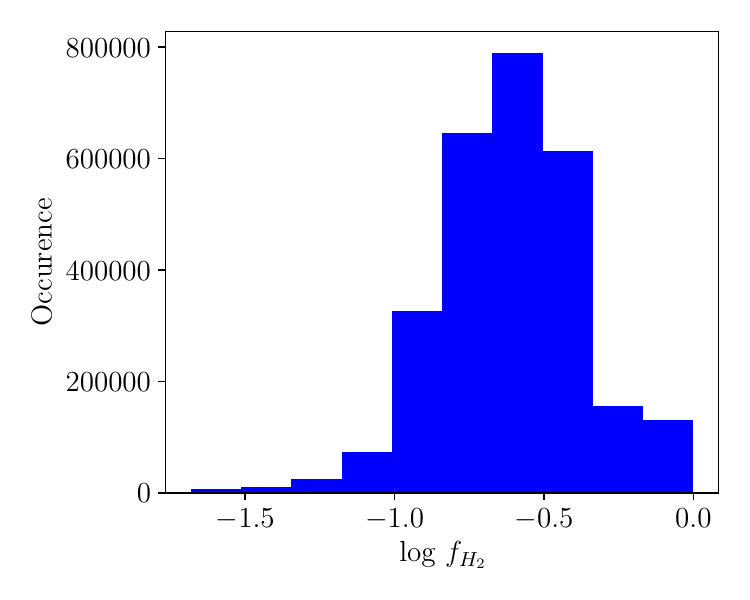}
    \caption{Distribution of values of \Hmol\ fractional abundance. Our mean \fhmol\ = 0.25 (with 1$\sigma$ range in linear scale 0.15 -- 0.43) is consistent with \cite{liszt_2023.H2.properties.diffuse.ISM.CO}, and \cite{shull_2021}, who found the average \fhmol\ to be 0.2 -- 0.4 and 0.20, respectively.}
    \label{fig:fh2_distribution}
\end{figure}

\begin{figure}
   \centering
   \includegraphics[width=\hsize]{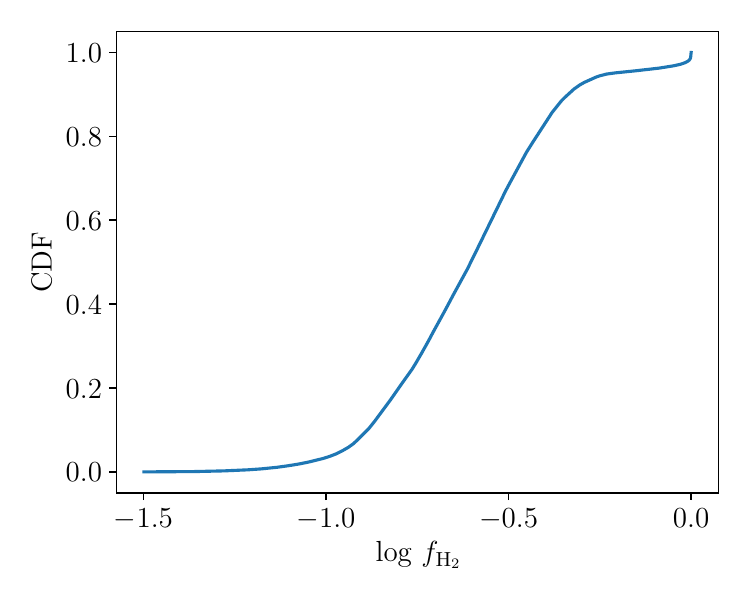}
    \caption{Cumulative distribution function of \fhmol. Atomic hydrogen is more abundant than molecular: $50\%$, and $80\%$ of the LOSs have \fhmol~$\lesssim 25\%$, and  \fhmol~$\lesssim 40\%$ respectively, while only $\sim 3\%$ have \fhmol~$\geq 80\%$.}
    \label{fig:fh2_cdf}
\end{figure}

\section{Characterization of the total gas properties of our Galaxy}
\label{sec:molecular_gas_properties}

Characterizing the total gas properties of our Galaxy is an important but challenging task. It is hard to constrain \NHmolecular, which is usually estimated using CO observations by assuming some global (and constant) \XCO\ value. However, \XCO\ varies by orders of magnitude between the various regions (Sect.~\ref{sec:XCO_comparison}), and in addition a significant portion of the molecular gas lies in diffuse clouds that are largely devoid of CO (CO-dark \Hmol). 

Dust traces the total hydrogen column density, including atomic and molecular in either CO-dark or CO-bright form. For the \NHmolecular\ estimate, we employed the non-linearities in the dust extinction (Sect.~\ref{sec:AvH2_NH2}). Thus our \NHmolecular\ map (Fig.~\ref{fig:NH2_XCO_full_sky_maps}) probes the total \Hmol\ column (CO-dark and CO-bright), which enables us to explore the various gas properties irrespective of the gas phase (atomic or molecular, CO-dark or CO-bright \Hmol). In the following sections, we use our constructed \NHmolecular\ map to explore the relative abundances between \HI\ and \Hmol\ gas, and we constrain the sky distribution of CO-dark and CO-bright \Hmol. 

\subsection{Relative abundance of atomic and molecular gas}

The molecular fractional abundance (\fhmol) is defined as:
\begin{equation}
    \label{eq:fh2}
    f_{{H}_{2}} = \frac{2N_{{H}_{2}}}{N_{{H}\,{\sc I}} + 2N_{{H}_{2}}}
\end{equation}
When \fhmol\ $\rightarrow 0$, then the total gas column is mostly in atomic form, while when \fhmol\ $\rightarrow 1$ gas is $100\%$ molecular. 

Some fraction of the dust is expected to be mixed with ionized hydrogen, which is omitted in Eq.~\ref{eq:fh2}. The contribution of $\rm H^{+}$ in dust extinction, if any, is only expected in low-extinction regions \citep[\Av~$\lesssim 0.093$~mag,][]{liszt_2014.NHI.NH2.NH.versus.EBV}. However, this is below the extinction range where our method is applicable, which is log~\NHmolecular~$\geq 19$ (Fig.~\ref{fig:AvH2_NH2}) or log~\Av~$\geq -0.5$ (Fig.~\ref{fig:gas_dust_variations}). Thus, we expect that our estimated \fhmol\ remains statistically unchanged even if some ionized hydrogen is present.

We used the \NHmolecular\ values from the map we constructed (Fig.~\ref{fig:NH2_XCO_full_sky_maps}, upper panel) and the \NHI\ data from the HI4PI survey to construct a full-sky \fhmol\ map (Fig.~\ref{fig:fh2_map}). We observe that close to the Galactic center ($\ell \lesssim 60\degr$ and $\ell \gtrsim 300\degr$, $b=0\degr$), gas tends to be fully molecular. On the other hand, \fhmol\ dramatically decreases in the outer parts of the Galactic plane ($300\degr \gtrsim \ell \gtrsim 60\degr$, $b=0\degr$), meaning that the relative abundance of \HI\ increases. 

The \fhmol\ full-sky map also shows several large-scale structures above the Galactic plane with enhanced molecular abundances ($\log$~\fhmol~$\gtrsim 0.5$ ): 1) the Taurus molecular cloud at $\ell, b \sim170\degr -15\degr$, 2) the Polaris Flare at $\ell, b \sim 125\degr, 30\degr$, 3) the North Polar Spur at $\ell, b \sim 27\degr, 10\degr$, and 4) Orion at $\ell, b \sim 170\degr, -15\degr$. Below we argue that the proximity of these molecular structures imply that the scale height of \NHmolecular\ of our Galaxy should be less than 300 pc.    

\cite{zucker_2021.cloud.distances} used the 3D extinction map of \cite{leike_2020} to derive the distances of Taurus and Orion, which are close to $150$ and $400$ pc respectively. The distance estimates to the Polaris Flare are more uncertain, ranging from 100 pc up to 400 pc \citep{heithausen_thaddeus_1990.Polaris.Flare.distance.stellar.extinction.measurements, zagury_1999.distance.polaris.flare.reddenning, Brunt_2003.molecular.clouds.distances.PCA.analysis.CO.data, schlafly_2014.cloud.distances.panstarrs.photometry, panopoulou_2016}. The North Polar Spur is a gigantic radio loop and there is a debate regarding its distance. Some Galactic models suggest that this structure is a few kpc away from the Sun. However, recently \cite{panopoulou_2021.distance.north.polar.spur} compared sub-mm and radio polarization data and found that the maximum distance of the North Polar Spur is approximately equal to 400 pc from the Sun. Altogether, all the aforementioned large-scale \Hmol\ structures above the Galactic plane are relatively close to the sun. 

To be conservative, we assume that the maximum distance of the aforementioned molecular clouds is 500 pc. Assuming that $b = 45 \degr$, which is the maximum Galactic Latitude where molecular clouds are observed in Fig.~\ref{fig:NH2_XCO_full_sky_maps}, we obtain a vertical distance from the Galactic plane equal to $350$ pc. This distance barely exceeds the outer edge of the Local Bubble \citep{pelgrims_2020.local.bubble.shape}. Therefore, most of the \Hmol\ gas of our Galaxy lies within a vertical distance of 350 pc of the Galactic plane. This is consistent with previous constraints of the \Hmol\ scale height of our Galaxy \citep{heyer_dame_2015.H2.clouds, marasco_2017}. 

Fig.~\ref{fig:fh2_distribution} shows the distribution of the logarithm of \fhmol. The distribution is nearly symmetric with an average equal to $\langle$\fhmol$\rangle$~$\approx 25 \%$, which is consistent with previous constraints derived with UV spectra of sparsely located point sources \citep{shull_2021}. From the cumulative distribution function of \fhmol, shown in Fig.~\ref{fig:fh2_cdf}, we calculate that a large fraction ($80\%$) of our Galaxy with \NHmolecular~$\geq 10^{20}$~\ColDens\ has \fhmol~$\lesssim 40 \%$. This indicates that either the majority of the molecular gas of our Galaxy lies in diffuse molecular clouds or that atomic clouds are more abundant than molecular clouds. Below we argue that the latter is more probable.

\begin{figure}
   \centering
   \includegraphics[width=\hsize]{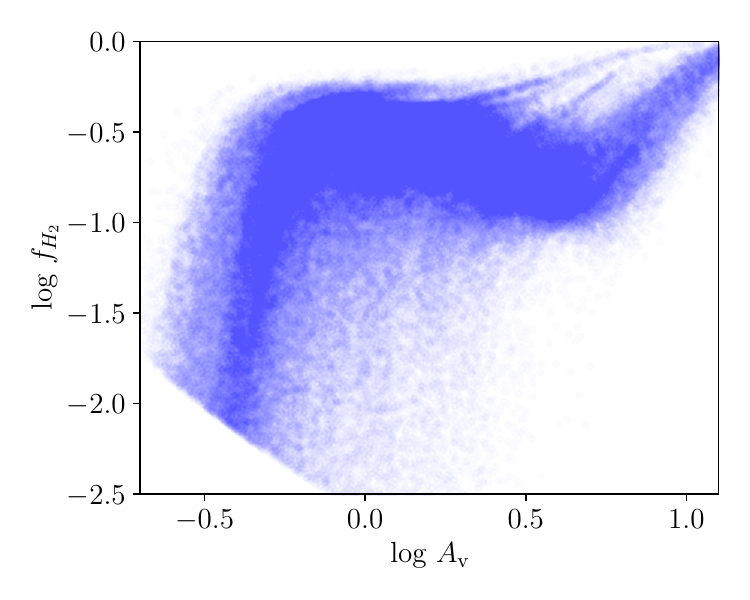}
    \caption{Molecular fractional abundance as a function of dust extinction. The nonlinear increase of \fhmol\ at $-0.5 \lesssim$~$\log$~\Av~(mag)~$\lesssim -0.2$ corresponds to the atomic to molecular gas transition. At larger \Av, \fhmol\ decreases by a factor of $1.5$, most probably because even along high-extinction sightlines there exist more atomic than molecular clouds.}
    \label{fig:fh2_AV}
\end{figure}

We display \fhmol\ versus \Av\ in Fig.~\ref{fig:fh2_AV}. For $-0.5 \lesssim$~$\log$~\Av~$\lesssim -0.2$, $\log$~\fhmol\ increases from -2.0 ($1 \%$ ) to -0.5 ($30\%$), because the atomic gas transitions to molecular in this range of extinction, given an ambient radiation field of $\sim 1$ Habing. The observed non-linear increase of \fhmol, is similar to the predicted behaviour of semi-analytical models of the \HI~$\rightarrow$~\Hmol\ transition \citep{draine_bertoldi_1996, krumholz_2008.HI.H2.transition, krumholz_2009.hi.h2.transition, mckee_krumholz_2010, Sternberg_2014, sternberg_2023, bialy_sternberg_2016.H2.formation}. For $-0.2 \lesssim$~$\log$~\Av~$\lesssim 0.7$, we observe that $\log$~\fhmol\ slightly decreases from -0.5 ($30\%$) to -0.7 ($20\%$). Although \Av\ increases by a factor of 5, \fhmol\ decreases by a factor of 1.5. This suppression of \fhmol\ could be due to young stars that are illuminating the surrounding material, hence responsible for photo-destruction of the molecular gas \citep{madden_2020.method.for.tracing.CO.dark.in.galaxies.using.cloudy.simulations}. We consider this scenario unlikely because the decline of \fhmol\ starts at \Av~$\approx 1$~mag which is typical of translucent (non star-forming) clouds, such as the Polaris Flare.

We conclude simply that there are more atomic
than molecular clouds in most of the LOSs of our Galaxy, even
where we observe molecular gas. The concatenation of atomic clouds increases \NHI, and consequently \Av, but not \NHmolecular, hence the reduction of \fhmol. 

This suppression of \fhmol\ is more prominent close to the Galactic plane: the majority of LOSs with $\log$~\fhmol~$\approx -1$ and $\log$~\Av~$\approx 0.7$ are observed at $|b| \approx 5 \degr$. However, even in regions at high Galactic latitudes ($|b| >40\degr$), which cover a large portion of the sky, we observed a similar trend. For $|b| > 40 \degr$, there are several LOSs with $\log$~\fhmol~$\approx -0.5$ and $\log$~\Av~$\approx 0$. 

This suppression is expected from the following approximate calculation. The average number of atomic clouds per LOS, as indicated by \HI\ emission line data, is close to three for $|b| > 40\degr$ \citep{panopoulou_lenz_2020.clouds.number}. On the other hand, the number of molecular clouds per LOS is a maximum two because: 1) molecular gas is associated with the \HI\ cloud having an exceptionally large column, usually observed in \HI\ emission data at low 
velocities measured with respect to the local standard of rest (LSR), and 2) molecular gas associated with \HI\ clouds at intermediate LSR velocities tend to have lower column densities than clouds with lower velocities and only occasionally show any sign of \Hmol\ \citep[e.g.][]{lenz_2015, rohser_2016.molecular.gas.IVCs, rohser_2016.all.sky.molecular.IVCs}. Thus, we expect the ratio of the number of molecular clouds to atomic clouds to be a maximum of $\approx 0.67$, implying that the high-Galactic-latitude sky has more atomic than molecular clouds. The majority of LOSs with $\log$~\fhmol~$\approx 0$ and $\log$~\Av~$\gtrsim 0.7$ are toward the Galacic midplane ($\ell \lesssim 30 \degr $ and $\ell \gtrsim 330 \degr$, $b = 0\degr$). 

Our conclusion about the decrease of \fhmol\ at high \Av\ is also evident in Fig.~ 20 of \cite{planck_collaboration_2011.diffuse.ISM.properties.molecular.hdyrogen}. These authors compare \fhmol\ and \NHtotal, where the \NHmolecular\ measurements used for the computation of \fhmol\ have been obtained from UV spectra \citep{rachford_2002, 2009ApJS_Rachford, Gillmon2006, wakker_2006.UV.survey}. This suggests that despite the significant beam difference ($16\arcmin$ in our study, and pencil beams for the UV data), the observed trend is robust. The concatenation of multiple clouds along the LOS can significantly complicate observational studies of the \HI~$\rightarrow$~\Hmol\ transition \citep{browning_2003.concatenation.H2.clouds}.

\begin{figure*}
   \centering
   \includegraphics[width=0.97\hsize]{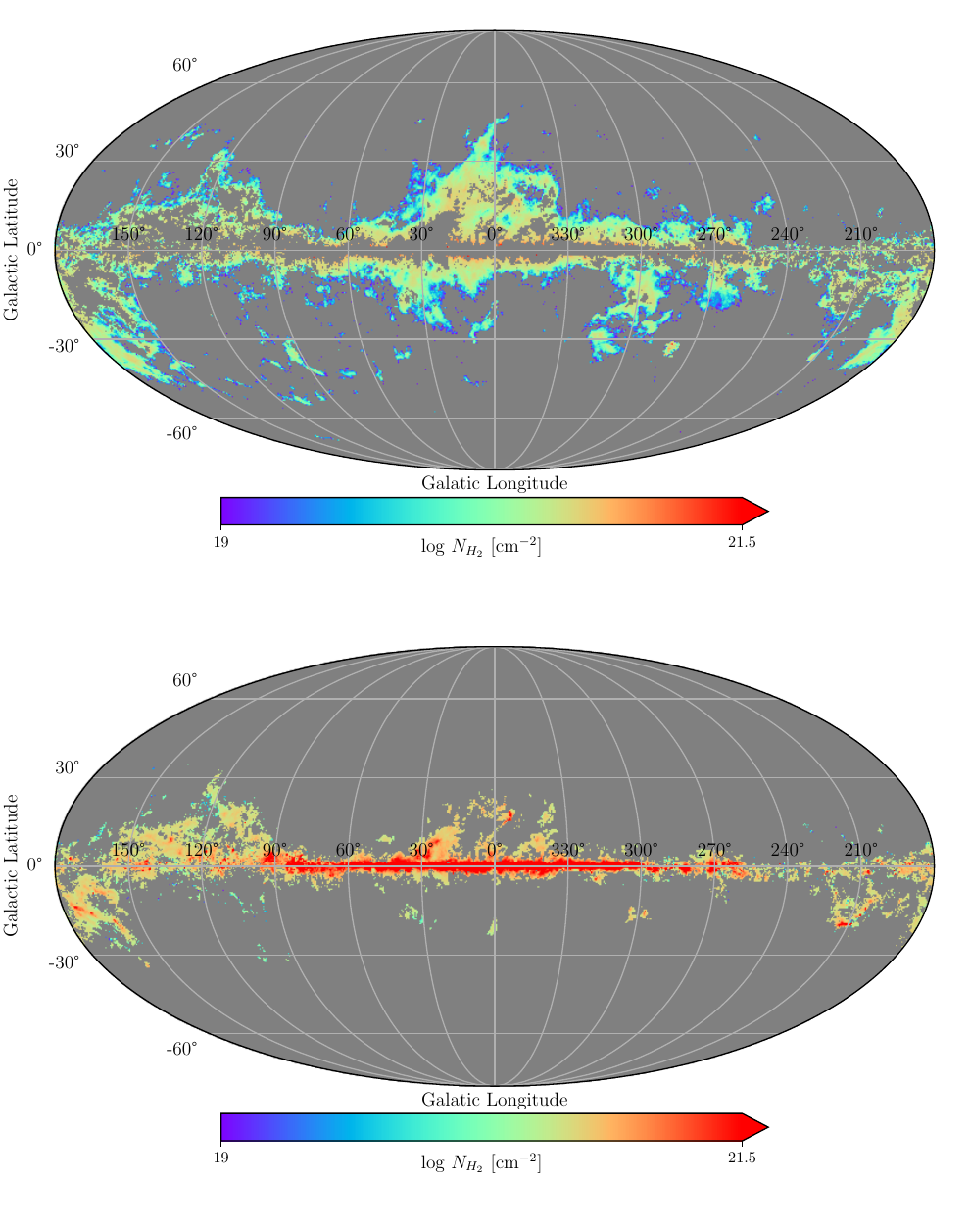}
    \caption{\textbf{Upper panel:} Full-sky CO-dark \NHmolecular\ map, showing sightlines with \NHmolecular~$\geq 10^{19}$~\ColDens\ but devoid of CO (\WCO~$\leq 1$~K~\kms). \textbf{Bottom panel:} CO-bright \NHmolecular\ map, corresponding to \NHmolecular~$\geq 10^{19}$~\ColDens\ and \WCO~$> 1$~K~\kms. The CO-dark \Hmol\ spans an area approximately equal to $17\%$ of the total sky.  This is a factor of two larger than that of the CO-bright \Hmol, and extends up to high Galactic Latitude. On the other hand, CO-bright \Hmol\ is mostly concentrated towards the Galactic plane. Moving away from the Galactic plane, we observe that \Hmol\ transitions from CO-bright to CO-dark form at $|b| \approx 5\deg$.  Although we still observe a few CO-bright \Hmol\ at higher latitude above the Galactic plane, most of the molecular gas is in CO-dark form there.}
    \label{fig:CO_components_full_sky_maps}
\end{figure*}

\subsection{The properties of CO-dark and CO-bright \Hmol}
\label{sec:co_dark_H2}

There is a significant fraction of molecular gas that is not traced by CO. This CO-dark \Hmol\ is observed in diffuse ISM clouds where the \HI\ $\rightarrow$ \Hmol\ transition is ongoing. In regions where the \HI\ $\rightarrow$ \Hmol\ transition is well advanced, carbon monoxide formation can also occur.
Between the extinctions that characterize the predominance of \Hmol\ and CO, carbon transitions from \Cplus\ to \Catomic. For typical ISM conditions, the onset of the \Hmol\ formation takes place at \Av~$\lesssim 0.5$ mag, the \Cplus~$\rightarrow$~\Catomic\ transition at $1 \lesssim$ \Av\ $\lesssim 3$ mag, while CO becomes the dominant carrier of carbon at \Av\ $\geq\ 3$ mag. During the transition from \Cplus\ to \Catomic, \Hmol, whose electronic transitions due to photon absorption (Lyman-Werner lines) start at 11.18 eV, has already started forming but carbon is still in atomic form. Sufficient shielding is required to reduce the energy of incident photons bellow the CO photo-dissociation threshold (11.09 eV), hence allowing the buildup of the abundance of CO. Until this energy threshold is reached, the \Hmol\ absorption lines are detectable but without any corresponding CO emission line. In this case, molecular hydrogen is primarily mixed with atomic carbon. 

The integrated intensity of the \CII\ line (at 158 \mum) is proportional to the gas density squared \citep{goldsmith_2018} at the densities characteristic of diffuse and translucent clouds. Towards several clouds with untraceable CO, the observed \CII\ intensity, even assuming all carbon is in the form of C$^+$, is larger than that expected solely from C-\HI\ collisions \citep{langer_2010}. The enhancement of the \CII\ line indicates the presence of \Hmol\ gas. 

The above discussion makes it clear that our ability to trace CO-dark \Hmol\ relies on the observational uncertainties in the CO line. Even in diffuse regions, there should be some CO emission, although very weak. Strictly speaking CO-dark \Hmol\ is located in regions where the atomic carbon is more abundant than CO, or in terms of observables, when the \CII\ or \CI\ lines are stronger than CO. In this work, we adopt a detectability threshold of the  CO line in order to define CO-dark \Hmol. The detectability limit is determined by the noise in the CO data of \cite{Dame2001}, which is 1~K~\kms. 

We adopt the definition that molecular gas is in CO-dark form when the following conditions are satisfied: 1) $\log$~\NHmolecular~$\geq 19$~\ColDens, and 2) \WCO~$\leq 1$~K~\kms. On the other hand, we define molecular gas that is traceable by CO, CO-bright \Hmol, as: 1) $\log$~\NHmolecular~$\geq 19$~\ColDens, and 2) \WCO~$> 1$~K~\kms.

Fig.~\ref{fig:CO_components_full_sky_maps} shows our inferred CO-dark (upper panel), and CO-bright (lower panel) \NHmolecular\ maps. We observe that CO-dark \Hmol\ extends to high Galactic latitudes ($|b| \approx 60 \degr$), and has generally lower column densities than CO-bright \Hmol, because CO-dark \Hmol\ is characteristic of regions having lower densities than CO-bright \Hmol. CO-dark \Hmol\ increases towards lower Galactic Latitudes, but in the Galactic plane almost all of the molecular gas is in CO-bright form due to the enhanced densities, and column densities there. CO-bright \Hmol\ lies close to the Galactic plane and only a few CO-emitting clouds can be seen at higher latitudes ($|b| \approx 30\degr$). 

We estimate that CO-dark \Hmol\ covers $\sim 17 \%$ of the total sky area, while CO-bright \Hmol\ $\sim 9\%$ of the sky. These values, however, should be considered with some caution because the \WCO\ map of \cite{Dame2001} is considered to be complete only for $|b| \leq 32\degr$, as indicated by the comparison between the CO intensities with far-infrared, and \HI\ data\footnote{\url{https://lambda.gsfc.nasa.gov/product/foreground/fg_wco_info.html}}. Thus, the \WCO\ map may miss some CO emission at high Galactic latitudes ($|b| > 32\degr$), although we do not expect the emission at such high $b$ to be significant. Our argument is supported by the CO survey of \cite{dame_2022.complete.CO.survey.northern.sky}, which is complete for the entire Northern sky, and shows that very few LOSs have detectable CO emission -- assuming the sensitivity of the \cite{Dame2001} survey -- at $|b| > 40\degr$. Therefore, our inferred fractional sky coverage of the CO-dark, and CO-bright \Hmol\ components should be considered as upper and lower limits respectively. Finally, we find that $66\%$ of the sightlines with $\log$~\NHmolecular~(\ColDens)~$\geq 19$ are in CO-dark form, while the remaining $34\%$ of sightlines are in CO-bright form. If in the definition of CO-dark \Hmol\ we increase the threshold to $\log$~\NHmolecular~(\ColDens)~$\geq 20$, we obtain that $57\%$ of the LOSs are in CO-dark and $43\%$ in CO-bright form. 

\begin{figure}
   \centering
   \includegraphics[width=\hsize]{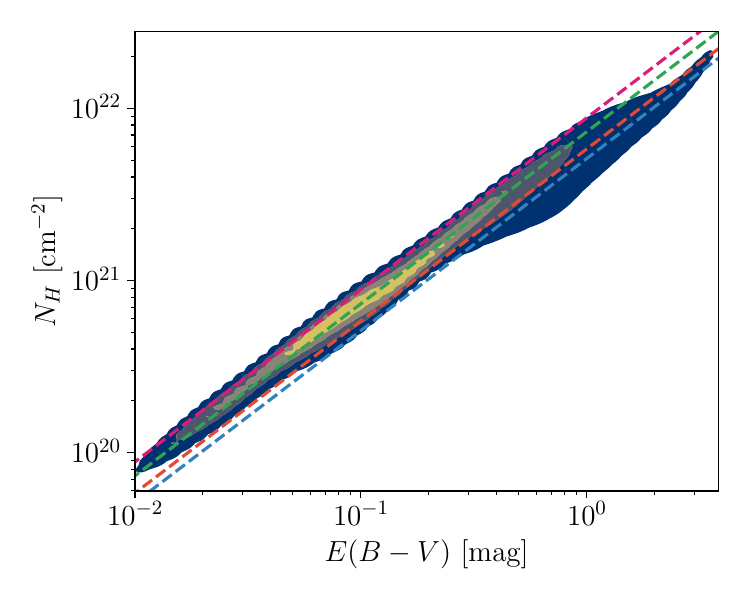}
    \caption{Total hydrogen column density versus dust reddening. The colored curves correspond to previously constrained dust-to-gas ratios. The blue, red, green, and magenta curves correspond to \NHtotal~/~\EBV~=~5.1 \citep{kalberla_2020}, 5.8 \citep{1978ApJ_BSD}, 7.0 \citep{planck_collaboration_2014.all_sky.dust.model.miville}, and 8.8 \citep{lenz_2017}  $\times 10^{21}$~\ColDens~mag$^{-1}$ respectively.}
    \label{fig:gas_correlations_NH}
\end{figure}

\section{Dust-to-gas ratio}
\label{sec:dust-to-gas_ratio}

\NHtotal~/~\EBV\ is a key ratio for ISM studies. Existing constraints on the Galactic average \NHtotal~/~\EBV\ 
vary significantly \citep{liszt_2023.H2.properties.diffuse.ISM.CO}. The constraint of \cite{1978ApJ_BSD}, \NHtotal~/~\EBV~$=5.8 \times 10^{21}$~\ColDens~mag$^{-1}$, has been the gold standard for several decades. But, several recent works \citep{lenz_2017, planck_collaboration_2014.all_sky.dust.model.miville, shull_2021, liszt_2023.H2.properties.diffuse.ISM.CO, liszt_2014.NHI.NH2.NH.versus.EBV, Gillmon2006, 2009ApJS_Rachford} have found larger values, 
although \cite{kalberla_2020} quote a smaller value for the Galactic average of \NHtotal~/~\EBV. However, when \cite{kalberla_2020} excluded regions close to the Galactic plane, they found a larger value than \cite{1978ApJ_BSD}. 

Variations in the \NHtotal~/~\EBV\ ratio can be induced by different factors: 1) environmental ISM conditions \citep{paradis_2012}, 2) dynamical effects \citep[e.g., imperfect coupling between dust and gas motions,][]{squire_hopkins_2018.drag.instablities, hopkins_squire_2018.drag.instabilities.analytical, hopkins_2020, hopkins_2022.dust.in.the.wind}, 3) the V--band could be less than optimal for measuring dust extinction \citep{Jura_1980.grain.size.growth.Orion, butler_2021.nearUV.extinction.correlates.better.with.NH.than.Av}. 

Atomic clouds at high Galactic Latitudes, where \NHtotal~$\approx$~\NHI, have larger dust-to-gas ratio, \NHtotal~/~\EBV\ $= 8 - 10 \times 10^{21}$~\ColDens~mag$^{-1}$ \citep{liszt_2014.ratio.NHI.EBV.high.latitudes,lenz_2017}, than the Galactic plane, where \NHtotal~/~\EBV\ lies in the range $5.8 - 6.1 \times 10^{21}$~\ColDens~mag$^{-1}$ \citep{1978ApJ_BSD, rachford_2002, 2009ApJS_Rachford, shull_2021}. Uncertainties in the employed techniques (UV or radio spectroscopy, far-infrared dust emission) could also induce variations in the constrained \NHtotal~/~\EBV\ \citep{shull_panopoulou_2023}.

The only direct way to constrain \NHtotal, hence the dust-to-gas ratio, is by the use of spectra of background objects \citep[e.g.,][]{1978ApJ_BSD, sheffer_2008, Gillmon2006, 2009ApJS_Rachford, shull_2021}. However, obtaining spectra is time consuming and only a limited number of LOSs can be targeted, usually toward relatively bright stars and quasars. The sky is sparsely sampled, which implies that the \NHtotal~/~\EBV\ estimates could be biased as spectroscopic surveys usually focus on the Galactic plane. Indirect methods for constraining the dust-to-gas ratio require some assumptions and approximations, but they have the advantage of uniformly sampling the Galaxy \citep{paradis_2012, planck_2011.dark.gas.estiamtes.bernard,planck_collaboration_2014.all_sky.dust.model.miville,kalberla_2020}. 

In Fig.~\ref{fig:gas_correlations_NH} we present our \NHtotal~/~\EBV\ constraint and compare it to previous work. The figure shows the total hydrogen column density, calculated as \NHtotal~=~\NHI~+~2\NHmolecular, including data covering the full celestial sphere; the data employed are the \NHI\ full-sky map from \cite{hi4pi}, and our constructed \NHmolecular\ map (Fig.~\ref{fig:NH2_XCO_full_sky_maps}). The colored dashed lines correspond to previous constraints. 

All the depicted lines are consistent, within variations, with our data. However, the green and red lines, which correspond to $ 7.0\times 10^{21}$~\ColDens~mag$^{-1}$ \citep{planck_collaboration_2014.all_sky.dust.model.miville}, and $5.8\times 10^{21}$~\ColDens~mag$^{-1}$ \citep{1978ApJ_BSD} respectively, agree more closely with the mean behaviour of our data.

We can approximate, to first order, the \NHtotal\ versus \EBV\ relation with a line. But, a closer inspection to Fig.~\ref{fig:gas_correlations_NH} suggests that some non-linearities arise. For \EBV~$\leq 0.1$~mag, the average behaviour of our data agrees well with the green line describing the results of \cite{planck_collaboration_2014.all_sky.dust.model.miville}, while for \EBV~$> 0.1$~mag the red profile of \cite{1978ApJ_BSD} seems to be a better fit. This indicates that at larger \EBV\ the dust-to-gas ratio decreases, which could happen due to dust coagulation. 

\section{Discussion}
\label{sec:discussion}

\subsection{A historical perspective on the extinction residuals}
\label{sec:discussion_historical_perspective_extinction_residuals}

\cite{boulanger_1996} found that dust intensity correlates with the \HI\ integrated intensity (\WHI). The correlation is linear for \WHI~$\leq\ 300$~K~\kms\ and non-linear for \WHI~$> 300$~K~\kms. Assuming that the \HI\ line is optically thin the \HI\ column density is given by \NHI\ = 1.823 $\times 10^{18}$ \WHI\ \citep{2011book_Draine}, yielding a transition at \NHI\ $\approx 5 \times 10^{20}$ \ColDens. \cite{boulanger_1996} argued that the non-linear excess between dust intensity and \WHI\ is due to the presence of molecular gas that is only probed by dust intensity which traces both the atomic and molecular gas phase, while the 21 cm emission line probes only the atomic phase. Their argument was based on the results of \cite{savage_1977}, who found, using the \Hmol\ absorption lines, that the molecular gas fraction rapidly increases when \NHI\ $\approx 5 \times 10^{20}$ \ColDens. An alternative explanation of the non-linear relationship between dust intensity and \WHI\ for \WHI~$> 300$~K~\kms\ 
is that at some point the emission line becomes optical thick and \WHI\ misses a significant faction of the \HI\ due to self-absorption, and reduction of the observed intensity \citep{Burton_1993}.

The same debate is found in more recent papers. \cite{fukui2014}, and \cite{fukui_2015} explored the correlation between \WHI\ and the dust opacity (\TauDust), but they also added one more free parameter in the analysis: the dust temperature (\Td). The dust temperature is anticorrelated with the total gas column. Warm dust traces the outer regions of ISM clouds which are primarily atomic, while cold dust traces the inner, primarily molecular, regions of the clouds. \cite{fukui_2015} showed that the correlation between \WHI\ and \TauDust\ is linear when \Td\ $> 20$ K. When dust is cold (\Td\ $< 20$ K), \TauDust\ increases non-linearly with respect to \WHI. \cite{fukui_2015} argued that this nonlinear excess of \TauDust\ with respect to \WHI\ is consistent with the presence of optically thick \HI\ gas, and although some \Hmol\ gas might be present, it is less abundant than \HI. At sufficiently high column densities, the increase in \TauDust\ would be dominated by the formation of the molecular gas, which is traceable by the CO emission line. However, \cite{fukui_2015} suggest that the excess of \TauDust\ at low \WHI\ is not due to CO-dark \Hmol, but due to optically thick \HI. 

\cite{murray_2018} addressed the question of whether the \HI\ optical depth effects can explain the excess of \TauDust\ when \WHI\ is high. They compiled a large catalogue of \TauHI\ absorption data, which provide the only direct way to estimate \TauHI. Their sample consisted of archival data towards 151 LOSs that are covered within the GALFA survey footprint. Following the analysis of \cite{fukui_2015}, \cite{murray_2018} found that as \TauHI\ increases, the \HI\ gas that is missed by the 21 cm emission line is less than a factor of 1.3; similar results have been found by \cite{liszt_2019.ebv.verus.HI}. This means that the increase in \TauHI\ cannot adequately explain the nonlinear relation between \WHI\ and \TauDust. 

We found that nonlinear deviations in the extinction with respect to \NHI\ correlate well with \NHmolecular. Our findings (Fig.~\ref{fig:AvH2_NH2}) support the claims of \cite{boulanger_1996}, and \cite{murray_2018} -- that extinction residuals are induced by molecular hydrogen. However, given the spread in our obtained \Av\ versus \NHmolecular\ relation (Fig.~\ref{fig:AvH2_NH2}), some, but not dominant, contribution of optically thick \HI\ gas is also plausible.

\subsection{Variations of \XCO}

The \XCO\ factor is an important observable for estimating the molecular gas content of molecular clouds and entire galaxies \citep{lee_2014.XCO.perseus.molecular.cloud, sun_phangs_2018.cloud.scale.properties.of.molecular.clouds.in.15.nearby.galaxies, sun_phangs_2020.molecular.cloud.properties.comparison.with.the.host.galaxies}. \XCO\ is expected to vary as a function of the local ISM properties, such as metallicity, density, and intensity of the UV background field \citep{gong_2017, gong_2018}. Observations of local ISM clouds suggest that the \XCO\ factor is relatively constant, with variations not exceeding a factor of two \citep{bolato_2013, liszt_2016.XCO}. 

The constancy of \XCO\ indicates that ISM clouds in our Galaxy have generally similar properties, such as temperature, and surface densities. This uniformity has important consequences regarding the nature of ISM clouds. For example, it explains the relation between non--thermal velocity linewidths and cloud size -- known as Larson's relation \citep{larson_1981.larson.relationI} -- as the outcome of clouds with magnetically--supported boundaries \citep{mouschovias_1995.larson.relation.Bfields}. A plethora of theoretical works address the question of why ISM clouds in the Milky Way are so similar \citep[e.g.,][]{narayanan.hopkins2013.XCO.is.constant.in.our.Galaxy, clark_glover2015.XCO.depends.on.SFR}.

We have found that CO-emitting regions are most probably observed when $20.5 \lesssim$~$\log$~\NHmolecular~(\ColDens)~$\lesssim 21$, and $3 \lesssim$~\WCO~$\lesssim 8$~K~\kms\ (Fig.~\ref{fig:gas_correlations_CO}). Our results, thus suggest, in agreement with previous studies, that our Galaxy favors clouds with relatively uniform CO and surface density properties. This is also supported by our inferred \fhmol\ distribution (Fig.~\ref{fig:fh2_distribution}) which is nearly symmetric and narrow; the 1 $\sigma$ of the distribution is 0.15 - 0.43. However, in individual molecular clouds (on angular scales of a few arcminutes), \XCO\ can vary by an order of magnitude from the outer to the inner parts, as atomic gas transitions to fully molecular form.

\subsection{Molecular hydrogen column densities as a function of the Galactocentric distance}

The Galactic plane contains most of the molecular gas of our Galaxy. Several authors employed data from the CO survey of \cite{Dame2001} to model the \Hmol\ surface as a function of distance from the Galactic center, but omitted any contribution from CO-dark \Hmol\  \citep{bronfman1988.CO.survey.radia.distribution.of.molecular.clouds, wouterloot1990.distribution.galactic.wraps, nakanishi_sofue2006.3d.distribution.HI.H2.gas.based.on.CO.data,nakanishi_sofue2016.3d.distribution.HI.H2.gas.based.on.CO.data, marasco_2017, miville_deschenes2017.H2.properties.of.Galaxy.disk}.

A common conclusion from these works is that the \Hmol\ surface density decreases nonlinearly at large Galactocentric distances, although the different studies differ on the exact cutoff distance. It would be interesting to see how the effect of the CO-dark \Hmol\ could affect the relative abundance between the \HI\ and \Hmol\ surface densities as a function of Galactocentric distance. The presence of CO-dark \Hmol\ could yield more extended \Hmol\ surface density profiles, as shown by \cite{pineda_2013} who estimated that the \Hmol\ gas in CO-dark form corresponds to  $\sim 20\%$ of the total molecular abundance at 4 kpc and $\sim 80 \%$ at 10 kpc. 

Our \fhmol\ map (Fig.~\ref{fig:fh2_map}) shows that in the Galactic plane gas is mostly molecular for $\ell \lesssim 60\degr$ and $\ell \gtrsim 300\degr$, while it becomes fully atomic for $ 90\degr \lesssim \ell \lesssim 270\degr$. Despite the presence of CO-dark \Hmol, the overall contribution of molecular gas at large Galactic latitudes is minimal; the abundance of CO-dark \Hmol\ becomes maximum at $150 \degr \lesssim \ell \lesssim 210\degr$, but there \fhmol\ is only $10\%$. Thus, the CO-dark \Hmol\ abundance seems to increase with Galactocentric distance, but this only affects the exact distance where \HI\ starts becoming more abundant than \Hmol.

\subsection{CO-dark \Hmol\ estimates in the Solar neighborhood}
\label{sec:co_dark_estimates_past_works}

Existing constraints on the relative abundance of CO-dark \Hmol\ in relation to total gas imply that a significant amount of molecular gas exists in CO-dark form \citep{planck_2011.dark.gas.estiamtes.bernard, paradis_2012, kalberla_2020}. As stated by \cite{DiLi_2018.OH.traces.dark.gas}, the derived abundances of CO-dark \Hmol\ depend on the detectability threshold of the CO line. This was demonstrated by \cite{donate_magnani_2017.dark.gas.detectability.threshold.decreases.dark.gass.fraction} who performed a high-sensitivity CO survey toward a region with existing, lower sensitivity CO data. The (new) high-sensitivity survey showed that the relative abundance of CO-dark with respect to the total \Hmol\ gas was approximately two times smaller than what had been inferred by the (previous) low-sensitivity CO survey. That said, the definition of CO-dark \Hmol\ is vague, making the comparison of the various results in the literature challenging.

Our definition of CO-dark and CO-bright \Hmol\ components differs from several past works \citep[e.g.,][]{paradis_2012, kalberla_2020}. In these works, the decomposition is based on the models of photo-dissociation regions \citep[PDRs,][]{wolfire_2010.dark.gas.theoretical.PDR.model, velusamy_2010}. In PDR models, molecular gas phases are organized into layers with CO-dark corresponding to a diffuse, and hot layer that surrounds the dense, and cold CO-bright \Hmol\ layer. 

We define CO-dark \Hmol, molecular hydrogen in LOSs devoid of CO. Using this definition, CO-dark and CO-bright \Hmol\ never co-exist (Fig.~\ref{fig:CO_components_full_sky_maps}). However, our definition allows us to emphasize that a significant fraction of the molecular sky ($\sim 60 \%$) is untraceable in CO maps, when a detectability threshold equal to 1~K~\kms\ is employed (Sect.~\ref{sec:co_dark_H2}). Our estimates should be treated as upper limits, given the incompleteness of the \cite{Dame2001} CO survey at high Galactic latitudes. However, we do not expect significant variations from these estimates because CO emission is minimal in that portion of the Milky Way. 

\section{Conclusions}
\label{sec:conclusions}

Molecular hydrogen is the most abundant molecule in the ISM, but it is hard to directly observe it and infer its column density. Several indirect methods have been used to probe \NHmolecular, usually by observing the CO emission lines and employing a CO -- \Hmol\ conversion factor (\XCO, Eq.~\ref{eq:xco_factor}). However, CO misses the diffuse molecular hydrogen that lies in translucent clouds, where densities are not sufficient to allow rapid formation of CO and column densities are insufficient to provide effective shielding against photdissociation.
The result is CO-dark \Hmol. We present a new indirect method for estimating \NHmolecular\ using \NHI\ and dust extinction data. 

We show that the nonlinear increments of dust extinction with respect to \NHI, %
\AvMolecular, 
observed for \NHI~$\geq 3\times 10^{20}$~\ColDens\ \citep{lenz_2017}, are due to the presence of molecular hydrogen. We measure the implied extinction due to molecular hydrogen, \AvMolecular, using publicly available extinction maps \citep{green_bayestar_2019.latest.update}, toward LOSs with direct \NHmolecular\ measurements that have been obtained with spectroscopic data. We show that \AvMolecular\ correlates with \NHmolecular\ (Fig.~\ref{fig:AvH2_NH2}). We obtain a best-fit model for the \AvMolecular\ -- \NHmolecular\ relation (Eq.~\ref{eq:Av_H2_fit}). The fitted relation is the basis of our methodology for estimating \NHmolecular. 

We employed the following assumptions, which are motivated by the literature: 1) \NHI~/~\EBV~$= 8.8 \times 10^{21}$~\ColDens~mag$^{-1}$  \citep{lenz_2017}, 2) $R_{V} = 3.1$ \citep{cardelli_1989.selective.extinction, schlafly_2014.cloud.distances.panstarrs.photometry}. We used the fitted relation (Eq.~\ref{eq:Av_H2_fit}) to construct a full-sky \NHmolecular\ map at $16\arcmin$ resolution (top panel in Fig.~\ref{fig:NH2_XCO_full_sky_maps}); our \NHmolecular\ map does not employ CO observations, and hence traces both the CO-dark (diffuse), and CO-bright \Hmol. The accuracy of the obtained \NHmolecular\ estimates is better than a factor of two, three, and five with a probability $68\%$, $95\%$, and $98\%$ respectively.

We compared our inferred \NHmolecular\ estimates with those of \cite{kalberla_2020} and found good agreement (Sect.~\ref{sec:comparison_kalberla_map}); there is some discrepancy which can be explained by the assumptions of the employed methods. We also compared our \NHmolecular\ map with the CO total intensity obtained from the composite survey of \cite{Dame2001} (Sect.~\ref{sec:XCO_comparison}). We constructed a map of \XCO, which is the ratio between \NHmolecular\ and \WCO\ (Fig.~\ref{fig:XCO_map}). We estimate that the Galactic value of \XCO\ is approximately equal to $2 \times 10^{20}$~\ColDens~(K~\kms)$^{-1}$, which is consistent with previous constraints \citep{bolato_2013, lada_dame_202.new.average.XCO}. However, we found that \XCO\ can vary by orders of magnitude on angular scales of a few arcminutes. 

In the diffuse (peripheral) parts of ISM clouds \XCO~$> 10^{21}$~\ColDens~(K~\kms)$^{-1}$, because a large portion of the molecular hydrogen there is not traced by CO, hence the large conversion factor. On the other hand, in the inner (denser) parts of the cloud, where CO is well mixed with molecular hydrogen, \XCO\ decreases to $\sim 5 \times 10^{19}$ \ColDens~(K~\kms)$^{-1}$. Toward LOSs with \Av\ $ \gtrsim 10$~mag -- mostly encountered in the Galactic plane -- the CO line saturates and \XCO~$> 10^{21}$~\ColDens~(K~\kms)$^{-1}$ (Fig. ~\ref{fig:gas_correlations_CO}).

We used our \NHmolecular\ map and archival \NHI\ data to construct a map (Fig.~\ref{fig:fh2_map}) of the molecular fractional abundance (\fhmol), which is the ratio between \NHmolecular\ and the total gas column (Eq.~\ref{eq:fh2}). For our Galaxy, we find an average \fhmol\ equal to $25\%$ (Fig.~\ref{fig:fh2_cdf}), which is consistent with previous estimates \citep{bellomi_2020, shull_2021}. In the Galactic plane ($b = 0\degr$), for $90\degr \lesssim \ell \lesssim 270 \degr$, \fhmol\ decreases abruptly from unity to 0.1. This agrees with previous constraints of the molecular gas distribution of our Galaxy \citep{miville_deschenes2017.H2.properties.of.Galaxy.disk, marasco_2017}. We find that \fhmol\ approaches unity, implying that gas is fully molecular, only toward $3\%$ of the LOSs with \NHmolecular~$\geq 10^{20}$~\ColDens. A large fraction of the sky ($\sim 66 \%$) with \NHmolecular~$\geq 10^{19}$~\ColDens\ is undetected in CO maps, when the sensitivity in the CO data is \WCO~=~1~K~\kms\ (Sect.~\ref{sec:co_dark_H2}). 

We explored the correlation between \NHtotal\ and \EBV\ (Sect.~\ref{sec:dust-to-gas_ratio}). Our data favor a \NHtotal~/~\EBV\ ratio close to $7 \times 10^{21}$~\ColDens~mag$^{-1}$ for \EBV~$\lesssim 0.1$ mag, consistent with the constraint of \cite{planck_collaboration_2014.all_sky.dust.model.miville}. For \EBV~$\gtrsim 0.1$ mag, our data fit better with the constraint of \cite{1978ApJ_BSD}, \NHtotal~/~\EBV~$\approx 5.8\times 10^{21}$~\ColDens~mag$^{-1}$. All the constructed maps will be made publicly available.

\begin{acknowledgements}
We are grateful to the reviewer, J. M. Shull, for a very constructive report which significantly improved the manuscript. We would like to thank  M. Heyer, P. Kalberla, K. Tassis, S. Stahler, and G. V. Panopoulou for useful comments on the draft and M. A Miville-Deschen\^{e}s, C. Murray, S. Bialy, A. Bracco, and B. Hensley for fruitful discussions. We also thank P. Kallemi for the graphic design in part of the figures. This work was supported by NSF grant AST-2109127. This work was performed in part at the Jet Propulsion Laboratory, California Institute of Technology, under contract with the National Aeronautics and Space Administration (80NM0018D0004). The authors acknowledge Interstellar Institute’s program “II6” and the Paris-Saclay University’s Institut Pascal for hosting discussions that nourished the development of the ideas behind this work.  
\end{acknowledgements}

\bibliographystyle{aa}
\bibliography{bibliography}

\begin{thebibliography}{155}
\expandafter\ifx\csname natexlab\endcsname\relax\def\natexlab#1{#1}\fi

\bibitem[{{Andr{\'e}} {et~al.}(2010){Andr{\'e}}, {Men'shchikov}, {Bontemps},
  {K{\"o}nyves}, {Motte}, {Schneider}, {Didelon}, {Minier}, {Saraceno},
  {Ward-Thompson}, {di Francesco}, {White}, {Molinari}, {Testi}, {Abergel},
  {Griffin}, {Henning}, {Royer}, {Mer{\'\i}n}, {Vavrek}, {Attard},
  {Arzoumanian}, {Wilson}, {Ade}, {Aussel}, {Baluteau}, {Benedettini},
  {Bernard}, {Blommaert}, {Cambr{\'e}sy}, {Cox}, {di Giorgio}, {Hargrave},
  {Hennemann}, {Huang}, {Kirk}, {Krause}, {Launhardt}, {Leeks}, {Le Pennec},
  {Li}, {Martin}, {Maury}, {Olofsson}, {Omont}, {Peretto}, {Pezzuto}, {Prusti},
  {Roussel}, {Russeil}, {Sauvage}, {Sibthorpe}, {Sicilia-Aguilar}, {Spinoglio},
  {Waelkens}, {Woodcraft}, \& {Zavagno}}]{andre_2010}
{Andr{\'e}}, P., {Men'shchikov}, A., {Bontemps}, S., {et~al.} 2010, \aap, 518,
  L102

\bibitem[{{Armus} {et~al.}(2023){Armus}, {Lai}, {U}, {Larson}, {Diaz-Santos},
  {Evans}, {Malkan}, {Rich}, {Medling}, {Law}, {Inami}, {Muller-Sanchez},
  {Charmandaris}, {van der Werf}, {Stierwalt}, {Linden}, {Privon},
  {Barcos-Mu{\~n}oz}, {Hayward}, {Song}, {Appleton}, {Aalto}, {Bohn},
  {B{\"o}ker}, {Brown}, {Finnerty}, {Howell}, {Iwasawa}, {Kemper}, {Marshall},
  {Mazzarella}, {McKinney}, {Murphy}, {Sanders}, \& {Surace}}]{armus_2023}
{Armus}, L., {Lai}, T., {U}, V., {et~al.} 2023, \apjl, 942, L37

\bibitem[{{Bailer-Jones} {et~al.}(2021){Bailer-Jones}, {Rybizki}, {Fouesneau},
  {Demleitner}, \& {Andrae}}]{bailer_jones_2021}
{Bailer-Jones}, C.~A.~L., {Rybizki}, J., {Fouesneau}, M., {Demleitner}, M., \&
  {Andrae}, R. 2021, \aj, 161, 147

\bibitem[{{Barriault} {et~al.}(2010){Barriault}, {Joncas}, {Lockman}, \&
  {Martin}}]{barriault_2010.UrsaMajorCirrus.OH.traces.H2.but.not.CO}
{Barriault}, L., {Joncas}, G., {Lockman}, F.~J., \& {Martin}, P.~G. 2010,
  \mnras, 407, 2645

\bibitem[{{Bellomi} {et~al.}(2020){Bellomi}, {Godard}, {Hennebelle},
  {Valdivia}, {Pineau des For{\^e}ts}, {Lesaffre}, \&
  {P{\'e}rault}}]{bellomi_2020}
{Bellomi}, E., {Godard}, B., {Hennebelle}, P., {et~al.} 2020, \aap, 643, A36

\bibitem[{{Bialy} \& {Sternberg}(2016)}]{bialy_sternberg_2016.H2.formation}
{Bialy}, S. \& {Sternberg}, A. 2016, \apj, 822, 83

\bibitem[{{Boccaletti} {et~al.}(2015){Boccaletti}, {Lagage}, {Baudoz},
  {Beichman}, {Bouchet}, {Cavarroc}, {Dubreuil}, {Glasse}, {Glauser}, {Hines},
  {Lajoie}, {Lebreton}, {Perrin}, {Pueyo}, {Reess}, {Rieke}, {Ronayette},
  {Rouan}, {Soummer}, \& {Wright}}]{boccaletti_2015.miri.predcted.performance}
{Boccaletti}, A., {Lagage}, P.~O., {Baudoz}, P., {et~al.} 2015, \pasp, 127, 633

\bibitem[{{Bohlin} {et~al.}(1978){Bohlin}, {Savage}, \& {Drake}}]{1978ApJ_BSD}
{Bohlin}, R.~C., {Savage}, B.~D., \& {Drake}, J.~F. 1978, \apj, 224, 132

\bibitem[{Bolatto {et~al.}(2013)Bolatto, Wolfire, \& Leroy}]{bolato_2013}
Bolatto, A.~D., Wolfire, M., \& Leroy, A.~K. 2013, Annual Review of Astronomy
  and Astrophysics, 51, 207

\bibitem[{{Borchert} {et~al.}(2022){Borchert}, {Walch}, {Seifried}, {Clarke},
  {Franeck}, \&
  {N{\"u}rnberger}}]{borchert_walch_2022.synthetic.CO.emission.XCO}
{Borchert}, E.~M.~A., {Walch}, S., {Seifried}, D., {et~al.} 2022, \mnras, 510,
  753

\bibitem[{{Bouchet} {et~al.}(2015){Bouchet}, {Garc{\'\i}a-Mar{\'\i}n},
  {Lagage}, {Amiaux}, {Augu{\'e}res}, {Bauwens}, {Blommaert}, {Chen}, {Detre},
  {Dicken}, {Dubreuil}, {Galdemard}, {Gastaud}, {Glasse}, {Gordon}, {Gougnaud},
  {Guillard}, {Justtanont}, {Krause}, {Leboeuf}, {Longval}, {Martin}, {Mazy},
  {Moreau}, {Olofsson}, {Ray}, {Rees}, {Renotte}, {Ressler}, {Ronayette},
  {Salasca}, {Scheithauer}, {Sykes}, {Thelen}, {Wells}, {Wright}, \&
  {Wright}}]{bouchet_2015.miri.instrumentation.methods}
{Bouchet}, P., {Garc{\'\i}a-Mar{\'\i}n}, M., {Lagage}, P.~O., {et~al.} 2015,
  \pasp, 127, 612

\bibitem[{{Boulanger} {et~al.}(1996){Boulanger}, {Abergel}, {Bernard},
  {Burton}, {Desert}, {Hartmann}, {Lagache}, \& {Puget}}]{boulanger_1996}
{Boulanger}, F., {Abergel}, A., {Bernard}, J.~P., {et~al.} 1996, \aap, 312, 256

\bibitem[{{Bronfman} {et~al.}(1988){Bronfman}, {Cohen}, {Alvarez}, {May}, \&
  {Thaddeus}}]{bronfman1988.CO.survey.radia.distribution.of.molecular.clouds}
{Bronfman}, L., {Cohen}, R.~S., {Alvarez}, H., {May}, J., \& {Thaddeus}, P.
  1988, \apj, 324, 248

\bibitem[{{Browning} {et~al.}(2003){Browning}, {Tumlinson}, \&
  {Shull}}]{browning_2003.concatenation.H2.clouds}
{Browning}, M.~K., {Tumlinson}, J., \& {Shull}, J.~M. 2003, \apj, 582, 810

\bibitem[{{Brunt}(2003)}]{Brunt_2003.molecular.clouds.distances.PCA.analysis.CO.data}
{Brunt}, C.~M. 2003, \apj, 584, 293

\bibitem[{Burton {et~al.}(1992)Burton, Elmegreen, Genzel, Pfenniger, \&
  Bartholdi}]{Burton_1993}
Burton, W.~B., Elmegreen, B.~G., Genzel, R., Pfenniger, D., \& Bartholdi, P.
  1992, in Astronomical Society of the Pacific Conference Series, Vol.~31,
  Astronomical Society of the Pacific Conference Series, 327

\bibitem[{{Butler} \&
  {Salim}(2021)}]{butler_2021.nearUV.extinction.correlates.better.with.NH.than.Av}
{Butler}, R.~E. \& {Salim}, S. 2021, \apj, 911, 40

\bibitem[{{Cardelli} {et~al.}(1989){Cardelli}, {Clayton}, \&
  {Mathis}}]{cardelli_1989.selective.extinction}
{Cardelli}, J.~A., {Clayton}, G.~C., \& {Mathis}, J.~S. 1989, \apj, 345, 245

\bibitem[{{Cartledge} {et~al.}(2004){Cartledge}, {Lauroesch}, {Meyer}, \&
  {Sofia}}]{cartledge_2004.abundance.using.HST}
{Cartledge}, S. I.~B., {Lauroesch}, J.~T., {Meyer}, D.~M., \& {Sofia}, U.~J.
  2004, \apj, 613, 1037

\bibitem[{{Clark} \& {Glover}(2015)}]{clark_glover2015.XCO.depends.on.SFR}
{Clark}, P.~C. \& {Glover}, S. C.~O. 2015, \mnras, 452, 2057

\bibitem[{{Cox} {et~al.}(2016){Cox}, {Arzoumanian}, {Andr{\'e}}, {Rygl},
  {Prusti}, {Men'shchikov}, {Royer}, {K{\'o}sp{\'a}l}, {Palmeirim}, {Ribas},
  {K{\"o}nyves}, {Bernard}, {Schneider}, {Bontemps}, {Merin}, {Vavrek}, {Alves
  de Oliveira}, {Didelon}, {Pilbratt}, \& {Waelkens}}]{cox_2016}
{Cox}, N.~L.~J., {Arzoumanian}, D., {Andr{\'e}}, P., {et~al.} 2016, \aap, 590,
  A110

\bibitem[{{Dame} {et~al.}(2001){Dame}, {Hartmann}, \& {Thaddeus}}]{Dame2001}
{Dame}, T.~M., {Hartmann}, D., \& {Thaddeus}, P. 2001, \apj, 547, 792

\bibitem[{{Dame} \&
  {Thaddeus}(2022)}]{dame_2022.complete.CO.survey.northern.sky}
{Dame}, T.~M. \& {Thaddeus}, P. 2022, \apjs, 262, 5

\bibitem[{{Donate} \&
  {Magnani}(2017)}]{donate_magnani_2017.dark.gas.detectability.threshold.decreases.dark.gass.fraction}
{Donate}, E. \& {Magnani}, L. 2017, \mnras, 472, 3169

\bibitem[{{Draine}(2011)}]{2011book_Draine}
{Draine}, B.~T. 2011, {Physics of the Interstellar and Intergalactic Medium}

\bibitem[{{Draine} \& {Bertoldi}(1996)}]{draine_bertoldi_1996}
{Draine}, B.~T. \& {Bertoldi}, F. 1996, \apj, 468, 269

\bibitem[{{Draine} {et~al.}(2007){Draine}, {Dale}, {Bendo}, {Gordon}, {Smith},
  {Armus}, {Engelbracht}, {Helou}, {Kennicutt}, {Li}, {Roussel}, {Walter},
  {Calzetti}, {Moustakas}, {Murphy}, {Rieke}, {Bot}, {Hollenbach}, {Sheth}, \&
  {Teplitz}}]{draine_2007}
{Draine}, B.~T., {Dale}, D.~A., {Bendo}, G., {et~al.} 2007, \apj, 663, 866

\bibitem[{{Foreman-Mackey} {et~al.}(2013){Foreman-Mackey}, {Hogg}, {Lang}, \&
  {Goodman}}]{emcee}
{Foreman-Mackey}, D., {Hogg}, D.~W., {Lang}, D., \& {Goodman}, J. 2013, \pasp,
  125, 306

\bibitem[{Fukui {et~al.}(2014)Fukui, Okamoto, Kaji, Yamamoto, Torii, Hayakawa,
  Tachihara, Dickey, Okuda, Ohama, Kuroda, \& Kuwahara}]{fukui2014}
Fukui, Y., Okamoto, R., Kaji, R., {et~al.} 2014, Astrophysical Journal, 796

\bibitem[{{Fukui} {et~al.}(2015){Fukui}, {Torii}, {Onishi}, {Yamamoto},
  {Okamoto}, {Hayakawa}, {Tachihara}, \& {Sano}}]{fukui_2015}
{Fukui}, Y., {Torii}, K., {Onishi}, T., {et~al.} 2015, \apj, 798, 6

\bibitem[{{Gaia Collaboration} {et~al.}(2021){Gaia Collaboration}, {Brown},
  {Vallenari}, {Prusti}, {de Bruijne}, {Babusiaux}, {Biermann}, {Creevey},
  {Evans}, {Eyer}, {Hutton}, {Jansen}, {Jordi}, {Klioner}, {Lammers},
  {Lindegren}, {Luri}, {Mignard}, {Panem}, {Pourbaix}, {Randich}, {Sartoretti},
  {Soubiran}, {Walton}, {Arenou}, {Bailer-Jones}, {Bastian}, {Cropper},
  {Drimmel}, {Katz}, {Lattanzi}, {van Leeuwen}, {Bakker}, {Cacciari},
  {Casta{\~n}eda}, {De Angeli}, {Ducourant}, {Fabricius}, {Fouesneau},
  {Fr{\'e}mat}, {Guerra}, {Guerrier}, {Guiraud}, {Jean-Antoine Piccolo},
  {Masana}, {Messineo}, {Mowlavi}, {Nicolas}, {Nienartowicz}, {Pailler},
  {Panuzzo}, {Riclet}, {Roux}, {Seabroke}, {Sordo}, {Tanga}, {Th{\'e}venin},
  {Gracia-Abril}, {Portell}, {Teyssier}, {Altmann}, {Andrae}, {Bellas-Velidis},
  {Benson}, {Berthier}, {Blomme}, {Brugaletta}, {Burgess}, {Busso}, {Carry},
  {Cellino}, {Cheek}, {Clementini}, {Damerdji}, {Davidson}, {Delchambre},
  {Dell'Oro}, {Fern{\'a}ndez-Hern{\'a}ndez}, {Galluccio}, {Garc{\'\i}a-Lario},
  {Garcia-Reinaldos}, {Gonz{\'a}lez-N{\'u}{\~n}ez}, {Gosset}, {Haigron},
  {Halbwachs}, {Hambly}, {Harrison}, {Hatzidimitriou}, {Heiter},
  {Hern{\'a}ndez}, {Hestroffer}, {Hodgkin}, {Holl}, {Jan{\ss}en}, {Jevardat de
  Fombelle}, {Jordan}, {Krone-Martins}, {Lanzafame}, {L{\"o}ffler}, {Lorca},
  {Manteiga}, {Marchal}, {Marrese}, {Moitinho}, {Mora}, {Muinonen}, {Osborne},
  {Pancino}, {Pauwels}, {Petit}, {Recio-Blanco}, {Richards}, {Riello},
  {Rimoldini}, {Robin}, {Roegiers}, {Rybizki}, {Sarro}, {Siopis}, {Smith},
  {Sozzetti}, {Ulla}, {Utrilla}, {van Leeuwen}, {van Reeven}, {Abbas}, {Abreu
  Aramburu}, {Accart}, {Aerts}, {Aguado}, {Ajaj}, {Altavilla}, {{\'A}lvarez},
  {{\'A}lvarez Cid-Fuentes}, {Alves}, {Anderson}, {Anglada Varela}, {Antoja},
  {Audard}, {Baines}, {Baker}, {Balaguer-N{\'u}{\~n}ez}, {Balbinot}, {Balog},
  {Barache}, {Barbato}, {Barros}, {Barstow}, {Bartolom{\'e}}, {Bassilana},
  {Bauchet}, {Baudesson-Stella}, {Becciani}, {Bellazzini}, {Bernet}, {Bertone},
  {Bianchi}, {Blanco-Cuaresma}, {Boch}, {Bombrun}, {Bossini}, {Bouquillon},
  {Bragaglia}, {Bramante}, {Breedt}, {Bressan}, {Brouillet}, {Bucciarelli},
  {Burlacu}, {Busonero}, {Butkevich}, {Buzzi}, {Caffau}, {Cancelliere},
  {C{\'a}novas}, {Cantat-Gaudin}, {Carballo}, {Carlucci}, {Carnerero},
  {Carrasco}, {Casamiquela}, {Castellani}, {Castro-Ginard}, {Castro Sampol},
  {Chaoul}, {Charlot}, {Chemin}, {Chiavassa}, {Cioni}, {Comoretto}, {Cooper},
  {Cornez}, {Cowell}, {Crifo}, {Crosta}, {Crowley}, {Dafonte}, {Dapergolas},
  {David}, {David}, {de Laverny}, {De Luise}, {De March}, {De Ridder}, {de
  Souza}, {de Teodoro}, {de Torres}, {del Peloso}, {del Pozo}, {Delbo},
  {Delgado}, {Delgado}, {Delisle}, {Di Matteo}, {Diakite}, {Diener},
  {Distefano}, {Dolding}, {Eappachen}, {Edvardsson}, {Enke}, {Esquej}, {Fabre},
  {Fabrizio}, {Faigler}, {Fedorets}, {Fernique}, {Fienga}, {Figueras},
  {Fouron}, {Fragkoudi}, {Fraile}, {Franke}, {Gai}, {Garabato},
  {Garcia-Gutierrez}, {Garc{\'\i}a-Torres}, {Garofalo}, {Gavras}, {Gerlach},
  {Geyer}, {Giacobbe}, {Gilmore}, {Girona}, {Giuffrida}, {Gomel}, {Gomez},
  {Gonzalez-Santamaria}, {Gonz{\'a}lez-Vidal}, {Granvik},
  {Guti{\'e}rrez-S{\'a}nchez}, {Guy}, {Hauser}, {Haywood}, {Helmi}, {Hidalgo},
  {Hilger}, {H{\l}adczuk}, {Hobbs}, {Holland}, {Huckle}, {Jasniewicz},
  {Jonker}, {Juaristi Campillo}, {Julbe}, {Karbevska}, {Kervella}, {Khanna},
  {Kochoska}, {Kontizas}, {Kordopatis}, {Korn}, {Kostrzewa-Rutkowska},
  {Kruszy{\'n}ska}, {Lambert}, {Lanza}, {Lasne}, {Le Campion}, {Le Fustec},
  {Lebreton}, {Lebzelter}, {Leccia}, {Leclerc}, {Lecoeur-Taibi}, {Liao},
  {Licata}, {Lindstr{\o}m}, {Lister}, {Livanou}, {Lobel}, {Madrero Pardo},
  {Managau}, {Mann}, {Marchant}, {Marconi}, {Marcos Santos}, {Marinoni},
  {Marocco}, {Marshall}, {Martin Polo}, {Mart{\'\i}n-Fleitas}, {Masip},
  {Massari}, {Mastrobuono-Battisti}, {Mazeh}, {McMillan}, {Messina},
  {Michalik}, {Millar}, {Mints}, {Molina}, {Molinaro}, {Moln{\'a}r},
  {Montegriffo}, {Mor}, {Morbidelli}, {Morel}, {Morris}, {Mulone}, {Munoz},
  {Muraveva}, {Murphy}, {Musella}, {Noval}, {Ord{\'e}novic}, {Orr{\`u}},
  {Osinde}, {Pagani}, {Pagano}, {Palaversa}, {Palicio}, {Panahi}, {Pawlak},
  {Pe{\~n}alosa Esteller}, {Penttil{\"a}}, {Piersimoni}, {Pineau}, {Plachy},
  {Plum}, {Poggio}, {Poretti}, {Poujoulet}, {Pr{\v{s}}a}, {Pulone}, {Racero},
  {Ragaini}, {Rainer}, {Raiteri}, {Rambaux}, {Ramos}, {Ramos-Lerate}, {Re
  Fiorentin}, {Regibo}, {Reyl{\'e}}, {Ripepi}, {Riva}, {Rixon}, {Robichon},
  {Robin}, {Roelens}, {Rohrbasser}, {Romero-G{\'o}mez}, {Rowell}, {Royer},
  {Rybicki}, {Sadowski}, {Sagrist{\`a} Sell{\'e}s}, {Sahlmann}, {Salgado},
  {Salguero}, {Samaras}, {Sanchez Gimenez}, {Sanna}, {Santove{\~n}a},
  {Sarasso}, {Schultheis}, {Sciacca}, {Segol}, {Segovia}, {S{\'e}gransan},
  {Semeux}, {Shahaf}, {Siddiqui}, {Siebert}, {Siltala}, {Slezak}, {Smart},
  {Solano}, {Solitro}, {Souami}, {Souchay}, {Spagna}, {Spoto}, {Steele},
  {Steidelm{\"u}ller}, {Stephenson}, {S{\"u}veges}, {Szabados}, {Szegedi-Elek},
  {Taris}, {Tauran}, {Taylor}, {Teixeira}, {Thuillot}, {Tonello}, {Torra},
  {Torra}, {Turon}, {Unger}, {Vaillant}, {van Dillen}, {Vanel}, {Vecchiato},
  {Viala}, {Vicente}, {Voutsinas}, {Weiler}, {Wevers}, {Wyrzykowski}, {Yoldas},
  {Yvard}, {Zhao}, {Zorec}, {Zucker}, {Zurbach}, \& {Zwitter}}]{gaia_dr3}
{Gaia Collaboration}, {Brown}, A.~G.~A., {Vallenari}, A., {et~al.} 2021, \aap,
  649, A1

\bibitem[{{Gillmon} {et~al.}(2006){Gillmon}, {Shull}, {Tumlinson}, \&
  {Danforth}}]{Gillmon2006}
{Gillmon}, K., {Shull}, J.~M., {Tumlinson}, J., \& {Danforth}, C. 2006, \apj,
  636, 891

\bibitem[{{Glasse} {et~al.}(2015){Glasse}, {Rieke}, {Bauwens},
  {Garc{\'\i}a-Mar{\'\i}n}, {Ressler}, {Rost}, {Tikkanen}, {Vandenbussche}, \&
  {Wright}}]{glasse_2015.sensitivity}
{Glasse}, A., {Rieke}, G.~H., {Bauwens}, E., {et~al.} 2015, \pasp, 127, 686

\bibitem[{Glover \& Clark(2016)}]{glover_2016}
Glover, S. C.~O. \& Clark, P.~C. 2016, Monthly Notices of the Royal
  Astronomical Society, 456, 3596

\bibitem[{{Glover} \& {Smith}(2016)}]{glover_2016.temperature.co.dark}
{Glover}, S. C.~O. \& {Smith}, R.~J. 2016, \mnras, 462, 3011

\bibitem[{{Goldsmith}(2013)}]{goldsmith_2013}
{Goldsmith}, P.~F. 2013, \apj, 774, 134

\bibitem[{{Goldsmith} {et~al.}(2018){Goldsmith}, {Pineda}, {Neufeld},
  {Wolfire}, {Risacher}, \& {Simon}}]{goldsmith_2018}
{Goldsmith}, P.~F., {Pineda}, J.~L., {Neufeld}, D.~A., {et~al.} 2018, \apj,
  856, 96

\bibitem[{{Gong} {et~al.}(2018){Gong}, {Ostriker}, \& {Kim}}]{gong_2018}
{Gong}, M., {Ostriker}, E.~C., \& {Kim}, C.-G. 2018, \apj, 858, 16

\bibitem[{{Gong} {et~al.}(2017){Gong}, {Ostriker}, \& {Wolfire}}]{gong_2017}
{Gong}, M., {Ostriker}, E.~C., \& {Wolfire}, M.~G. 2017, \apj, 843, 38

\bibitem[{{Gordon} {et~al.}(2015){Gordon}, {Chen}, {Anderson}, {Azzollini},
  {Bergeron}, {Bouchet}, {Bouwman}, {Cracraft}, {Fischer}, {Friedman},
  {Garc{\'\i}a-Mar{\'\i}n}, {Glasse}, {Glauser}, {Goodson}, {Greene}, {Hines},
  {Khorrami}, {Lahuis}, {Lajoie}, {Meixner}, {Morrison}, {O'Sullivan},
  {Pontoppidan}, {Regan}, {Ressler}, {Rieke}, {Scheithauer}, {Walker}, \&
  {Wright}}]{gordon_2015.data.reduction}
{Gordon}, K.~D., {Chen}, C.~H., {Anderson}, R.~E., {et~al.} 2015, \pasp, 127,
  696

\bibitem[{{Green} {et~al.}(2019){Green}, {Schlafly}, {Zucker}, {Speagle}, \&
  {Finkbeiner}}]{green_bayestar_2019.latest.update}
{Green}, G.~M., {Schlafly}, E., {Zucker}, C., {Speagle}, J.~S., \&
  {Finkbeiner}, D. 2019, \apj, 887, 93

\bibitem[{{Grenier} {et~al.}(2005){Grenier}, {Casandjian}, \&
  {Terrier}}]{Grenier_2005}
{Grenier}, I.~A., {Casandjian}, J.-M., \& {Terrier}, R. 2005, Science, 307,
  1292

\bibitem[{{Gudennavar} {et~al.}(2012){Gudennavar}, {Bubbly}, {Preethi}, \&
  {Murthy}}]{Gudennavar_2012}
{Gudennavar}, S.~B., {Bubbly}, S.~G., {Preethi}, K., \& {Murthy}, J. 2012,
  \apjs, 199, 8

\bibitem[{{Heiles} \&
  {Troland}(2003{\natexlab{a}})}]{heiles_troland_2003.ISM.phases.Gaussian.decomposition}
{Heiles}, C. \& {Troland}, T.~H. 2003{\natexlab{a}}, \apjs, 145, 329

\bibitem[{{Heiles} \&
  {Troland}(2003{\natexlab{b}})}]{heiles_troland_2003.diffuse.ISM.sonic.Mach.number}
{Heiles}, C. \& {Troland}, T.~H. 2003{\natexlab{b}}, \apj, 586, 1067

\bibitem[{{Heithausen} \&
  {Thaddeus}(1990)}]{heithausen_thaddeus_1990.Polaris.Flare.distance.stellar.extinction.measurements}
{Heithausen}, A. \& {Thaddeus}, P. 1990, \apjl, 353, L49

\bibitem[{{Heyer} \& {Dame}(2015)}]{heyer_dame_2015.H2.clouds}
{Heyer}, M. \& {Dame}, T.~M. 2015, \araa, 53, 583

\bibitem[{{HI4PI Collaboration} {et~al.}(2016){HI4PI Collaboration}, {Ben
  Bekhti}, {Fl{\"o}er}, {Keller}, {Kerp}, {Lenz}, {Winkel}, {Bailin},
  {Calabretta}, {Dedes}, {Ford}, {Gibson}, {Haud}, {Janowiecki}, {Kalberla},
  {Lockman}, {McClure-Griffiths}, {Murphy}, {Nakanishi}, {Pisano}, \&
  {Staveley-Smith}}]{hi4pi}
{HI4PI Collaboration}, {Ben Bekhti}, N., {Fl{\"o}er}, L., {et~al.} 2016, \aap,
  594, A116

\bibitem[{{Hildebrand}(1983)}]{hildebrand_1983.review.determination.ISM.mass.using.far.infrared.data}
{Hildebrand}, R.~H. 1983, \qjras, 24, 267

\bibitem[{{Hopkins} {et~al.}(2022){Hopkins}, {Rosen}, {Squire}, {Panopoulou},
  {Soliman}, {Seligman}, \& {Steinwandel}}]{hopkins_2022.dust.in.the.wind}
{Hopkins}, P.~F., {Rosen}, A.~L., {Squire}, J., {et~al.} 2022, \mnras, 517,
  1491

\bibitem[{{Hopkins} \&
  {Squire}(2018)}]{hopkins_squire_2018.drag.instabilities.analytical}
{Hopkins}, P.~F. \& {Squire}, J. 2018, \mnras, 480, 2813

\bibitem[{{Hopkins} {et~al.}(2020){Hopkins}, {Squire}, \&
  {Seligman}}]{hopkins_2020}
{Hopkins}, P.~F., {Squire}, J., \& {Seligman}, D. 2020, \mnras, 496, 2123

\bibitem[{{Jaschek} \& {Egret}(1982)}]{IAU_Be.stars.caltagoue}
{Jaschek}, M. \& {Egret}, D. 1982, in Be Stars, ed. M.~{Jaschek} \& H.~G.
  {Groth}, Vol.~98, 261

\bibitem[{{Jura}(1980)}]{Jura_1980.grain.size.growth.Orion}
{Jura}, M. 1980, \apj, 235, 63

\bibitem[{{Juvela} {et~al.}(2018){Juvela}, {He}, {Pattle}, {Liu}, {Bendo},
  {Eden}, {Feh{\'e}r}, {Michel}, {Fuller}, {Hirano}, {Kim}, {Li}, {Liu},
  {Malinen}, {Marshall}, {Paradis}, {Parsons}, {Pelkonen}, {Rawlings},
  {Ristorcelli}, {Samal}, {Tatematsu}, {Thompson}, {Traficante}, {Wang},
  {Ward-Thompson}, {Wu}, {Yi}, \& {Yoo}}]{juvela_2018.SED.fitting}
{Juvela}, M., {He}, J., {Pattle}, K., {et~al.} 2018, \aap, 612, A71

\bibitem[{{Kalberla} \&
  {Haud}(2015)}]{kalberla_2015.treatment.of.GASS.systematics}
{Kalberla}, P.~M.~W. \& {Haud}, U. 2015, \aap, 578, A78

\bibitem[{{Kalberla} \&
  {Haud}(2018)}]{kalberla_2018.phase.properties.HI.decomposition}
{Kalberla}, P.~M.~W. \& {Haud}, U. 2018, \aap, 619, A58

\bibitem[{{Kalberla} {et~al.}(2020){Kalberla}, {Kerp}, \&
  {Haud}}]{kalberla_2020}
{Kalberla}, P.~M.~W., {Kerp}, J., \& {Haud}, U. 2020, \aap, 639, A26

\bibitem[{{Kendrew} {et~al.}(2015){Kendrew}, {Scheithauer}, {Bouchet},
  {Amiaux}, {Azzollini}, {Bouwman}, {Chen}, {Dubreuil}, {Fischer}, {Glasse},
  {Greene}, {Lagage}, {Lahuis}, {Ronayette}, {Wright}, \&
  {Wright}}]{kendrew_2015.miri.low.res.spectrometer}
{Kendrew}, S., {Scheithauer}, S., {Bouchet}, P., {et~al.} 2015, \pasp, 127, 623

\bibitem[{{Kirk} {et~al.}(2010){Kirk}, {Polehampton}, {Anderson}, {Baluteau},
  {Bontemps}, {Joblin}, {Jones}, {Naylor}, {Ward-Thompson}, {White}, {Abergel},
  {Ade}, {Andr{\'e}}, {Arab}, {Bernard}, {Blagrave}, {Boulanger}, {Cohen},
  {Compiegne}, {Cox}, {Dartois}, {Davis}, {Emery}, {Fulton}, {Gry}, {Habart},
  {Huang}, {Lagache}, {Lim}, {Madden}, {Makiwa}, {Martin},
  {Miville-Desch{\^e}nes}, {Molinari}, {Moseley}, {Motte}, {Okumura}, {Pinheiro
  Gon{\c{c}}alves}, {Rod{\'o}n}, {Russeil}, {Saraceno}, {Sidher}, {Spencer},
  {Swinyard}, \& {Zavagno}}]{kirk_2010}
{Kirk}, J.~M., {Polehampton}, E., {Anderson}, L.~D., {et~al.} 2010, \aap, 518,
  L82

\bibitem[{{K{\"o}nyves} {et~al.}(2015){K{\"o}nyves}, {Andr{\'e}},
  {Men'shchikov}, {Palmeirim}, {Arzoumanian}, {Schneider}, {Roy}, {Didelon},
  {Maury}, {Shimajiri}, {Di Francesco}, {Bontemps}, {Peretto}, {Benedettini},
  {Bernard}, {Elia}, {Griffin}, {Hill}, {Kirk}, {Ladjelate}, {Marsh}, {Martin},
  {Motte}, {Nguy{\^e}n Luong}, {Pezzuto}, {Roussel}, {Rygl}, {Sadavoy},
  {Schisano}, {Spinoglio}, {Ward-Thompson}, \&
  {White}}]{konyves_2015.dense.cores.Aquila}
{K{\"o}nyves}, V., {Andr{\'e}}, P., {Men'shchikov}, A., {et~al.} 2015, \aap,
  584, A91

\bibitem[{{K{\"o}nyves} {et~al.}(2010){K{\"o}nyves}, {Andr{\'e}},
  {Men'shchikov}, {Schneider}, {Arzoumanian}, {Bontemps}, {Attard}, {Motte},
  {Didelon}, {Maury}, {Abergel}, {Ali}, {Baluteau}, {Bernard}, {Cambr{\'e}sy},
  {Cox}, {di Francesco}, {di Giorgio}, {Griffin}, {Hargrave}, {Huang}, {Kirk},
  {Li}, {Martin}, {Minier}, {Molinari}, {Olofsson}, {Pezzuto}, {Russeil},
  {Roussel}, {Saraceno}, {Sauvage}, {Sibthorpe}, {Spinoglio}, {Testi},
  {Ward-Thompson}, {White}, {Wilson}, {Woodcraft}, \&
  {Zavagno}}]{konyves_2010.presterllar.cores.Aquilla}
{K{\"o}nyves}, V., {Andr{\'e}}, P., {Men'shchikov}, A., {et~al.} 2010, \aap,
  518, L106

\bibitem[{{Krumholz} {et~al.}(2008){Krumholz}, {McKee}, \&
  {Tumlinson}}]{krumholz_2008.HI.H2.transition}
{Krumholz}, M.~R., {McKee}, C.~F., \& {Tumlinson}, J. 2008, \apj, 689, 865

\bibitem[{{Krumholz} {et~al.}(2009){Krumholz}, {McKee}, \&
  {Tumlinson}}]{krumholz_2009.hi.h2.transition}
{Krumholz}, M.~R., {McKee}, C.~F., \& {Tumlinson}, J. 2009, \apj, 693, 216

\bibitem[{{Lada} \& {Dame}(2020)}]{lada_dame_202.new.average.XCO}
{Lada}, C.~J. \& {Dame}, T.~M. 2020, \apj, 898, 3

\bibitem[{{Langer} {et~al.}(2015){Langer}, {Goldsmith}, {Pineda}, {Velusamy},
  {Requena-Torres}, \& {Wiesemeyer}}]{langer_2015.CO.dark.CMZ}
{Langer}, W.~D., {Goldsmith}, P.~F., {Pineda}, J.~L., {et~al.} 2015, \aap, 576,
  A1

\bibitem[{{Langer} {et~al.}(2010){Langer}, {Velusamy}, {Pineda}, {Goldsmith},
  {Li}, \& {Yorke}}]{langer_2010}
{Langer}, W.~D., {Velusamy}, T., {Pineda}, J.~L., {et~al.} 2010, \aap, 521, L17

\bibitem[{{Langer} {et~al.}(2014){Langer}, {Velusamy}, {Pineda}, {Willacy}, \&
  {Goldsmith}}]{langer_2014.co.dark.GOTC+}
{Langer}, W.~D., {Velusamy}, T., {Pineda}, J.~L., {Willacy}, K., \&
  {Goldsmith}, P.~F. 2014, \aap, 561, A122

\bibitem[{{Larson}(1981)}]{larson_1981.larson.relationI}
{Larson}, R.~B. 1981, \mnras, 194, 809

\bibitem[{{Lebouteiller} {et~al.}(2019){Lebouteiller}, {Cormier}, {Madden},
  {Galametz}, {Hony}, {Galliano}, {Chevance}, {Lee}, {Braine}, {Polles},
  {Reque{\~n}a-Torres}, {Indebetouw}, {Hughes}, \&
  {Abel}}]{lebouteiller_2019.co.dark.magellanic.clouds}
{Lebouteiller}, V., {Cormier}, D., {Madden}, S.~C., {et~al.} 2019, \aap, 632,
  A106

\bibitem[{{Lee} {et~al.}(2014){Lee}, {Stanimirovi{\'c}}, {Wolfire}, {Shetty},
  {Glover}, {Molina}, \& {Klessen}}]{lee_2014.XCO.perseus.molecular.cloud}
{Lee}, M.-Y., {Stanimirovi{\'c}}, S., {Wolfire}, M.~G., {et~al.} 2014, \apj,
  784, 80

\bibitem[{{Leike} {et~al.}(2020){Leike}, {Glatzle}, \&
  {En{\ss}lin}}]{leike_2020}
{Leike}, R.~H., {Glatzle}, M., \& {En{\ss}lin}, T.~A. 2020, \aap, 639, A138

\bibitem[{{Lenz} {et~al.}(2017){Lenz}, {Hensley}, \& {Dor{\'e}}}]{lenz_2017}
{Lenz}, D., {Hensley}, B.~S., \& {Dor{\'e}}, O. 2017, \apj, 846, 38

\bibitem[{{Lenz} {et~al.}(2015){Lenz}, {Kerp}, {Fl{\"o}er}, {Winkel},
  {Boulanger}, \& {Lagache}}]{lenz_2015}
{Lenz}, D., {Kerp}, J., {Fl{\"o}er}, L., {et~al.} 2015, \aap, 573, A83

\bibitem[{{Li} {et~al.}(2018){Li}, {Tang}, {Nguyen}, {Dawson}, {Heiles}, {Xu},
  {Pan}, {Goldsmith}, {Gibson}, {Murray}, {Robishaw}, {McClure-Griffiths},
  {Dickey}, {Pineda}, {Stanimirovi{\'c}}, {Bronfman}, {Troland}, \& {PRIMO
  Collaboration}}]{DiLi_2018.OH.traces.dark.gas}
{Li}, D., {Tang}, N., {Nguyen}, H., {et~al.} 2018, \apjs, 235, 1

\bibitem[{{Li} {et~al.}(2015){Li}, {Xu}, {Heiles}, {Pan}, \&
  {Tang}}]{DiLi_2015.dark.gas}
{Li}, D., {Xu}, D., {Heiles}, C., {Pan}, Z., \& {Tang}, N. 2015, Publication of
  Korean Astronomical Society, 30, 75

\bibitem[{Lindner {et~al.}(2015)Lindner, Vera-Ciro, Murray, Stanimirovi{\'c},
  Babler, Heiles, Hennebelle, Goss, \& Dickey}]{Lindner_2015}
Lindner, R.~R., Vera-Ciro, C., Murray, C.~E., {et~al.} 2015, The Astronomical
  Journal, 149, 138

\bibitem[{{Liszt}(2014{\natexlab{a}})}]{liszt_2014.NHI.NH2.NH.versus.EBV}
{Liszt}, H. 2014{\natexlab{a}}, \apj, 783, 17

\bibitem[{{Liszt}(2014{\natexlab{b}})}]{liszt_2014.ratio.NHI.EBV.high.latitudes}
{Liszt}, H. 2014{\natexlab{b}}, \apj, 780, 10

\bibitem[{{Liszt}(2019)}]{liszt_2019.ebv.verus.HI}
{Liszt}, H. 2019, \apj, 881, 29

\bibitem[{{Liszt} \& {Gerin}(2023)}]{liszt_2023.H2.properties.diffuse.ISM.CO}
{Liszt}, H. \& {Gerin}, M. 2023, \apj, 943, 172

\bibitem[{{Liszt} \& {Gerin}(2016)}]{liszt_2016.XCO}
{Liszt}, H.~S. \& {Gerin}, M. 2016, \aap, 585, A80

\bibitem[{{Lombardi} {et~al.}(2006){Lombardi}, {Alves}, \&
  {Lada}}]{lombardi_2006.pipe.nebula.XCO}
{Lombardi}, M., {Alves}, J., \& {Lada}, C.~J. 2006, \aap, 454, 781

\bibitem[{{Madden} {et~al.}(2020){Madden}, {Cormier}, {Hony}, {Lebouteiller},
  {Abel}, {Galametz}, {De Looze}, {Chevance}, {Polles}, {Lee}, {Galliano},
  {Lambert-Huyghe}, {Hu}, \&
  {Ramambason}}]{madden_2020.method.for.tracing.CO.dark.in.galaxies.using.cloudy.simulations}
{Madden}, S.~C., {Cormier}, D., {Hony}, S., {et~al.} 2020, \aap, 643, A141

\bibitem[{{Mangum} \& {Shirley}(2015)}]{mangum_2015.how.to.observe.H2}
{Mangum}, J.~G. \& {Shirley}, Y.~L. 2015, \pasp, 127, 266

\bibitem[{{Marasco} {et~al.}(2017){Marasco}, {Fraternali}, {van der Hulst}, \&
  {Oosterloo}}]{marasco_2017}
{Marasco}, A., {Fraternali}, F., {van der Hulst}, J.~M., \& {Oosterloo}, T.
  2017, \aap, 607, A106

\bibitem[{{Marchal} {et~al.}(2019){Marchal}, {Miville-Desch{\^e}nes}, {Orieux},
  {Gac}, {Soussen}, {Lesot}, {d'Allonnes}, \& {Salom{\'e}}}]{rohsa_2019}
{Marchal}, A., {Miville-Desch{\^e}nes}, M.-A., {Orieux}, F., {et~al.} 2019,
  \aap, 626, A101

\bibitem[{{Martin} {et~al.}(2012){Martin}, {Roy}, {Bontemps},
  {Miville-Desch{\^e}nes}, {Ade}, {Bock}, {Chapin}, {Devlin}, {Dicker},
  {Griffin}, {Gundersen}, {Halpern}, {Hargrave}, {Hughes}, {Klein}, {Marsden},
  {Mauskopf}, {Netterfield}, {Olmi}, {Patanchon}, {Rex}, {Scott}, {Semisch},
  {Truch}, {Tucker}, {Tucker}, {Viero}, \& {Wiebe}}]{martinPG_2012}
{Martin}, P.~G., {Roy}, A., {Bontemps}, S., {et~al.} 2012, \apj, 751, 28

\bibitem[{{McKee} \& {Krumholz}(2010)}]{mckee_krumholz_2010}
{McKee}, C.~F. \& {Krumholz}, M.~R. 2010, \apj, 709, 308

\bibitem[{{Miville-Desch{\^e}nes} {et~al.}(2002){Miville-Desch{\^e}nes},
  {Boulanger}, {Joncas}, \& {Falgarone}}]{miville_2002}
{Miville-Desch{\^e}nes}, M.~A., {Boulanger}, F., {Joncas}, G., \& {Falgarone},
  E. 2002, \aap, 381, 209

\bibitem[{{Miville-Desch{\^e}nes} {et~al.}(2010){Miville-Desch{\^e}nes},
  {Martin}, {Abergel}, {Bernard}, {Boulanger}, {Lagache}, {Anderson},
  {Andr{\'e}}, {Arab}, {Baluteau}, {Blagrave}, {Bontemps}, {Cohen},
  {Compiegne}, {Cox}, {Dartois}, {Davis}, {Emery}, {Fulton}, {Gry}, {Habart},
  {Huang}, {Joblin}, {Jones}, {Kirk}, {Lim}, {Madden}, {Makiwa}, {Menshchikov},
  {Molinari}, {Moseley}, {Motte}, {Naylor}, {Okumura}, {Pinheiro
  Gon{\c{c}}alves}, {Polehampton}, {Rod{\'o}n}, {Russeil}, {Saraceno},
  {Schneider}, {Sidher}, {Spencer}, {Swinyard}, {Ward-Thompson}, {White}, \&
  {Zavagno}}]{miville_2010_polaris_PS}
{Miville-Desch{\^e}nes}, M.~A., {Martin}, P.~G., {Abergel}, A., {et~al.} 2010,
  \aap, 518, L104

\bibitem[{{Miville-Desch{\^e}nes} {et~al.}(2017){Miville-Desch{\^e}nes},
  {Murray}, \& {Lee}}]{miville_deschenes2017.H2.properties.of.Galaxy.disk}
{Miville-Desch{\^e}nes}, M.-A., {Murray}, N., \& {Lee}, E.~J. 2017, \apj, 834,
  57

\bibitem[{{Mouschovias} \&
  {Psaltis}(1995)}]{mouschovias_1995.larson.relation.Bfields}
{Mouschovias}, T.~C. \& {Psaltis}, D. 1995, \apjl, 444, L105

\bibitem[{{Murray} {et~al.}(2014){Murray}, {Lindner}, {Stanimirovi{\'c}},
  {Goss}, {Heiles}, {Dickey}, {Pingel}, {Lawrence}, {Jencson}, {Babler}, \&
  {Hennebelle}}]{murray_2014.Texc.lyman.alpha.scattering}
{Murray}, C.~E., {Lindner}, R.~R., {Stanimirovi{\'c}}, S., {et~al.} 2014,
  \apjl, 781, L41

\bibitem[{{Murray} {et~al.}(2018){Murray}, {Peek}, {Lee}, \&
  {Stanimirovi{\'c}}}]{murray_2018}
{Murray}, C.~E., {Peek}, J.~E.~G., {Lee}, M.-Y., \& {Stanimirovi{\'c}}, S.
  2018, \apj, 862, 131

\bibitem[{{Nakanishi} \&
  {Sofue}(2006)}]{nakanishi_sofue2006.3d.distribution.HI.H2.gas.based.on.CO.data}
{Nakanishi}, H. \& {Sofue}, Y. 2006, \pasj, 58, 847

\bibitem[{{Nakanishi} \&
  {Sofue}(2016)}]{nakanishi_sofue2016.3d.distribution.HI.H2.gas.based.on.CO.data}
{Nakanishi}, H. \& {Sofue}, Y. 2016, \pasj, 68, 5

\bibitem[{{Narayanan} \&
  {Hopkins}(2013)}]{narayanan.hopkins2013.XCO.is.constant.in.our.Galaxy}
{Narayanan}, D. \& {Hopkins}, P.~F. 2013, \mnras, 433, 1223

\bibitem[{{Nguyen} {et~al.}(2018){Nguyen}, {Dawson}, {Miville-Desch{\^e}nes},
  {Tang}, {Li}, {Heiles}, {Murray}, {Stanimirovi{\'c}}, {Gibson},
  {McClure-Griffiths}, {Troland}, {Bronfman}, \&
  {Finger}}]{Nguyen_2018.dgr.variation}
{Nguyen}, H., {Dawson}, J.~R., {Miville-Desch{\^e}nes}, M.~A., {et~al.} 2018,
  \apj, 862, 49

\bibitem[{{Palmeirim} {et~al.}(2013){Palmeirim}, {Andr{\'e}}, {Kirk},
  {Ward-Thompson}, {Arzoumanian}, {K{\"o}nyves}, {Didelon}, {Schneider},
  {Benedettini}, {Bontemps}, {Di Francesco}, {Elia}, {Griffin}, {Hennemann},
  {Hill}, {Martin}, {Men'shchikov}, {Molinari}, {Motte}, {Nguyen Luong},
  {Nutter}, {Peretto}, {Pezzuto}, {Roy}, {Rygl}, {Spinoglio}, \&
  {White}}]{palmeirim_2013_striations}
{Palmeirim}, P., {Andr{\'e}}, P., {Kirk}, J., {et~al.} 2013, \aap, 550, A38

\bibitem[{{Panopoulou} {et~al.}(2021){Panopoulou}, {Dickinson}, {Readhead},
  {Pearson}, \& {Peel}}]{panopoulou_2021.distance.north.polar.spur}
{Panopoulou}, G.~V., {Dickinson}, C., {Readhead}, A.~C.~S., {Pearson}, T.~J.,
  \& {Peel}, M.~W. 2021, \apj, 922, 210

\bibitem[{{Panopoulou} \& {Lenz}(2020)}]{panopoulou_lenz_2020.clouds.number}
{Panopoulou}, G.~V. \& {Lenz}, D. 2020, \apj, 902, 120

\bibitem[{{Panopoulou} {et~al.}(2016){Panopoulou}, {Psaradaki}, \&
  {Tassis}}]{panopoulou_2016}
{Panopoulou}, G.~V., {Psaradaki}, I., \& {Tassis}, K. 2016, \mnras, 462, 1517

\bibitem[{{Paradis} {et~al.}(2012){Paradis}, {Dobashi}, {Shimoikura},
  {Kawamura}, {Onishi}, {Fukui}, \& {Bernard}}]{paradis_2012}
{Paradis}, D., {Dobashi}, K., {Shimoikura}, T., {et~al.} 2012, \aap, 543, A103

\bibitem[{{Paradis} {et~al.}(2023){Paradis}, {M{\'e}ny}, {Demyk},
  {Noriega-Crespo}, \& {Ristorcelli}}]{paradis_2023}
{Paradis}, D., {M{\'e}ny}, C., {Demyk}, K., {Noriega-Crespo}, A., \&
  {Ristorcelli}, I. 2023, \aap, 674, A141

\bibitem[{{Paradis} {et~al.}(2010){Paradis}, {Veneziani}, {Noriega-Crespo},
  {Paladini}, {Piacentini}, {Bernard}, {de Bernardis}, {Calzoletti},
  {Faustini}, {Martin}, {Masi}, {Montier}, {Natoli}, {Ristorcelli}, {Thompson},
  {Traficante}, \& {Molinari}}]{paradis_2010.spectral.index.variations}
{Paradis}, D., {Veneziani}, M., {Noriega-Crespo}, A., {et~al.} 2010, \aap, 520,
  L8

\bibitem[{{Payne} {et~al.}(1980){Payne}, {Salpeter}, \&
  {Terzian}}]{payne_1980.doppler.temperature}
{Payne}, H.~E., {Salpeter}, E.~E., \& {Terzian}, Y. 1980, \apj, 240, 499

\bibitem[{{Peek} \& {Schiminovich}(2013)}]{peek_stanimirovic_2013}
{Peek}, J.~E.~G. \& {Schiminovich}, D. 2013, \apj, 771, 68

\bibitem[{{Pelgrims} {et~al.}(2020){Pelgrims}, {Ferri{\`e}re}, {Boulanger},
  {Lallement}, \& {Montier}}]{pelgrims_2020.local.bubble.shape}
{Pelgrims}, V., {Ferri{\`e}re}, K., {Boulanger}, F., {Lallement}, R., \&
  {Montier}, L. 2020, \aap, 636, A17

\bibitem[{{Pineda} {et~al.}(2008){Pineda}, {Caselli}, \&
  {Goodman}}]{pineda_2008.perseus.XCO.variations}
{Pineda}, J.~E., {Caselli}, P., \& {Goodman}, A.~A. 2008, \apj, 679, 481

\bibitem[{{Pineda} {et~al.}(2010){Pineda}, {Goldsmith}, {Chapman}, {Snell},
  {Li}, {Cambr{\'e}sy}, \& {Brunt}}]{pineda_2010}
{Pineda}, J.~L., {Goldsmith}, P.~F., {Chapman}, N., {et~al.} 2010, \apj, 721,
  686

\bibitem[{{Pineda} {et~al.}(2017){Pineda}, {Langer}, {Goldsmith}, {Horiuchi},
  {Kuiper}, {Muller}, {Hughes}, {Ott}, {Requena-Torres}, {Velusamy}, \&
  {Wong}}]{pineda_2017}
{Pineda}, J.~L., {Langer}, W.~D., {Goldsmith}, P.~F., {et~al.} 2017, \apj, 839,
  107

\bibitem[{{Pineda} {et~al.}(2013){Pineda}, {Langer}, {Velusamy}, \&
  {Goldsmith}}]{pineda_2013}
{Pineda}, J.~L., {Langer}, W.~D., {Velusamy}, T., \& {Goldsmith}, P.~F. 2013,
  \aap, 554, A103

\bibitem[{{Planck Collaboration} {et~al.}(2014){Planck Collaboration},
  {Abergel}, {Ade}, {Aghanim}, {Alves}, {Aniano}, {Armitage-Caplan}, {Arnaud},
  {Ashdown}, {Atrio-Barandela}, {Aumont}, {Baccigalupi}, {Banday}, {Barreiro},
  {Bartlett}, {Battaner}, {Benabed}, {Beno{\^\i}t}, {Benoit-L{\'e}vy},
  {Bernard}, {Bersanelli}, {Bielewicz}, {Bobin}, {Bock}, {Bonaldi}, {Bond},
  {Borrill}, {Bouchet}, {Boulanger}, {Bridges}, {Bucher}, {Burigana}, {Butler},
  {Cardoso}, {Catalano}, {Chamballu}, {Chary}, {Chiang}, {Chiang},
  {Christensen}, {Church}, {Clemens}, {Clements}, {Colombi}, {Colombo},
  {Combet}, {Couchot}, {Coulais}, {Crill}, {Curto}, {Cuttaia}, {Danese},
  {Davies}, {Davis}, {de Bernardis}, {de Rosa}, {de Zotti}, {Delabrouille},
  {Delouis}, {D{\'e}sert}, {Dickinson}, {Diego}, {Dole}, {Donzelli},
  {Dor{\'e}}, {Douspis}, {Draine}, {Dupac}, {Efstathiou}, {En{\ss}lin},
  {Eriksen}, {Falgarone}, {Finelli}, {Forni}, {Frailis}, {Fraisse},
  {Franceschi}, {Galeotta}, {Ganga}, {Ghosh}, {Giard}, {Giardino},
  {Giraud-H{\'e}raud}, {Gonz{\'a}lez-Nuevo}, {G{\'o}rski}, {Gratton},
  {Gregorio}, {Grenier}, {Gruppuso}, {Guillet}, {Hansen}, {Hanson}, {Harrison},
  {Helou}, {Henrot-Versill{\'e}}, {Hern{\'a}ndez-Monteagudo}, {Herranz},
  {Hildebrandt}, {Hivon}, {Hobson}, {Holmes}, {Hornstrup}, {Hovest},
  {Huffenberger}, {Jaffe}, {Jaffe}, {Jewell}, {Joncas}, {Jones}, {Juvela},
  {Keih{\"a}nen}, {Keskitalo}, {Kisner}, {Knoche}, {Knox}, {Kunz},
  {Kurki-Suonio}, {Lagache}, {L{\"a}hteenm{\"a}ki}, {Lamarre}, {Lasenby},
  {Laureijs}, {Lawrence}, {Leonardi}, {Le{\'o}n-Tavares}, {Lesgourgues},
  {Levrier}, {Liguori}, {Lilje}, {Linden-V{\o}rnle}, {L{\'o}pez-Caniego},
  {Lubin}, {Mac{\'\i}as-P{\'e}rez}, {Maffei}, {Maino}, {Mandolesi}, {Maris},
  {Marshall}, {Martin}, {Mart{\'\i}nez-Gonz{\'a}lez}, {Masi}, {Massardi},
  {Matarrese}, {Matthai}, {Mazzotta}, {McGehee}, {Melchiorri}, {Mendes},
  {Mennella}, {Migliaccio}, {Mitra}, {Miville-Desch{\^e}nes}, {Moneti},
  {Montier}, {Morgante}, {Mortlock}, {Munshi}, {Murphy}, {Naselsky}, {Nati},
  {Natoli}, {Netterfield}, {N{\o}rgaard-Nielsen}, {Noviello}, {Novikov},
  {Novikov}, {Osborne}, {Oxborrow}, {Paci}, {Pagano}, {Pajot}, {Paladini},
  {Paoletti}, {Pasian}, {Patanchon}, {Perdereau}, {Perotto}, {Perrotta},
  {Piacentini}, {Piat}, {Pierpaoli}, {Pietrobon}, {Plaszczynski},
  {Pointecouteau}, {Polenta}, {Ponthieu}, {Popa}, {Poutanen}, {Pratt},
  {Pr{\'e}zeau}, {Prunet}, {Puget}, {Rachen}, {Reach}, {Rebolo}, {Reinecke},
  {Remazeilles}, {Renault}, {Ricciardi}, {Riller}, {Ristorcelli}, {Rocha},
  {Rosset}, {Roudier}, {Rowan-Robinson}, {Rubi{\~n}o-Mart{\'\i}n}, {Rusholme},
  {Sandri}, {Santos}, {Savini}, {Scott}, {Seiffert}, {Shellard}, {Spencer},
  {Starck}, {Stolyarov}, {Stompor}, {Sudiwala}, {Sunyaev}, {Sureau}, {Sutton},
  {Suur-Uski}, {Sygnet}, {Tauber}, {Tavagnacco}, {Terenzi}, {Toffolatti},
  {Tomasi}, {Tristram}, {Tucci}, {Tuovinen}, {T{\"u}rler}, {Umana},
  {Valenziano}, {Valiviita}, {Van Tent}, {Verstraete}, {Vielva}, {Villa},
  {Vittorio}, {Wade}, {Wandelt}, {Welikala}, {Ysard}, {Yvon}, {Zacchei}, \&
  {Zonca}}]{planck_collaboration_2014.all_sky.dust.model.miville}
{Planck Collaboration}, {Abergel}, A., {Ade}, P.~A.~R., {et~al.} 2014, \aap,
  571, A11

\bibitem[{{Planck Collaboration} {et~al.}(2011{\natexlab{a}}){Planck
  Collaboration}, {Abergel}, {Ade}, {Aghanim}, {Arnaud}, {Ashdown}, {Aumont},
  {Baccigalupi}, {Balbi}, {Banday}, {Barreiro}, {Bartlett}, {Battaner},
  {Benabed}, {Beno{\^\i}t}, {Bernard}, {Bersanelli}, {Bhatia}, {Blagrave},
  {Bock}, {Bonaldi}, {Bond}, {Borrill}, {Bouchet}, {Boulanger}, {Bucher},
  {Burigana}, {Cabella}, {Cantalupo}, {Cardoso}, {Catalano}, {Cay{\'o}n},
  {Challinor}, {Chamballu}, {Chiang}, {Chiang}, {Christensen}, {Clements},
  {Colombi}, {Couchot}, {Coulais}, {Crill}, {Cuttaia}, {Danese}, {Davies},
  {Davis}, {de Bernardis}, {de Gasperis}, {de Rosa}, {de Zotti},
  {Delabrouille}, {Delouis}, {D{\'e}sert}, {Dickinson}, {Donzelli}, {Dor{\'e}},
  {D{\"o}rl}, {Douspis}, {Dupac}, {Efstathiou}, {En{\ss}lin}, {Eriksen},
  {Finelli}, {Forni}, {Frailis}, {Franceschi}, {Galeotta}, {Ganga}, {Giard},
  {Giardino}, {Giraud-H{\'e}raud}, {Gonz{\'a}lez-Nuevo}, {G{\'o}rski},
  {Gratton}, {Gregorio}, {Gruppuso}, {Hansen}, {Harrison}, {Helou},
  {Henrot-Versill{\'e}}, {Herranz}, {Hildebrandt}, {Hivon}, {Hobson}, {Holmes},
  {Hovest}, {Hoyland}, {Huffenberger}, {Jaffe}, {Joncas}, {Jones}, {Jones},
  {Juvela}, {Keih{\"a}nen}, {Keskitalo}, {Kisner}, {Kneissl}, {Knox},
  {Kurki-Suonio}, {Lagache}, {Lamarre}, {Lasenby}, {Laureijs}, {Lawrence},
  {Leach}, {Leonardi}, {Leroy}, {Linden-V{\o}rnle}, {Lockman},
  {L{\'o}pez-Caniego}, {Lubin}, {Mac{\'\i}as-P{\'e}rez}, {MacTavish}, {Maffei},
  {Maino}, {Mandolesi}, {Mann}, {Maris}, {Marshall}, {Martin},
  {Mart{\'\i}nez-Gonz{\'a}lez}, {Masi}, {Matarrese}, {Matthai}, {Mazzotta},
  {McGehee}, {Meinhold}, {Melchiorri}, {Mendes}, {Mennella},
  {Miville-Desch{\^e}nes}, {Moneti}, {Montier}, {Morgante}, {Mortlock},
  {Munshi}, {Murphy}, {Naselsky}, {Nati}, {Natoli}, {Netterfield},
  {N{\o}rgaard-Nielsen}, {Noviello}, {Novikov}, {Novikov}, {O'Dwyer},
  {Osborne}, {Pajot}, {Paladini}, {Pasian}, {Patanchon}, {Perdereau},
  {Perotto}, {Perrotta}, {Piacentini}, {Piat}, {Pinheiro Gon{\c{c}}alves},
  {Plaszczynski}, {Pointecouteau}, {Polenta}, {Ponthieu}, {Poutanen},
  {Pr{\'e}zeau}, {Prunet}, {Puget}, {Rachen}, {Reach}, {Reinecke}, {Renault},
  {Ricciardi}, {Riller}, {Ristorcelli}, {Rocha}, {Rosset}, {Rowan-Robinson},
  {Rubi{\~n}o-Mart{\'\i}n}, {Rusholme}, {Sandri}, {Santos}, {Savini}, {Scott},
  {Seiffert}, {Shellard}, {Smoot}, {Starck}, {Stivoli}, {Stolyarov}, {Stompor},
  {Sudiwala}, {Sygnet}, {Tauber}, {Terenzi}, {Toffolatti}, {Tomasi}, {Torre},
  {Tristram}, {Tuovinen}, {Umana}, {Valenziano}, {Vielva}, {Villa}, {Vittorio},
  {Wade}, {Wandelt}, {Wilkinson}, {Yvon}, {Zacchei}, \&
  {Zonca}}]{planck_collaboration_2011.diffuse.ISM.properties.molecular.hdyrogen}
{Planck Collaboration}, {Abergel}, A., {Ade}, P.~A.~R., {et~al.}
  2011{\natexlab{a}}, \aap, 536, A24

\bibitem[{{Planck Collaboration} {et~al.}(2016){Planck Collaboration}, {Ade},
  {Aghanim}, {Alves}, {Arnaud}, {Arzoumanian}, {Ashdown}, {Aumont},
  {Baccigalupi}, {Band ay}, {Barreiro}, {Bartolo}, {Battaner}, {Benabed},
  {Beno{\^\i}t}, {Benoit-L{\'e}vy}, {Bernard}, {Bersanelli}, {Bielewicz},
  {Bock}, {Bonavera}, {Bond}, {Borrill}, {Bouchet}, {Boulanger}, {Bracco},
  {Burigana}, {Calabrese}, {Cardoso}, {Catalano}, {Chiang}, {Christensen},
  {Colombo}, {Combet}, {Couchot}, {Crill}, {Curto}, {Cuttaia}, {Danese},
  {Davies}, {Davis}, {de Bernardis}, {de Rosa}, {de Zotti}, {Delabrouille},
  {Dickinson}, {Diego}, {Dole}, {Donzelli}, {Dor{\'e}}, {Douspis}, {Ducout},
  {Dupac}, {Efstathiou}, {Elsner}, {En{\ss}lin}, {Eriksen},
  {Falceta-Gon{\c{c}}alves}, {Falgarone}, {Ferri{\`e}re}, {Finelli}, {Forni},
  {Frailis}, {Fraisse}, {Franceschi}, {Frejsel}, {Galeotta}, {Galli}, {Ganga},
  {Ghosh}, {Giard}, {Gjerl{\o}w}, {Gonz{\'a}lez-Nuevo}, {G{\'o}rski},
  {Gregorio}, {Gruppuso}, {Gudmundsson}, {Guillet}, {Harrison}, {Helou},
  {Hennebelle}, {Henrot-Versill{\'e}}, {Hern{\'a}ndez-Monteagudo}, {Herranz},
  {Hildebrand t}, {Hivon}, {Holmes}, {Hornstrup}, {Huffenberger}, {Hurier},
  {Jaffe}, {Jaffe}, {Jones}, {Juvela}, {Keih{\"a}nen}, {Keskitalo}, {Kisner},
  {Knoche}, {Kunz}, {Kurki-Suonio}, {Lagache}, {Lamarre}, {Lasenby},
  {Lattanzi}, {Lawrence}, {Leonardi}, {Levrier}, {Liguori}, {Lilje},
  {Linden-V{\o}rnle}, {L{\'o}pez-Caniego}, {Lubin}, {Mac{\'\i}as-P{\'e}rez},
  {Maino}, {Mandolesi}, {Mangilli}, {Maris}, {Martin},
  {Mart{\'\i}nez-Gonz{\'a}lez}, {Masi}, {Matarrese}, {Melchiorri}, {Mendes},
  {Mennella}, {Migliaccio}, {Miville-Desch{\^e}nes}, {Moneti}, {Montier},
  {Morgante}, {Mortlock}, {Munshi}, {Murphy}, {Naselsky}, {Nati},
  {Netterfield}, {Noviello}, {Novikov}, {Novikov}, {Oppermann}, {Oxborrow},
  {Pagano}, {Pajot}, {Paladini}, {Paoletti}, {Pasian}, {Perotto}, {Pettorino},
  {Piacentini}, {Piat}, {Pierpaoli}, {Pietrobon}, {Plaszczynski},
  {Pointecouteau}, {Polenta}, {Ponthieu}, {Pratt}, {Prunet}, {Puget}, {Rachen},
  {Reinecke}, {Remazeilles}, {Renault}, {Renzi}, {Ristorcelli}, {Rocha},
  {Rossetti}, {Roudier}, {Rubi{\~n}o-Mart{\'\i}n}, {Rusholme}, {Sandri},
  {Santos}, {Savelainen}, {Savini}, {Scott}, {Soler}, {Stolyarov}, {Sudiwala},
  {Sutton}, {Suur-Uski}, {Sygnet}, {Tauber}, {Terenzi}, {Toffolatti}, {Tomasi},
  {Tristram}, {Tucci}, {Umana}, {Valenziano}, {Valiviita}, {Van Tent},
  {Vielva}, {Villa}, {Wade}, {Wandelt}, {Wehus}, {Ysard}, {Yvon}, \&
  {Zonca}}]{planck_collaboration_2016}
{Planck Collaboration}, {Ade}, P.~A.~R., {Aghanim}, N., {et~al.} 2016, \aap,
  586, A138

\bibitem[{{Planck Collaboration} {et~al.}(2011{\natexlab{b}}){Planck
  Collaboration}, {Ade}, {Aghanim}, {Arnaud}, {Ashdown}, {Aumont},
  {Baccigalupi}, {Balbi}, {Banday}, {Barreiro}, {Bartlett}, {Battaner},
  {Benabed}, {Beno{\^\i}t}, {Bernard}, {Bersanelli}, {Bhatia}, {Bock},
  {Bonaldi}, {Bond}, {Borrill}, {Bouchet}, {Boulanger}, {Bucher}, {Burigana},
  {Cabella}, {Cardoso}, {Catalano}, {Cay{\'o}n}, {Challinor}, {Chamballu},
  {Chiang}, {Chiang}, {Christensen}, {Clements}, {Colombi}, {Couchot},
  {Coulais}, {Crill}, {Cuttaia}, {Dame}, {Danese}, {Davies}, {Davis}, {de
  Bernardis}, {de Gasperis}, {de Rosa}, {de Zotti}, {Delabrouille}, {Delouis},
  {D{\'e}sert}, {Dickinson}, {Dobashi}, {Donzelli}, {Dor{\'e}}, {D{\"o}rl},
  {Douspis}, {Dupac}, {Efstathiou}, {En{\ss}lin}, {Eriksen}, {Falgarone},
  {Finelli}, {Forni}, {Fosalba}, {Frailis}, {Franceschi}, {Fukui}, {Galeotta},
  {Ganga}, {Giard}, {Giardino}, {Giraud-H{\'e}raud}, {Gonz{\'a}lez-Nuevo},
  {G{\'o}rski}, {Gratton}, {Gregorio}, {Grenier}, {Gruppuso}, {Hansen},
  {Harrison}, {Helou}, {Henrot-Versill{\'e}}, {Herranz}, {Hildebrandt},
  {Hivon}, {Hobson}, {Holmes}, {Hovest}, {Hoyland}, {Huffenberger}, {Jaffe},
  {Jones}, {Juvela}, {Kawamura}, {Keih{\"a}nen}, {Keskitalo}, {Kisner},
  {Kneissl}, {Knox}, {Kurki-Suonio}, {Lagache}, {Lamarre}, {Lasenby},
  {Laureijs}, {Lawrence}, {Leach}, {Leonardi}, {Leroy}, {Lilje},
  {Linden-V{\o}rnle}, {L{\'o}pez-Caniego}, {Lubin}, {Mac{\'\i}as-P{\'e}rez},
  {MacTavish}, {Maffei}, {Maino}, {Mandolesi}, {Mann}, {Maris}, {Martin},
  {Mart{\'\i}nez-Gonz{\'a}lez}, {Masi}, {Matarrese}, {Matthai}, {Mazzotta},
  {McGehee}, {Meinhold}, {Melchiorri}, {Mendes}, {Mennella},
  {Miville-Desch{\^e}nes}, {Moneti}, {Montier}, {Morgante}, {Mortlock},
  {Munshi}, {Murphy}, {Naselsky}, {Natoli}, {Netterfield},
  {N{\o}rgaard-Nielsen}, {Noviello}, {Novikov}, {Novikov}, {O'Dwyer}, {Onishi},
  {Osborne}, {Pajot}, {Paladini}, {Paradis}, {Pasian}, {Patanchon},
  {Perdereau}, {Perotto}, {Perrotta}, {Piacentini}, {Piat}, {Plaszczynski},
  {Pointecouteau}, {Polenta}, {Ponthieu}, {Poutanen}, {Pr{\'e}zeau}, {Prunet},
  {Puget}, {Reach}, {Reinecke}, {Renault}, {Ricciardi}, {Riller},
  {Ristorcelli}, {Rocha}, {Rosset}, {Rowan-Robinson}, {Rubi{\~n}o-Mart{\'\i}n},
  {Rusholme}, {Sandri}, {Santos}, {Savini}, {Scott}, {Seiffert}, {Shellard},
  {Smoot}, {Starck}, {Stivoli}, {Stolyarov}, {Stompor}, {Sudiwala}, {Sygnet},
  {Tauber}, {Terenzi}, {Toffolatti}, {Tomasi}, {Torre}, {Tristram}, {Tuovinen},
  {Umana}, {Valenziano}, {Vielva}, {Villa}, {Vittorio}, {Wade}, {Wandelt},
  {Wilkinson}, {Yvon}, {Zacchei}, \&
  {Zonca}}]{planck_2011.dark.gas.estiamtes.bernard}
{Planck Collaboration}, {Ade}, P.~A.~R., {Aghanim}, N., {et~al.}
  2011{\natexlab{b}}, \aap, 536, A19

\bibitem[{{Planck Collaboration} {et~al.}(2020){Planck Collaboration},
  {Aghanim}, {Akrami}, {Alves}, {Ashdown}, {Aumont}, {Baccigalupi},
  {Ballardini}, {Banday}, {Barreiro}, {Bartolo}, {Basak}, {Benabed}, {Bernard},
  {Bersanelli}, {Bielewicz}, {Bock}, {Bond}, {Borrill}, {Bouchet}, {Boulanger},
  {Bracco}, {Bucher}, {Burigana}, {Calabrese}, {Cardoso}, {Carron}, {Chary},
  {Chiang}, {Colombo}, {Combet}, {Crill}, {Cuttaia}, {de Bernardis}, {de
  Zotti}, {Delabrouille}, {Delouis}, {Di Valentino}, {Dickinson}, {Diego},
  {Dor{\'e}}, {Douspis}, {Ducout}, {Dupac}, {Efstathiou}, {Elsner},
  {En{\ss}lin}, {Eriksen}, {Falgarone}, {Fantaye}, {Fernandez-Cobos},
  {Ferri{\`e}re}, {Finelli}, {Forastieri}, {Frailis}, {Fraisse}, {Franceschi},
  {Frolov}, {Galeotta}, {Galli}, {Ganga}, {G{\'e}nova-Santos}, {Gerbino},
  {Ghosh}, {Gonz{\'a}lez-Nuevo}, {G{\'o}rski}, {Gratton}, {Green}, {Gruppuso},
  {Gudmundsson}, {Guillet}, {Handley}, {Hansen}, {Helou}, {Herranz}, {Hivon},
  {Huang}, {Jaffe}, {Jones}, {Keih{\"a}nen}, {Keskitalo}, {Kiiveri}, {Kim},
  {Krachmalnicoff}, {Kunz}, {Kurki-Suonio}, {Lagache}, {Lamarre}, {Lasenby},
  {Lattanzi}, {Lawrence}, {Le Jeune}, {Levrier}, {Liguori}, {Lilje},
  {Lindholm}, {L{\'o}pez-Caniego}, {Lubin}, {Ma}, {Mac{\'\i}as-P{\'e}rez},
  {Maggio}, {Maino}, {Mandolesi}, {Mangilli}, {Marcos-Caballero}, {Maris},
  {Martin}, {Mart{\'\i}nez-Gonz{\'a}lez}, {Matarrese}, {Mauri}, {McEwen},
  {Melchiorri}, {Mennella}, {Migliaccio}, {Miville-Desch{\^e}nes}, {Molinari},
  {Moneti}, {Montier}, {Morgante}, {Moss}, {Natoli}, {Pagano}, {Paoletti},
  {Patanchon}, {Perrotta}, {Pettorino}, {Piacentini}, {Polastri}, {Polenta},
  {Puget}, {Rachen}, {Reinecke}, {Remazeilles}, {Renzi}, {Ristorcelli},
  {Rocha}, {Rosset}, {Roudier}, {Rubi{\~n}o-Mart{\'\i}n}, {Ruiz-Granados},
  {Salvati}, {Sandri}, {Savelainen}, {Scott}, {Sirignano}, {Sunyaev},
  {Suur-Uski}, {Tauber}, {Tavagnacco}, {Tenti}, {Toffolatti}, {Tomasi},
  {Trombetti}, {Valiviita}, {Vansyngel}, {Van Tent}, {Vielva}, {Villa},
  {Vittorio}, {Wandelt}, {Wehus}, {Zacchei}, \& {Zonca}}]{planck_coll_2020_ebv}
{Planck Collaboration}, {Aghanim}, N., {Akrami}, Y., {et~al.} 2020, \aap, 641,
  A12

\bibitem[{{Rachford} {et~al.}(2009){Rachford}, {Snow}, {Destree}, {Ross},
  {Ferlet}, {Friedman}, {Gry}, {Jenkins}, {Morton}, {Savage}, {Shull},
  {Sonnentrucker}, {Tumlinson}, {Vidal-Madjar}, {Welty}, \&
  {York}}]{2009ApJS_Rachford}
{Rachford}, B.~L., {Snow}, T.~P., {Destree}, J.~D., {et~al.} 2009, \apjs, 180,
  125

\bibitem[{{Rachford} {et~al.}(2002){Rachford}, {Snow}, {Tumlinson}, {Shull},
  {Blair}, {Ferlet}, {Friedman}, {Gry}, {Jenkins}, {Morton}, {Savage},
  {Sonnentrucker}, {Vidal-Madjar}, {Welty}, \& {York}}]{rachford_2002}
{Rachford}, B.~L., {Snow}, T.~P., {Tumlinson}, J., {et~al.} 2002, \apj, 577,
  221

\bibitem[{{Rieke} {et~al.}(2015){Rieke}, {Wright}, {B{\"o}ker}, {Bouwman},
  {Colina}, {Glasse}, {Gordon}, {Greene}, {G{\"u}del}, {Henning}, {Justtanont},
  {Lagage}, {Meixner}, {N{\o}rgaard-Nielsen}, {Ray}, {Ressler}, {van Dishoeck},
  \& {Waelkens}}]{rieke_2015.miri.intro}
{Rieke}, G.~H., {Wright}, G.~S., {B{\"o}ker}, T., {et~al.} 2015, \pasp, 127,
  584

\bibitem[{{Riener} {et~al.}(2019){Riener}, {Kainulainen}, {Henshaw}, {Orkisz},
  {Murray}, \& {Beuther}}]{gausspy_2019}
{Riener}, M., {Kainulainen}, J., {Henshaw}, J.~D., {et~al.} 2019, \aap, 628,
  A78

\bibitem[{{Ripple} {et~al.}(2013){Ripple}, {Heyer}, {Gutermuth}, {Snell}, \&
  {Brunt}}]{ripple_heyer_2013.XCO.variations}
{Ripple}, F., {Heyer}, M.~H., {Gutermuth}, R., {Snell}, R.~L., \& {Brunt},
  C.~M. 2013, \mnras, 431, 1296

\bibitem[{{R{\"o}hser} {et~al.}(2016{\natexlab{a}}){R{\"o}hser}, {Kerp}, {Ben
  Bekhti}, \& {Winkel}}]{rohser_2016.molecular.gas.IVCs}
{R{\"o}hser}, T., {Kerp}, J., {Ben Bekhti}, N., \& {Winkel}, B.
  2016{\natexlab{a}}, \aap, 592, A142

\bibitem[{{R{\"o}hser} {et~al.}(2016{\natexlab{b}}){R{\"o}hser}, {Kerp},
  {Lenz}, \& {Winkel}}]{rohser_2016.all.sky.molecular.IVCs}
{R{\"o}hser}, T., {Kerp}, J., {Lenz}, D., \& {Winkel}, B. 2016{\natexlab{b}},
  \aap, 596, A94

\bibitem[{{Savage} {et~al.}(1977){Savage}, {Bohlin}, {Drake}, \&
  {Budich}}]{savage_1977}
{Savage}, B.~D., {Bohlin}, R.~C., {Drake}, J.~F., \& {Budich}, W. 1977, \apj,
  216, 291

\bibitem[{{Savage} \&
  {Sembach}(1996)}]{savage_sembach_1996.review.on.ISM.abundances.using.HST}
{Savage}, B.~D. \& {Sembach}, K.~R. 1996, \araa, 34, 279

\bibitem[{{Schlafly} \& {Finkbeiner}(2011)}]{schlafly_finkbeiner_2011}
{Schlafly}, E.~F. \& {Finkbeiner}, D.~P. 2011, \apj, 737, 103

\bibitem[{{Schlafly} {et~al.}(2010){Schlafly}, {Finkbeiner}, {Schlegel},
  {Juri{\'c}}, {Ivezi{\'c}}, {Gibson}, {Knapp}, \&
  {Weaver}}]{schlafly_2010.renormalizing.schlegel.map}
{Schlafly}, E.~F., {Finkbeiner}, D.~P., {Schlegel}, D.~J., {et~al.} 2010, \apj,
  725, 1175

\bibitem[{{Schlafly} {et~al.}(2014){Schlafly}, {Green}, {Finkbeiner}, {Rix},
  {Bell}, {Burgett}, {Chambers}, {Draper}, {Hodapp}, {Kaiser}, {Magnier},
  {Martin}, {Metcalfe}, {Price}, \&
  {Tonry}}]{schlafly_2014.cloud.distances.panstarrs.photometry}
{Schlafly}, E.~F., {Green}, G., {Finkbeiner}, D.~P., {et~al.} 2014, \apj, 786,
  29

\bibitem[{{Schlegel} {et~al.}(1998){Schlegel}, {Finkbeiner}, \&
  {Davis}}]{schlegel_1998}
{Schlegel}, D.~J., {Finkbeiner}, D.~P., \& {Davis}, M. 1998, \apj, 500, 525

\bibitem[{{Schnee} {et~al.}(2006){Schnee}, {Bethell}, \&
  {Goodman}}]{schnee_2006.NH2.using.dust.far.infrared}
{Schnee}, S., {Bethell}, T., \& {Goodman}, A. 2006, \apjl, 640, L47

\bibitem[{{Seifried} {et~al.}(2020){Seifried}, {Haid}, {Walch}, {Borchert}, \&
  {Bisbas}}]{seifried_2020.SILC.CO_dark.properties}
{Seifried}, D., {Haid}, S., {Walch}, S., {Borchert}, E.~M.~A., \& {Bisbas},
  T.~G. 2020, \mnras, 492, 1465

\bibitem[{{Sheffer} {et~al.}(2008){Sheffer}, {Rogers}, {Federman}, {Abel},
  {Gredel}, {Lambert}, \& {Shaw}}]{sheffer_2008}
{Sheffer}, Y., {Rogers}, M., {Federman}, S.~R., {et~al.} 2008, \apj, 687, 1075

\bibitem[{{Shull} {et~al.}(2021){Shull}, {Danforth}, \&
  {Anderson}}]{shull_2021}
{Shull}, J.~M., {Danforth}, C.~W., \& {Anderson}, K.~L. 2021, \apj, 911, 55

\bibitem[{{Shull} \& {Panopoulou}(2023)}]{shull_panopoulou_2023}
{Shull}, J.~M. \& {Panopoulou}, G.~V. 2023, arXiv e-prints, arXiv:2310.12205

\bibitem[{{Shull} {et~al.}(2000){Shull}, {Tumlinson}, {Jenkins}, {Moos},
  {Rachford}, {Savage}, {Sembach}, {Snow}, {Sonneborn}, {York}, {Blair},
  {Green}, {Friedman}, \& {Sahnow}}]{shull_2000.ISM.NH2.using.UV.spectra}
{Shull}, J.~M., {Tumlinson}, J., {Jenkins}, E.~B., {et~al.} 2000, \apjl, 538,
  L73

\bibitem[{{Skalidis} {et~al.}(2022){Skalidis}, {Tassis}, {Panopoulou},
  {Pineda}, {Gong}, {Mandarakas}, {Blinov}, {Kiehlmann}, \&
  {Kypriotakis}}]{skalidis_2022}
{Skalidis}, R., {Tassis}, K., {Panopoulou}, G.~V., {et~al.} 2022, \aap, 665,
  A77

\bibitem[{{Sofia} {et~al.}(2005){Sofia}, {Wolff}, {Rachford}, {Gordon},
  {Clayton}, {Cartledge}, {Martin}, {Draine}, {Mathis}, {Snow}, \&
  {Whittet}}]{sofia_2005.galactic.sightlines}
{Sofia}, U.~J., {Wolff}, M.~J., {Rachford}, B., {et~al.} 2005, \apj, 625, 167

\bibitem[{{Sota} {et~al.}(2014){Sota}, {Ma{\'\i}z Apell{\'a}niz}, {Morrell},
  {Barb{\'a}}, {Walborn}, {Gamen}, {Arias}, \&
  {Alfaro}}]{sota_2014.goss.survey}
{Sota}, A., {Ma{\'\i}z Apell{\'a}niz}, J., {Morrell}, N.~I., {et~al.} 2014,
  \apjs, 211, 10

\bibitem[{{Squire} \& {Hopkins}(2018)}]{squire_hopkins_2018.drag.instablities}
{Squire}, J. \& {Hopkins}, P.~F. 2018, \apjl, 856, L15

\bibitem[{{Sternberg} {et~al.}(2023){Sternberg}, {Bialy}, \&
  {Gurman}}]{sternberg_2023}
{Sternberg}, A., {Bialy}, S., \& {Gurman}, A. 2023, arXiv e-prints,
  arXiv:2308.13889

\bibitem[{Sternberg {et~al.}(2014)Sternberg, Petit, Roueff, \&
  Bourlot}]{Sternberg_2014}
Sternberg, A., Petit, F.~L., Roueff, E., \& Bourlot, J.~L. 2014, The
  Astrophysical Journal, 790, 10

\bibitem[{{Sun} {et~al.}(2020){Sun}, {Leroy}, {Schinnerer}, {Hughes},
  {Rosolowsky}, {Querejeta}, {Schruba}, {Liu}, {Saito}, {Herrera}, {Faesi},
  {Usero}, {Pety}, {Kruijssen}, {Ostriker}, {Bigiel}, {Blanc}, {Bolatto},
  {Boquien}, {Chevance}, {Dale}, {Deger}, {Emsellem}, {Glover}, {Grasha},
  {Groves}, {Henshaw}, {Jimenez-Donaire}, {Kim}, {Klessen}, {Kreckel}, {Lee},
  {Meidt}, {Sandstrom}, {Sardone}, {Utomo}, \&
  {Williams}}]{sun_phangs_2020.molecular.cloud.properties.comparison.with.the.host.galaxies}
{Sun}, J., {Leroy}, A.~K., {Schinnerer}, E., {et~al.} 2020, \apjl, 901, L8

\bibitem[{{Sun} {et~al.}(2018){Sun}, {Leroy}, {Schruba}, {Rosolowsky},
  {Hughes}, {Kruijssen}, {Meidt}, {Schinnerer}, {Blanc}, {Bigiel}, {Bolatto},
  {Chevance}, {Groves}, {Herrera}, {Hygate}, {Pety}, {Querejeta}, {Usero}, \&
  {Utomo}}]{sun_phangs_2018.cloud.scale.properties.of.molecular.clouds.in.15.nearby.galaxies}
{Sun}, J., {Leroy}, A.~K., {Schruba}, A., {et~al.} 2018, \apj, 860, 172

\bibitem[{{Velusamy} {et~al.}(2010){Velusamy}, {Langer}, {Pineda}, {Goldsmith},
  {Li}, \& {Yorke}}]{velusamy_2010}
{Velusamy}, T., {Langer}, W.~D., {Pineda}, J.~L., {et~al.} 2010, \aap, 521, L18

\bibitem[{{Visser} {et~al.}(2009){Visser}, {van Dishoeck}, \&
  {Black}}]{visser_2009}
{Visser}, R., {van Dishoeck}, E.~F., \& {Black}, J.~H. 2009, \aap, 503, 323

\bibitem[{{Wakker}(2006)}]{wakker_2006.UV.survey}
{Wakker}, B.~P. 2006, \apjs, 163, 282

\bibitem[{{Wells} {et~al.}(2015){Wells}, {Pel}, {Glasse}, {Wright},
  {Aitink-Kroes}, {Azzollini}, {Beard}, {Brandl}, {Gallie}, {Geers}, {Glauser},
  {Hastings}, {Henning}, {Jager}, {Justtanont}, {Kruizinga}, {Lahuis}, {Lee},
  {Martinez-Delgado}, {Mart{\'\i}nez-Galarza}, {Meijers}, {Morrison},
  {M{\"u}ller}, {Nakos}, {O'Sullivan}, {Oudenhuysen}, {Parr-Burman}, {Pauwels},
  {Rohloff}, {Schmalzl}, {Sykes}, {Thelen}, {van Dishoeck}, {Vandenbussche},
  {Venema}, {Visser}, {Waters}, \&
  {Wright}}]{wells_2015.medium.res.spectrometer}
{Wells}, M., {Pel}, J.~W., {Glasse}, A., {et~al.} 2015, \pasp, 127, 646

\bibitem[{{Wolfire} {et~al.}(2010){Wolfire}, {Hollenbach}, \&
  {McKee}}]{wolfire_2010.dark.gas.theoretical.PDR.model}
{Wolfire}, M.~G., {Hollenbach}, D., \& {McKee}, C.~F. 2010, \apj, 716, 1191

\bibitem[{{Wouterloot} {et~al.}(1990){Wouterloot}, {Brand}, {Burton}, \&
  {Kwee}}]{wouterloot1990.distribution.galactic.wraps}
{Wouterloot}, J.~G.~A., {Brand}, J., {Burton}, W.~B., \& {Kwee}, K.~K. 1990,
  \aap, 230, 21

\bibitem[{{Wright} {et~al.}(2023){Wright}, {Rieke}, {Glasse}, {Ressler},
  {Garc{\'\i}a Mar{\'\i}n}, {Aguilar}, {Alberts}, {{\'A}lvarez-M{\'a}rquez},
  {Argyriou}, {Banks}, {Baudoz}, {Boccaletti}, {Bouchet}, {Bouwman}, {Brandl},
  {Breda}, {Bright}, {Cale}, {Colina}, {Cossou}, {Coulais}, {Cracraft}, {De
  Meester}, {Dicken}, {Engesser}, {Etxaluze}, {Fox}, {Friedman}, {Fu},
  {Gasman}, {G{\'a}sp{\'a}r}, {Gastaud}, {Geers}, {Glauser}, {Gordon},
  {Greene}, {Greve}, {Grundy}, {G{\"u}del}, {Guillard}, {Haderlein},
  {Hashimoto}, {Henning}, {Hines}, {Holler}, {Detre}, {Jahromi}, {James},
  {Jones}, {Justtanont}, {Kavanagh}, {Kendrew}, {Klaassen}, {Krause},
  {Labiano}, {Lagage}, {Lambros}, {Larson}, {Law}, {Lee}, {Libralato}, {Lorenzo
  Alverez}, {Meixner}, {Morrison}, {Mueller}, {Murray}, {Mycroft}, {Myers},
  {Nayak}, {Naylor}, {Nickson}, {Noriega-Crespo}, {{\"O}stlin}, {O'Sullivan},
  {Ottens}, {Patapis}, {Penanen}, {Pietraszkiewicz}, {Ray}, {Regan},
  {Roteliuk}, {Royer}, {Samara-Ratna}, {Samuelson}, {Sargent}, {Scheithauer},
  {Schneider}, {Schreiber}, {Shaughnessy}, {Sheehan}, {Shivaei}, {Sloan},
  {Tamas}, {Teague}, {Temim}, {Tikkanen}, {Tustain}, {van Dishoeck},
  {Vandenbussche}, {Weilert}, {Whitehouse}, \&
  {Wolff}}]{wright_2023.miri.inflight.performance}
{Wright}, G.~S., {Rieke}, G.~H., {Glasse}, A., {et~al.} 2023, \pasp, 135,
  048003

\bibitem[{{Zagury} {et~al.}(1999){Zagury}, {Boulanger}, \&
  {Banchet}}]{zagury_1999.distance.polaris.flare.reddenning}
{Zagury}, F., {Boulanger}, F., \& {Banchet}, V. 1999, \aap, 352, 645

\bibitem[{{Zhang} {et~al.}(2023){Zhang}, {Yuan}, \&
  {Chen}}]{zhang_2023.Rv.2d.full.sky.map}
{Zhang}, R., {Yuan}, H., \& {Chen}, B. 2023, \apjs, 269, 6

\bibitem[{{Zucker} {et~al.}(2021){Zucker}, {Goodman}, {Alves}, {Bialy}, {Koch},
  {Speagle}, {Foley}, {Finkbeiner}, {Leike}, {En{\ss}lin}, {Peek}, \&
  {Edenhofer}}]{zucker_2021.cloud.distances}
{Zucker}, C., {Goodman}, A., {Alves}, J., {et~al.} 2021, \apj, 919, 35

\end{thebibliography}

\begin{appendix}
\section{Full-sky \NHmolecular\ properties of our Galaxy using the Planck extinction map}
\label{sec:appendix_nh2_different_av_maps}

\begin{figure*}[p]
   \centering
   \includegraphics[width=0.8\hsize]{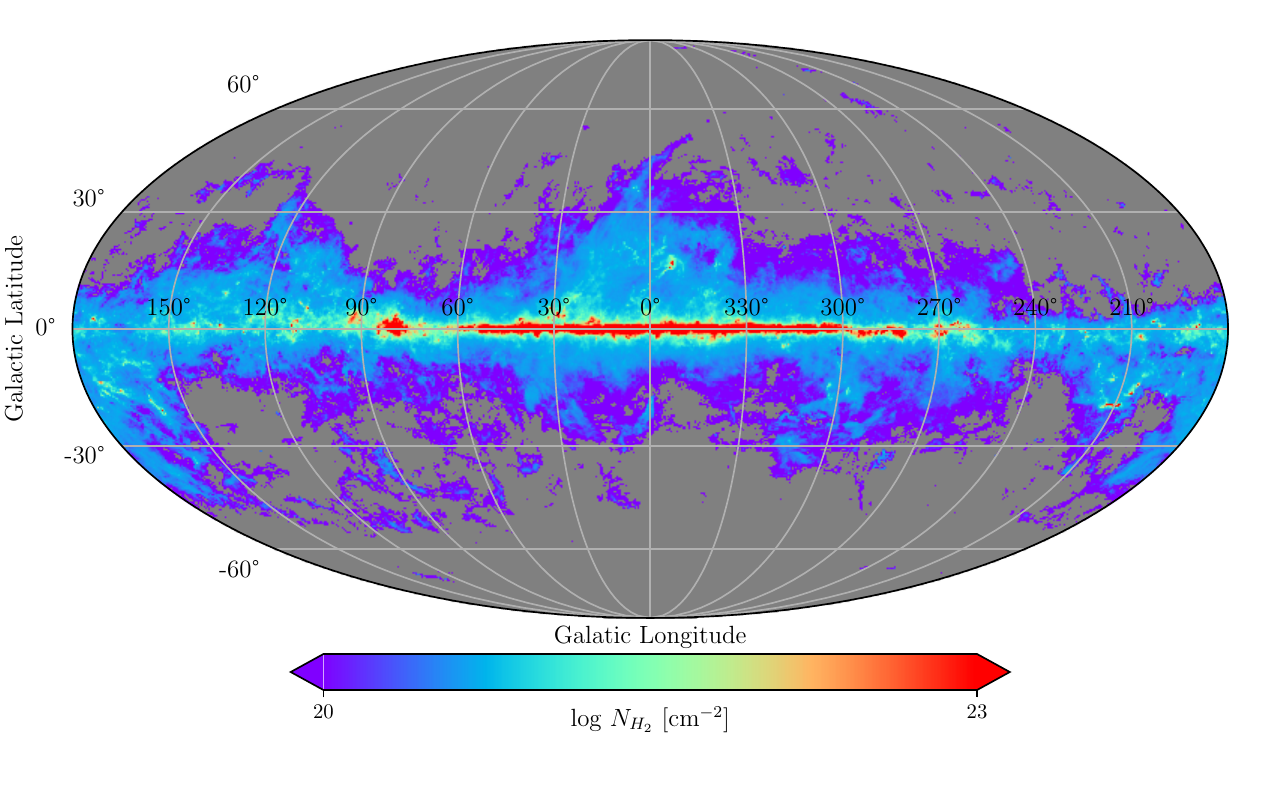}
    \caption{Full-sky \NHmolecular\ map of our Galaxy constructed with the extinction map of \cite{planck_coll_2020_ebv}. Resolution and grid properties are the same as in Fig.~\ref{fig:NH2_XCO_full_sky_maps}.}
    \label{fig:NH2_map_planck}
\end{figure*}

We constructed a full-sky \NHmolecular\ map of our Galaxy using the extinction data of \cite{planck_coll_2020_ebv} (Fig.~\ref{fig:NH2_map_planck}). We compared against the \NHmolecular\ map that we constructed with the extinction data of \cite{schlegel_1998} (Fig.~\ref{fig:NH2_XCO_full_sky_maps}).

Fig.~\ref{fig:NH2_map_planck} shows more extended low-\NHmolecular\ regions at high Galactic latitudes than Fig.~\ref{fig:NH2_XCO_full_sky_maps}. This happens because \Av\ in \cite{planck_coll_2020_ebv} tends to be, overall, larger than in \cite{schlegel_1998}. As we show below, the relatively enhanced \Av\ values of \cite{planck_coll_2020_ebv} may affect some of the estimated molecular gas properties of our Galaxy. 

We use the \NHmolecular\ data shown in Fig.~\ref{fig:NH2_map_planck} to calculate the Galactic average \XCO, and \fhmol. We obtain that $\langle$\XCO$\rangle \approx 4\times 10^{20}$~\ColDens~(K~\kms)$^{-1}$, and $\langle$\fhmol$\rangle \approx 30 \%$, which are $\sim 2$ larger than what we inferred in Sects.~\ref{sec:XCO_comparison}, and~\ref{sec:molecular_gas_properties}, but both consistent with the range of values in the literature. In addition, the estimated sky fraction of CO-dark with respect to the total \Hmol\ becomes $\sim 79\%$, while that of CO-bright $\sim 21\%$ for LOSs with \NHmolecular~$\geq 10^{20}$~\ColDens. In this case the relative abundance of CO-dark \Hmol\ is almost two times larger than what we get from the \cite{schlegel_1998} map (Sect.~\ref{sec:co_dark_H2}). We conclude that differences in the extinction maps can induce at most factor of two deviations in the estimated molecular hydrogen properties.

\end{appendix}
\end{document}